\newcommand\beq{\begin{eqnarray}}
\newcommand\eeq{\end{eqnarray}}

\newcommand\ETmiss{E_T^{\rm miss}}
\newcommand\epsu{\epsilon_{\scriptscriptstyle U}}
\newcommand\epsup{\epsilon'_{\scriptscriptstyle U}}
\newcommand\epsd{\epsilon_{\scriptscriptstyle D}}
\newcommand\epse{\epsilon_{\scriptscriptstyle E}}

\def\sigmabar{\overline\sigma}
\def\lsim{\mathrel{\rlap{\lower4pt\hbox{$\sim$}}
    \raise1pt\hbox{$<$}}}                
\def\gsim{\mathrel{\rlap{\lower4pt\hbox{$\sim$}}
    \raise1pt\hbox{$>$}}}

\def\Mmess{{M_{\rm mess}}}
\def\gev{\, {\rm GeV}}
\def\tev{\, {\rm TeV}}
\def\xfb{\, {\rm fb}^{-1}}
\def\hmass{\sim\!\!125\, {\rm GeV}}
\def\asubsection{\subsection}

\documentclass[
amsmath,
prd,nofootinbib,floatfix,11pt,
]{revtex4}

\allowdisplaybreaks
\usepackage{graphicx}
\usepackage{setspace}

\begin{document}
\renewcommand{\theequation}{\arabic{section}.\arabic{equation}}
\renewcommand{\thefigure}{\arabic{section}.\arabic{figure}}
\renewcommand{\thetable}{\arabic{section}.\arabic{table}}

\begin{flushright}
CERN-TH-2012-147
\end{flushright}
\vspace{-0.4cm}

\title{\Large
Implications of gauge-mediated 
supersymmetry breaking with vector-like quarks and a $\sim$125 GeV Higgs 
boson}

\author{Stephen P.~Martin$^a$ and James D.~Wells$^b$}
\affiliation{
{\it $^{(a)}$Department of Physics, Northern Illinois University, DeKalb 
IL 60115,}  and \\
{\it Fermi National Accelerator Laboratory, P.O. Box 500, Batavia 
IL 60510,}
\\
{\it $^{(b)}$CERN Theoretical Physics (PH-TH), CH-1211 Geneva 23, 
Switzerland}, and \\
{\it MCTP, Department of Physics, University of Michigan, Ann 
Arbor MI 48109}.
}

\begin{abstract}\normalsize \baselineskip=14pt 
We investigate the implications of models that achieve a Standard 
Model-like Higgs boson of mass near 125 GeV by introducing additional 
TeV-scale supermultiplets in the vector-like ${\bf 10}+\overline{\bf 10}$ 
representation of $SU(5)$, within the context of 
gauge-mediated supersymmetry breaking. We study the resulting mass 
spectrum of superpartners, comparing and contrasting to the usual 
gauge-mediated and CMSSM scenarios, and discuss implications for LHC 
supersymmetry searches. This approach implies that exotic vector-like 
fermions $t'_{1,2}$, $b'$,and $\tau'$ should be within the reach of the 
LHC. We discuss the masses, the couplings to electroweak bosons, and the 
decay branching ratios of the exotic fermions, with and without various 
unification assumptions for the mass and mixing parameters. We comment on 
LHC prospects for discovery of the exotic fermion states, both for 
decays that are prompt and non-prompt on detector-crossing time scales.
\end{abstract}


\maketitle

\vspace{-1cm}

\baselineskip=12.0pt

\tableofcontents

\baselineskip=13.9pt

\setcounter{footnote}{1}
\setcounter{figure}{0}
\setcounter{table}{0}

\section{Introduction\label{sec:intro}}
\setcounter{equation}{0}
\setcounter{figure}{0}
\setcounter{table}{0}
\setcounter{footnote}{1}

Recently the ATLAS and CMS experiments at the LHC have put forward data 
analysis results that suggest the Higgs boson mass could be close to 
$125\gev$~\cite{ATLAScombined,CMScombined}. The statistical significance is 
not at the `discovery level', nor is it enough to determine if the 
putative Higgs boson signal is really that of the Standard Model (SM) 
Higgs boson, or some close cousin that may have somewhat different 
couplings and rates. Nevertheless, we wish to investigate the supposition 
that the Higgs boson exists at this mass and is SM-like in its 
couplings.

Stipulating the above, supersymmetry is an ideal theoretical 
framework to accommodate the results. The many favorable features of 
supersymmetry are well-known~\cite{primer}, but the one most applicable 
here is its generic prediction for a SM-like Higgs boson with mass less 
than about $130\gev$.  Within some frameworks of supersymmetry, such as 
`natural' versions of minimal supergravity (mSUGRA) or minimal gauge mediated supersymmetry breaking (GMSB), a 
Higgs mass value of $\hmass$\footnote{In this article, $\hmass$ always 
means any value calculated theoretically to be 
between $122\gev <M_h<128\gev$, consistent with 
the LHC results taking into account experimental uncertainty (notably the lack of a definitive signal) 
as well as 
theoretical errors in calculating the Higgs mass from the 
supersymmetric input parameters.} seems perhaps uncomfortably high.  
Within other frameworks, such as `unnatural' 
PeV-scale supersymmetry \cite{PeVSUSY}
or 
split supersymmetry \cite{splitSUSY}, such a mass value seems perhaps uncomfortably low. 
Nevertheless, almost any approach to supersymmetry allows one to easily 
absorb this Higgs mass into the list of defining data and then present 
the resulting allowed parameter space.

In this article we wish to see how well one can explain a $\hmass$ Higgs 
boson mass using `natural' supersymmetry. There are many good discussions of 
this already present in the literature~\cite{lightstopscenario,otherscenarios}, but the approach we 
take here is to use extra vector-like matter supermultiplets
to raise the Higgs mass \cite{Moroi:1991mg}-\cite{Nakayama:2012zc}. 
As shown in detail in \cite{Martin:2009bg},
the Yukawa coupling of the vector-like quarks to the Higgs
has a fixed point at a value large enough to substantially increase the
lightest Higgs mass while giving a fit to precision electroweak oblique observables
that is as good as, or slightly better than, the SM.
This 
can be done in various different scenarios for the soft terms, but here we choose to investigate
within the context of GMSB; earlier studies of
this can be found in
\cite{Endo:2011mc,Evans:2011uq,Endo:2011xq,Endo:2012rd,Nakayama:2012zc}. The details of the specific
model we study will be discussed in the next section. We like this 
approach because the superpartner masses are not required to become 
extremely heavy to raise the light Higgs mass through large logarithms in 
the radiative corrections, nor does one need to invoke very large Higgs-stop-antistop
supersymmetry-breaking couplings. Instead, the extra vector-like states, 
interacting with the Higgs boson, make extra contributions to the Higgs 
boson mass in a natural way.  This approach has been reemphasized 
recently also by~\cite{Li:2011ab}-\cite{Nakayama:2012zc} within the context of 
the $\hmass$ Higgs 
boson signal, and our study confirms some previous results and extends 
the understanding by investigating correlations within a unified theory 
and detailing the phenomenological implications that can be useful for the 
LHC experiments to confirm or reject this hypothesized explanation for 
the Higgs boson mass value.

\section{Minimal GMSB model with extra vector-like particles\label{sec:model}}
\setcounter{equation}{0}
\setcounter{figure}{0}
\setcounter{table}{0}
\setcounter{footnote}{1}

\asubsection{Theory definition, parameters and spectrum\label{subsec:theorydef}}

The theory under consideration here is 
a minimal GMSB theory with one $SU(5)$
${\bf 5}+\overline{\bf 5}$ messenger multiplet pair, along with a 
${\bf 10}+\overline{\bf 10}$ multiplet pair at the TeV scale.  
We choose this model because it is minimal, illustrates the 
key phenomenological features of this broad class of theories, 
and maintains perturbative gauge coupling unification at the high scale. The same model has also been considered in \cite{Endo:2011mc,Evans:2011uq,Endo:2011xq,Endo:2012rd,Nakayama:2012zc}.
The unification scale (defined as 
the scale where $g_1$ and $g_2$ meet) turns out to be larger than
the corresponding scale in the MSSM by a factor of 2-4, depending on
the sparticle thresholds and the GMSB messenger scale. As in ref.~\cite{Martin:2009bg},
we use 3-loop beta functions for the gauge couplings and gaugino masses, and 2-loop beta functions for all other parameters. These renormalization group equations are not given explicitly here, because they can be obtained in a straightforward and automated way from the general results given in 
refs.~\cite{betas:1,betas:2,betas:3}.

To set the notation, the MSSM fields are defined below 
along with their $SU(3)_C \times SU(2)_L \times U(1)_Y$ quantum numbers:
\beq
&&
q_i = ({\bf 3}, {\bf 2}, 1/6), 
~~~~
\overline{u}_i = ({\bf \overline{3}}, {\bf 1}, -2/3), 
~~~~
\overline{d}_i = ({\bf \overline{3}}, {\bf 1}, 1/3), 
~~~~
\nonumber \\ &&
\ell_i = ({\bf 1}, {\bf 2}, -1/2), 
~~~~
\overline{e}_i = ({\bf 1}, {\bf 1}, 1), 
~~~~
H_u = ({\bf 1}, {\bf 2}, 1/2), 
~~~~
H_d = ({\bf 1}, {\bf 2}, -1/2).
\eeq
with $i = 1,2,3$ denoting the three families.
The MSSM superpotential, in the approximation that only third-family 
Yukawa couplings are included, is
\beq
W_{\rm MSSM} = \mu H_u H_d + y_t H_u q_3 \overline u_3 - y_b H_d q_3 \overline d_3 
- y_\tau H_d \ell_3 \overline e_3.
\label{eq:WMSSM}
\eeq

The ${\bf 10}$ and $\overline{\bf 10}$ $SU(5)$ multiplets 
are comprised of $Q$, $\overline U$, $\overline E$
and $\overline Q$, $U$, $E$ supermultiplets, respectively, with
\beq
&&
Q = ({\bf 3}, {\bf 2}, 1/6), 
~~~~~~
U = ({\bf 3}, {\bf 1}, 2/3), 
~~~~~~
E = ({\bf 1}, {\bf 1}, -1), \\ &&
\overline{Q} = (\overline{\bf 3}, {\bf 2}, -1/6), 
~~~~~~ 
\overline{U} = (\overline{\bf 3}, {\bf 1}, -2/3), 
~~~~~~
\overline{E} = ({\bf 1}, {\bf 1}, 1).
\label{eq:extraslist}
\eeq
These extra fields interact with the MSSM Higgs bosons at the 
renormalizable level. The relevant superpotential is
\beq
W_{QUE} = M_Q Q \overline Q + M_U U \overline U + M_E E \overline E +
k H_u Q \overline U - k' H_d \overline Q U .
\label{eq:WQUE}
\eeq
The extra superfields of the ${\bf 10}+\overline{\bf 10}$ give rise to 
additional exotic particles beyond the MSSM:  charge $+2/3$ quarks 
$t'_{1,2}$ (plus scalar superpartners $\tilde t'_{1,2,3,4}$), a charge 
$-1/3$ quark $b'$ (plus scalar superpartners $\tilde b'_{1,2}$), and a 
charged lepton $\tau'$ (plus scalar superpartners $\tilde \tau'_{1,2}$).

As noted in \cite{Martin:2009bg}, the Yukawa interaction $k$ 
is subject to an infrared-stable quasi-fixed point \cite{Hill:1980sq} slightly above $k=1.0$ at the TeV
scale. This value is both natural (since a large range of high-scale input values
closely approaches it), and is easily large enough to mediate
a correction to the lightest Higgs boson mass $M_h$ that can
accommodate $M_h \hmass$ or larger, depending of course on the other parameters of the theory.
In this paper, we will always assume that $k$ is near its (strongly attractive) 
quasi-fixed point, by arbitrarily taking $k=1$ near the apparent scale of gauge coupling unification and evolving it down. Taking larger values at the high scale would only increase the TeV-scale 
value of $k$ by about 2\% at most, although it should be kept in mind that the contribution to the Higgs squared mass correction scales like $k^4$.
For simplicity, we will take $k'$ to be small, since it does not help to raise the $h^0$ mass,
although a small non-zero value would not affect the results below very much.
The superpartner spectrum of this theory is determined by 
the normal procedures for minimal GMSB. The input parameters needed are 
$\tan\beta$, ${\rm sign}(\mu)$, the mass scale for the 
$\bf{5}+\overline{\bf 5}$ messenger masses $\Mmess$ and the supersymmetry 
breaking transmission scale $\Lambda$ which is equal to $\langle 
F_S\rangle/\langle S\rangle$ where $\langle F_S\rangle$ and $\langle 
S\rangle$ are vacuum expectation values of the $F$-component and scalar 
component of the chiral superfield $S$ that couples directly to the 
messenger sector. Using standard techniques~\cite{primer} one can then 
compute the superpartner spectrum and Higgs boson mass spectrum.
Corrections to the lightest Higgs boson mass $M_h$ are obtained using the full 
one-loop effective potential approximation, as in~\cite{Martin:2009bg}. (We have checked the MSSM 
contributions against FeynHiggs \cite{FeynHiggs}
and we find agreement to within expected uncertainties of 1-2 GeV.)
One-loop corrections to the pole masses of all strongly interacting
particles are also included; these are particularly important for the gluino.

If the exotic states only interacted among themselves and the Higgs 
fields, then a $Z_2$ quantum number could be defined on the 
superpotential with odd assignments to $Q,\overline Q, U, \overline U, E, 
\overline E$ and even assignments for everything else, leading to 
stability of the lightest new fermion state. At the renormalizable level, 
the only way the lightest new quark $t_1'$ and the $\tau'$ can decay is 
by breaking this $Z_2$ symmetry via superpotential
mixing interactions with MSSM states,
\beq
W_{\rm mix} = 
  \epsu H_u q_3 \overline U 
+ \epsup H_u Q \overline u_3 
- \epsd H_d Q \overline d_3 
-\epse H_d\ell_3\overline{E}
,
\label{eq:WQUEmix}
\eeq
where $\epsu$, $\epsup$, $\epsd$, and $\epse$ are new Yukawa couplings. Note that this is consistent
with matter parity provided that the supermultiplets
$Q,\overline Q, U, \overline U, E, \overline E$
are assigned odd matter parity, so that the new fermions have even $R$-parity. 
We 
assume that the mixing Yukawa couplings are confined to the 
third-family MSSM fields $q_3, \overline u_3, \overline d_3, \ell_3$, in order to 
avoid dangerous flavor violating effects; the bounds on 
third-family mixings with new heavy states are much 
less stringent than for first and second-family quarks and 
leptons~\cite{fourthflavor,KPST}. As we will see in section \ref{sec:precisiontests},
couplings less than $0.1$ to third generation quarks and leptons are 
easily small enough to avoid all flavor constraints. Assuming this for simplicity, 
then $\epsu$, $\epsup$, $\epsd$, and $\epse$ are small enough to be neglected in 
wave function renormalizations, and so do not 
contribute to other couplings' renormalization group equations, and
only contribute linearly to their own. Furthermore, their effects on the mass
eigenstates of the new particles
can be treated as small perturbations.

It is interesting to consider the case of $SU(5)$-symmetric interactions near the unification 
scale. If one assigns $H_u$ and $H_d$ to the ${\bf 5}$ and ${\bf \overline{5}}$ 
representations
respectively, and $Q, \overline U, \overline E$ to the ${\bf 10}$ and 
$\overline Q, U, E$ to the ${\bf \overline{10}}$, then one has
\beq
&&M_Q = M_U = M_E,
\label{eq:GUTMrelations}
\\
&&\epsu = \epsup,\qquad\quad \epsd = \epse .
\eeq
at the unification scale. The further unification in $SO(10)$ implies the stronger condition
\beq
\epsu = \epsup = \epsd = \epse .
\label{eq:GUTepsrelations}
\eeq
A logical guess is that
the origin of the masses $M_Q, M_U, M_E$ is similar to that of the MSSM $\mu$ term,
and might occur well below the unification scale. For example, one can imagine that
they arise from 
non-renormalizable superpotential operators like
\beq
W = \frac{1}{M_{P}} S \overline S (\lambda_\mu H_u H_d + \lambda_Q Q \overline Q +
\lambda_U U \overline U + \lambda_E E \overline E),
\eeq
where $S$, $\overline S$ are SM singlet fields (possibly the same) which 
carry a Peccei-Quinn charge and 
get vacuum expectation values (VEVs) at an intermediate scale, as recently
proposed in this context by \cite{Nakayama:2012zc}, giving rise to masses
$\mu = \lambda_\mu\langle S \overline S \rangle/M_P$
and $M_{Q,U,E} = \lambda_{Q,U,E} \langle S \overline S \rangle/M_P$. 
Note that if the dimensionless couplings
$\lambda_{Q,U,E}$ are small,  then their 
renormalization group evolution from the apparent unification
scale down to the scale at which $S,\overline S$ get VEVs is the same as that of the corresponding
masses $M_{Q,U,E}$, depending only on the wavefunction renormalization anomalous dimensions of 
the chiral superfields $Q,\overline Q, U, \overline U, E, \overline E$. 
In this case, it is sensible to evolve the masses as if they were the same at the
scale of apparent gauge coupling unification, based on an assumed unification of the corresponding
superpotential couplings $\lambda_{Q,U,E}$.
Of course, the relations (\ref{eq:GUTMrelations}) and (\ref{eq:GUTepsrelations}) 
are certainly not mandatory. 
The tree-level relations between couplings (or masses) implied by 
GUT groups can be greatly modified by non-renormalizable terms, alternative assignments of the 
Higgs fields, and mixing effects near the GUT scale. However, eqs.~(\ref{eq:GUTMrelations})
and (\ref{eq:GUTepsrelations}) do constitute a plausible and 
useful benchmark case that we will use for some of the explorations in this 
paper. At the TeV scale, typical values obtained from the renormalization group running are then:
\beq
M_Q: M_U : M_E &\approx & 1.8 : 1 : 0.45,
\label{eq:unifiedMF}
\eeq
with some variation at the $<$20\% level due to the choice of GMSB 
messenger scale and $\Lambda$. (The ratios $M_Q/M_U$ and $M_U/M_E$ at the 
TeV scale 
tend to decrease with larger $M_{\rm mess}$ and $\Lambda$.) 
The ratios of mixing couplings also 
exhibit a pattern when the unification condition 
eq.~(\ref{eq:GUTepsrelations}) is assumed, but with a strong dependence 
on the trajectory for $k$. In general one finds $\epsup$ slightly larger 
than $\epsu$, and $\epsd$ larger than $\epsu$ by a factor of 3.5 to 6. In 
the following, we will sometimes consider the typical case
\beq
\epsu : \epsup : \epsd &\approx & 1 : 1.15 : 4.5
\label{eq:unifiedeps}
\eeq
as a benchmark for illustration when considering the branching ratios of $t_1'$ and $b'$.

The model we study here is not the unique extension of GMSB models to 
include vector-like quarks that raise the Higgs mass. One can replace the 
$U+\overline{U}$ fields by $D+\overline D+E +\overline{E}$ fields without
changing the prospects for perturbative gauge coupling unification, as 
discussed in \cite{Martin:2009bg}.
In that case, a Yukawa coupling $H_u \overline Q D$ will raise the Higgs 
mass, and the gross features of the
superpartner mass spectrum will be unchanged. The exotic 
fermions will consist of      
$b_{1,2}'$, $t'$, and $\tau'_{1,2}$, with decays discussed in 
\cite{Martin:2009bg}. This
model is arguably somewhat less motivated, in that it does not have 
complete GUT multiplets. Another variation
replaces the ${\bf 10} + {\bf \overline{10}}$ at the TeV scale by a ${\bf 
5} + {\bf \overline 5} + {\bf 1} +    
{\bf 1} = L +
\overline L + D + \overline D + N + \overline N$, with a
Yukawa coupling $H_u \overline L N$ doing the work of raising the Higgs 
mass. This model has a
larger set of possibilities for the GMSB messenger fields consistent with 
gauge coupling unification.
However, it also results in a much smaller contribution to $M_h$, unless 
one includes a larger hierarchy
between the exotic leptons and their scalar superpartners.
In order to keep the present paper bounded, we will not pursue those 
approaches further here.

\asubsection{Mass spectra for sample models\label{subsec:spectra}}

\begin{figure}[!ptb]
\begin{minipage}[]{0.41\linewidth}
\begin{center}
\includegraphics[height=8.45cm,angle=0]{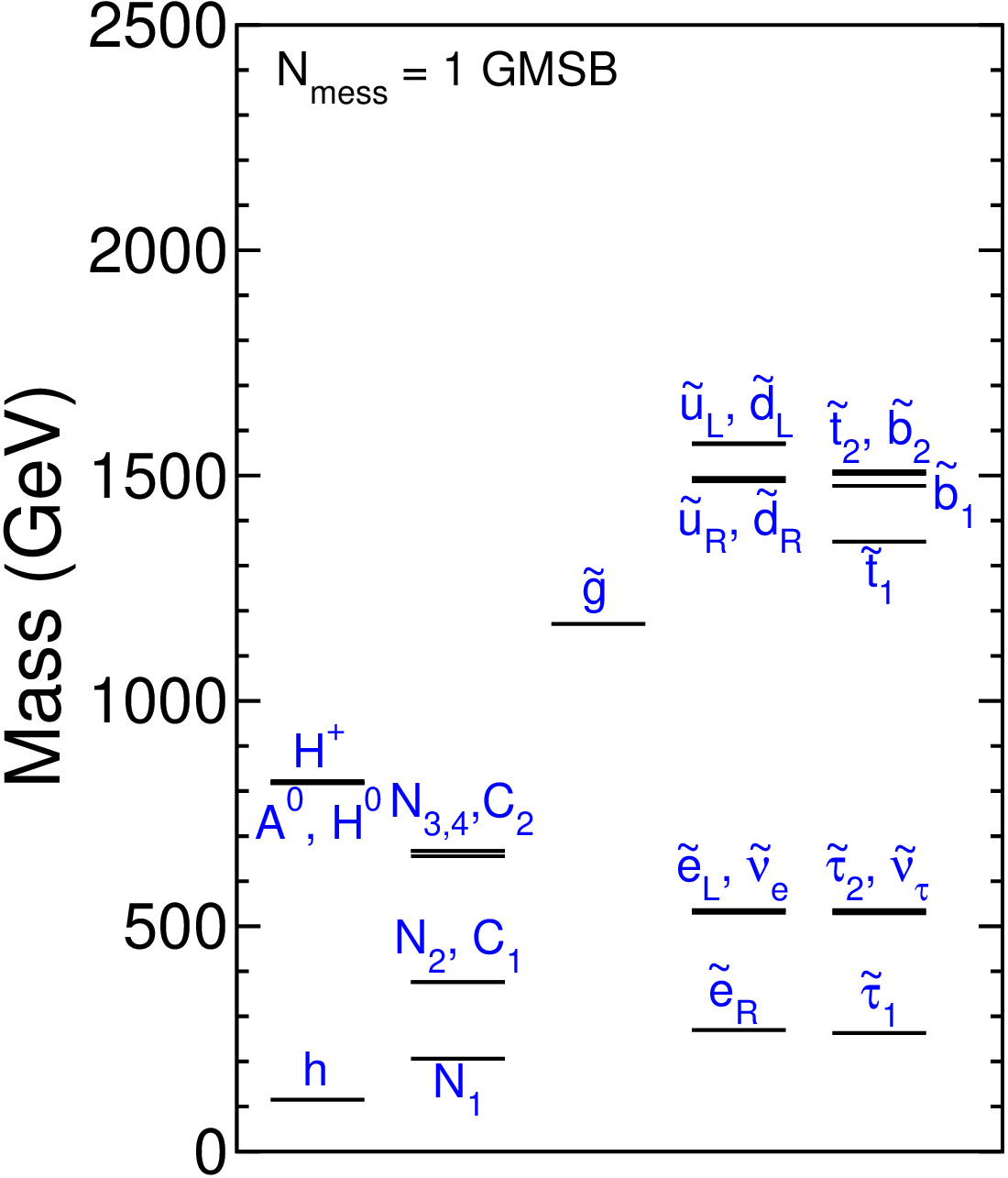}
\end{center}
\end{minipage}
\begin{minipage}[]{0.58\linewidth}
\begin{flushright}
\includegraphics[height=8.45cm,angle=0]{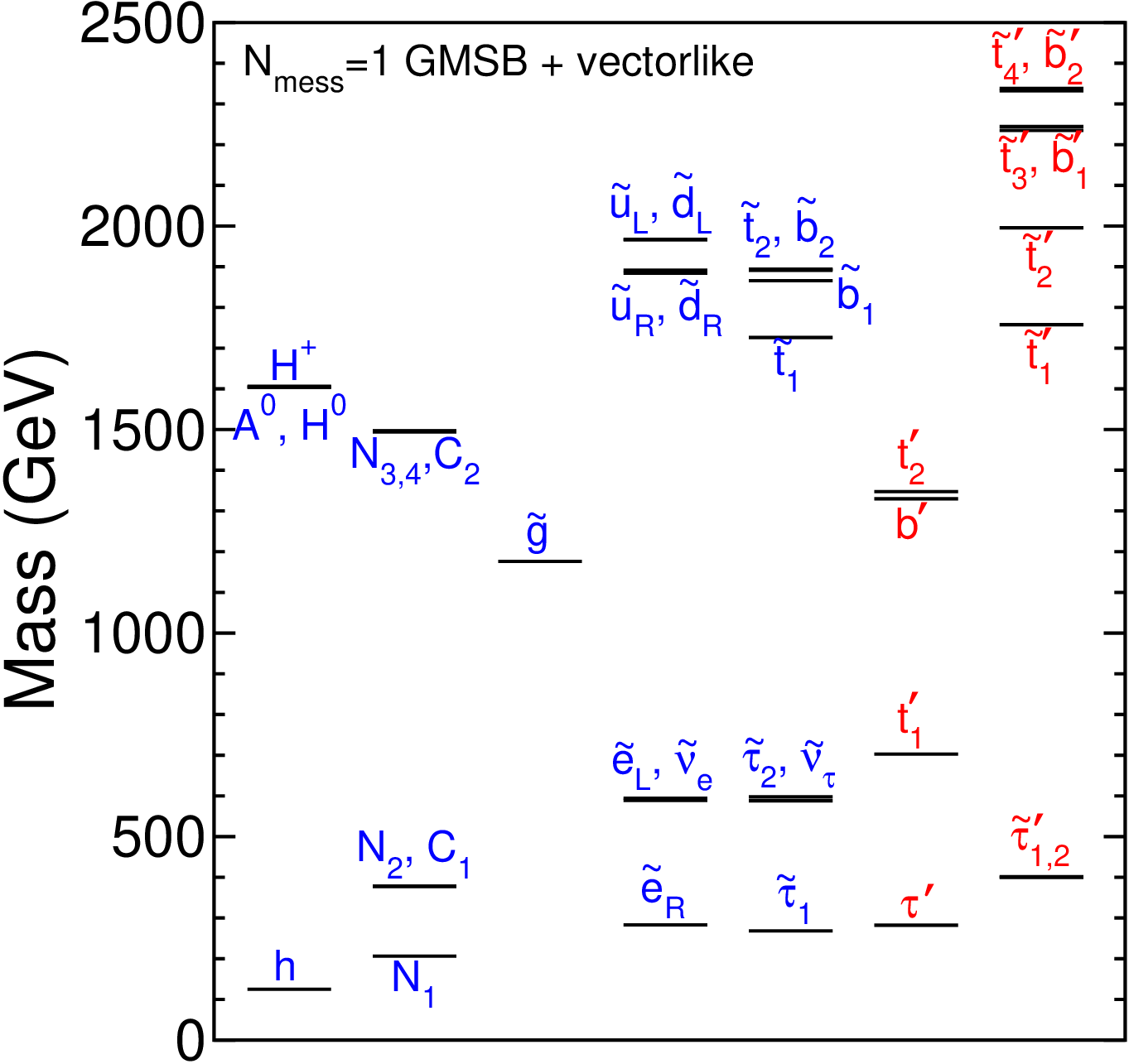}
\end{flushright}
\end{minipage}
\caption{\label{fig:samples1} The mass spectra for new particles in
a minimal GMSB model (left) with light messengers 
and a similar model with additional
chiral supermultiplets in a  
${\bf 10} + {\bf \overline{10}}$ of $SU(5)$ (right). In both cases,
$M_{\rm mess} = 1500$ TeV, $\Lambda = 150$ TeV,
$\tan\beta = 15$, and $\mu>0$. In the model with extra vector-like 
fields, $k=1$, and $M_Q=M_U=M_E = 215$ GeV at the unification scale. The lightest 
Higgs mass is 114.9 GeV (left) and 124.9 GeV (right).}
\end{figure}

In Figure \ref{fig:samples1}, we show the mass spectrum
 of all new particles
in a sample model with $M_{\rm mess} = 1500$ TeV, $\Lambda = 150$ TeV, $\tan\beta=15$,
$\mu>0$. The left panel shows the result for the minimal GMSB model with these parameters,
and the right panel the model of interest extended by the ${\bf 10} + {\bf \overline{10}}$
fields. 
The minimal GMSB model in the left panel can only manage $M_h=115$ GeV,
and is therefore clearly ruled out if $M_h\hmass$. In the right panel, we choose $M_Q=M_U=M_E = 215$ GeV at the unification scale, as this leads
to $M_h = 125$ GeV. 
Comparing the two models, we see that in both cases $M_{\tilde g}$ is
close to 1160 GeV; this is significant because the gluino mass is the most important parameter 
pertaining to the discovery of the odd $R$-parity sector at the LHC when squarks are much heavier,
as here.
However, in this model at least,
the lightest new strongly interacting particle is actually the $t_1'$ with mass near 700 GeV;
it is much lighter than the other vector-like quarks $b'$ and $t_2'$, and their superpartners,
as well as the MSSM squarks and gluino.
The lightest new particle from the ${\bf 10} + {\bf \overline{10}}$ sector is the $\tau'$,
which if quasi-stable could also be a candidate for the first beyond-the-SM discovery despite 
lacking strong interactions, as we will discuss below.
The model with vector-like supermultiplets also produces squarks that are significantly heavier than
the prediction for minimal GMSB. The Higgsino-like neutralinos and charginos 
$\tilde N_3, \tilde N_4, \tilde C_2$ are also more than a factor of 2 heavier than the prediction
of minimal GMSB, corresponding to a much larger $|\mu|$. If $|\mu|$ is treated as a proxy for the
amount of fine tuning in the model, we are forced to accept that the model with extra vector-like
supermultiplets is more unnatural than the minimal GMSB model, but this psychological 
price must be paid if $M_h\hmass$.

\begin{figure}[!ptb]
\begin{minipage}[]{0.41\linewidth}
\begin{center}
\includegraphics[height=8.45cm,angle=0]{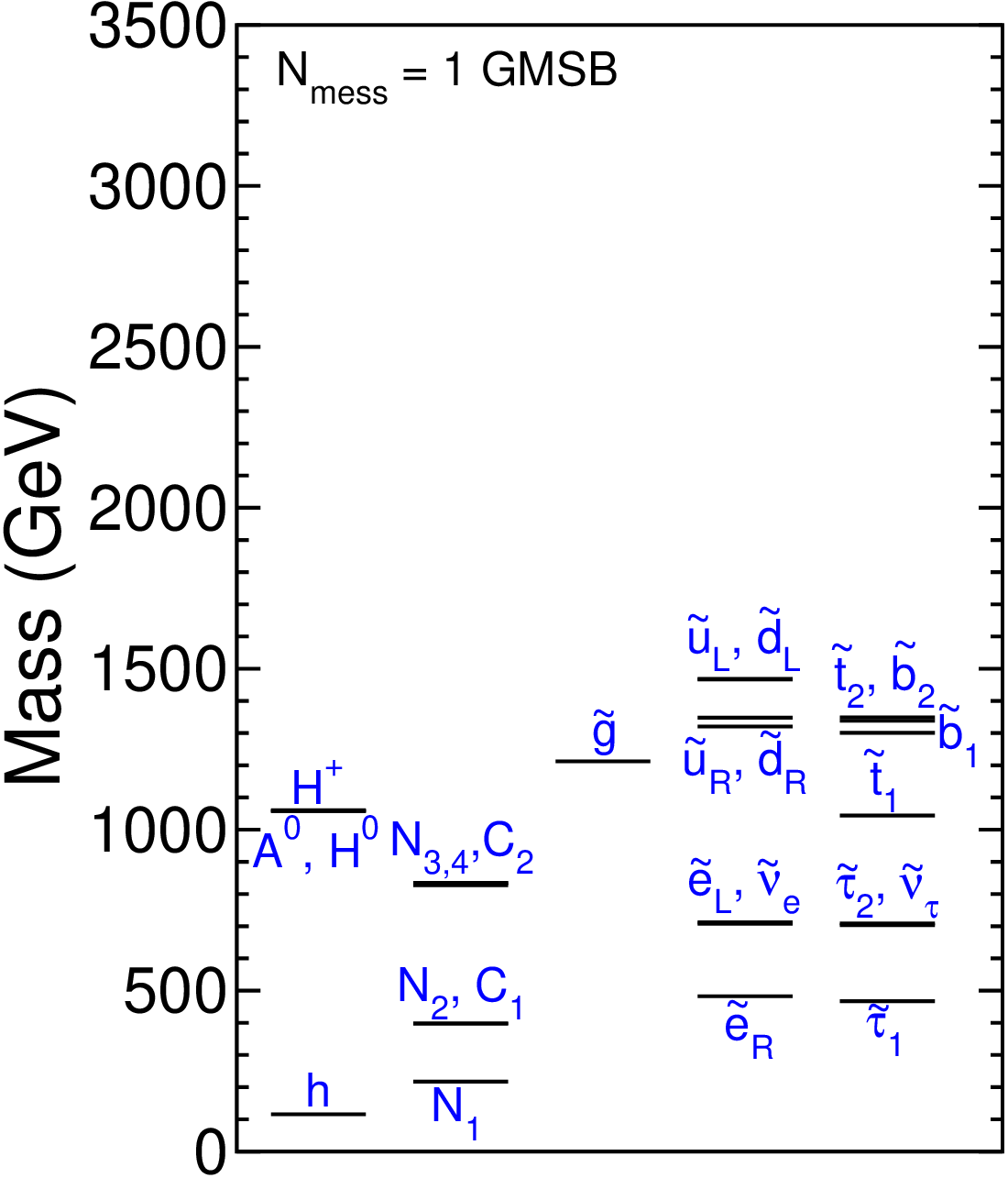}
\end{center}
\end{minipage}
\begin{minipage}[]{0.58\linewidth}
\begin{flushright}
\includegraphics[height=8.45cm,angle=0]{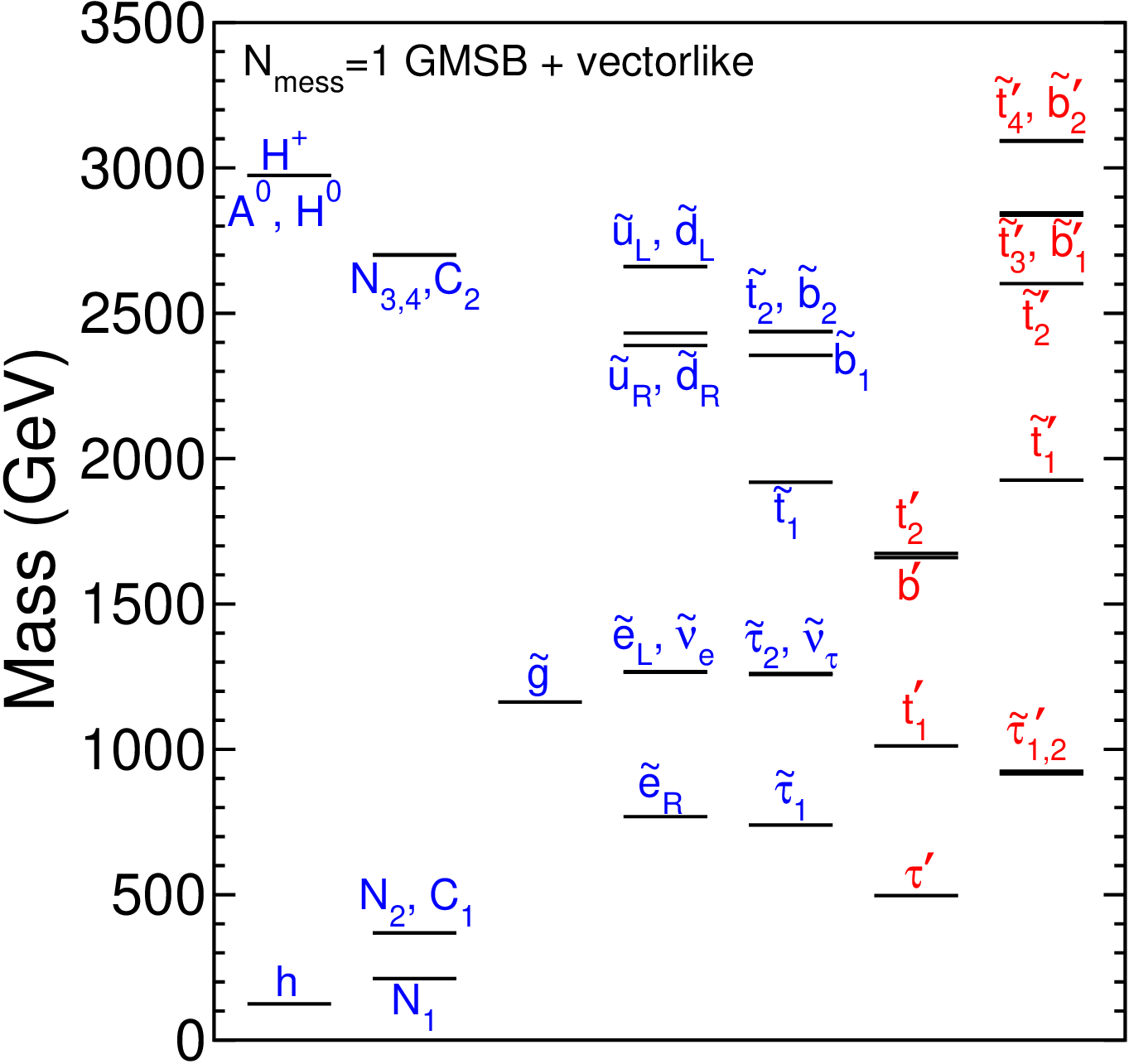}
\end{flushright}
\end{minipage}
\caption{\label{fig:samples2} As in Figure \ref{fig:samples2}, but with 
very heavy messengers of supersymmetry breaking at $M_{\rm mess} = 10^{14}$ GeV.
The other parameters are $\Lambda = 160$ TeV,
$\tan\beta = 15$, $\mu>0$, and $k=1$ and $M_Q=M_U=M_E = 400$ GeV at the unification scale
in the model with extra vector-like particles.
The lightest 
Higgs masses are 115.5 GeV (left) and 124.6 GeV (right).}
\end{figure}

Figure \ref{fig:samples2} shows a similar comparison, but for a much higher messenger scale
$M_{\rm mess} = 10^{14}$ GeV. The effect of raising the messenger scale is to further increase
the squark and slepton masses for the model with extra vector-like matter, both in an absolute sense 
and compared to the minimal GMSB model. 
The Higgsino-like neutralinos and charginos are also much heavier 
in the extended model, pointing to more fine tuning needed in the electroweak
symmetry breaking potential, as noted above.
For the same input parameters, the gluino mass is suppressed in the extended model
on the right compared to the minimal model, but only by about 4\%.
In both Figures \ref{fig:samples1} and \ref{fig:samples2}, the heavier Higgs bosons $A^0$,
$H^0$, and $H^\pm$ have their masses substantially increased when the model is extended to include
vector-like supermultiplets.

If $\mu M_2$ is positive, there will be a positive correction to the anomalous magnetic moment of the
muon, bringing the theoretical prediction into better agreement with the experimental result
\cite{Bennett:2006fi}, as has been emphasized in the present context by \cite{Endo:2011mc}. However, because
we are not willing to interpret the present $\sim 3 \sigma$ discrepancy as evidence against the SM, we simply take $\mu M_2 > 0$
and do not impose any constraint from $(g-2)_\mu$. 
It is also useful to note that for all models of this type, the effect of the vectorlike quarks is to bring slightly closer agreement with
precision electroweak oblique corrections
than in the SM, but not by a statistically significant amount \cite{Martin:2009bg}. 

\asubsection{Achieving $M_h\hmass$\label{subsec:achieving125}}

\begin{figure}[!tb]
\begin{minipage}[]{0.49\linewidth}
\includegraphics[width=\linewidth,angle=0]{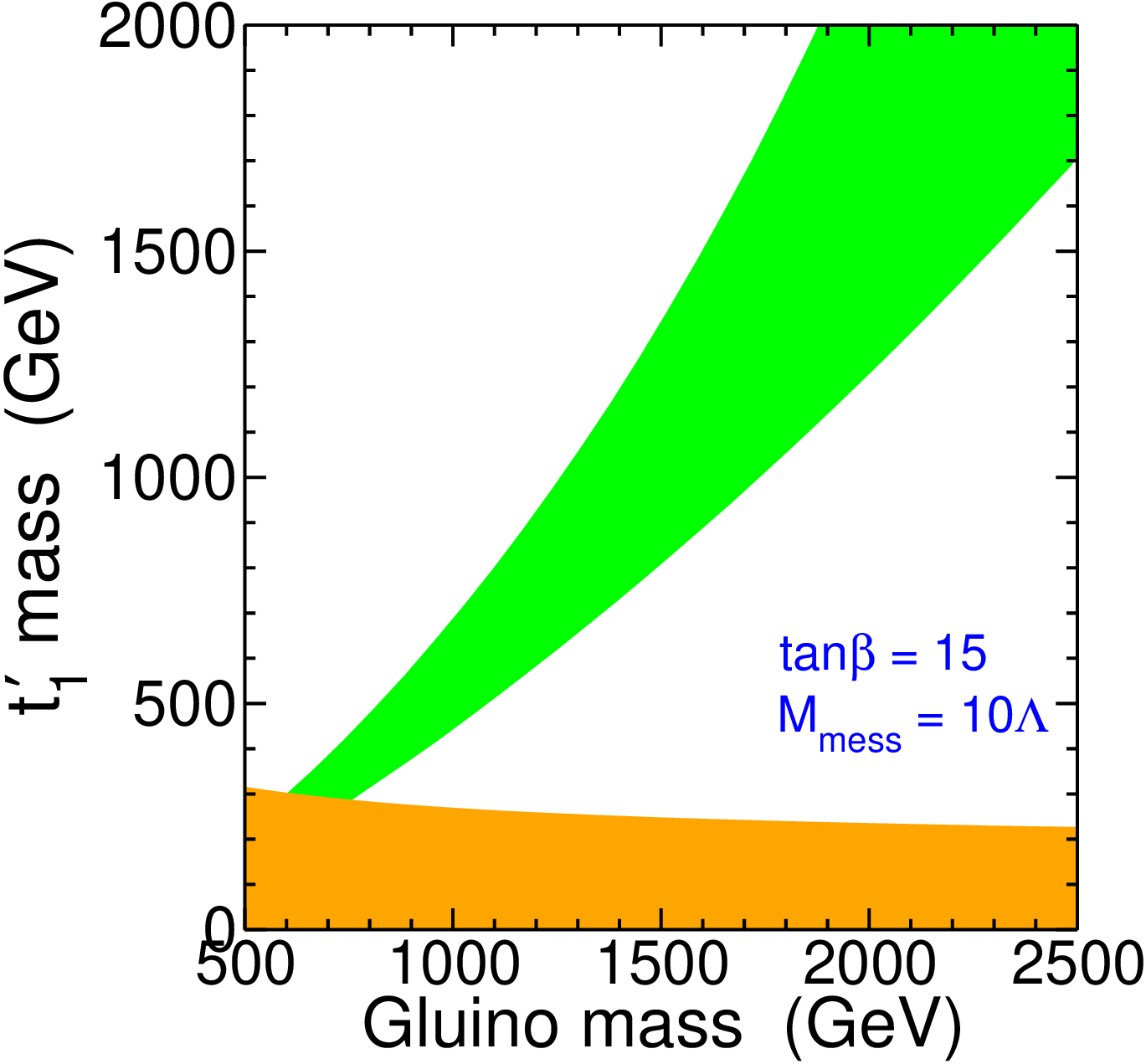}
\end{minipage}
\begin{minipage}[]{0.49\linewidth}
\begin{flushright}
\includegraphics[width=\linewidth,angle=0]{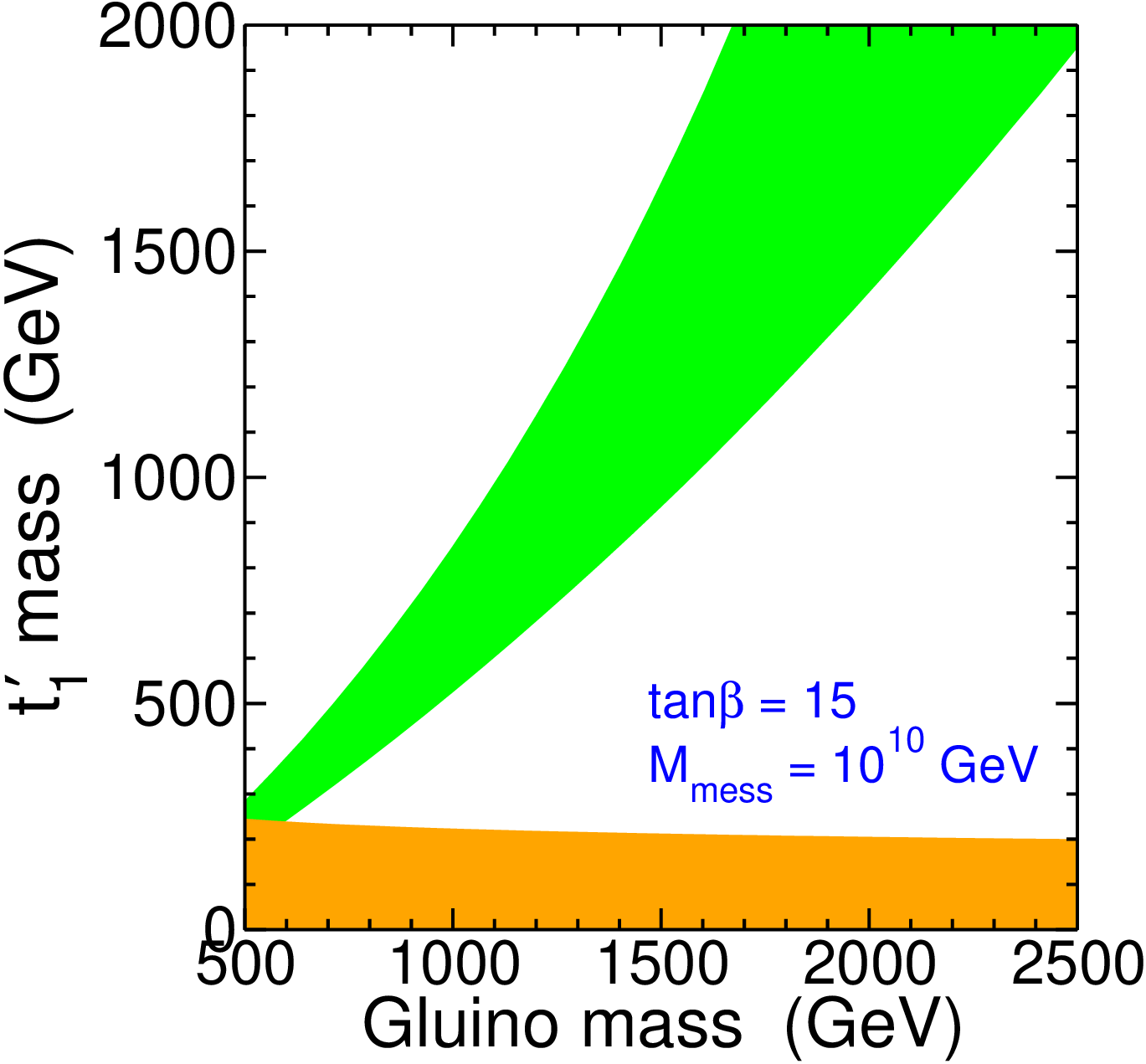}
\end{flushright}
\end{minipage}
\begin{minipage}[]{0.49\linewidth}
\begin{flushright}
\includegraphics[width=\linewidth,angle=0]{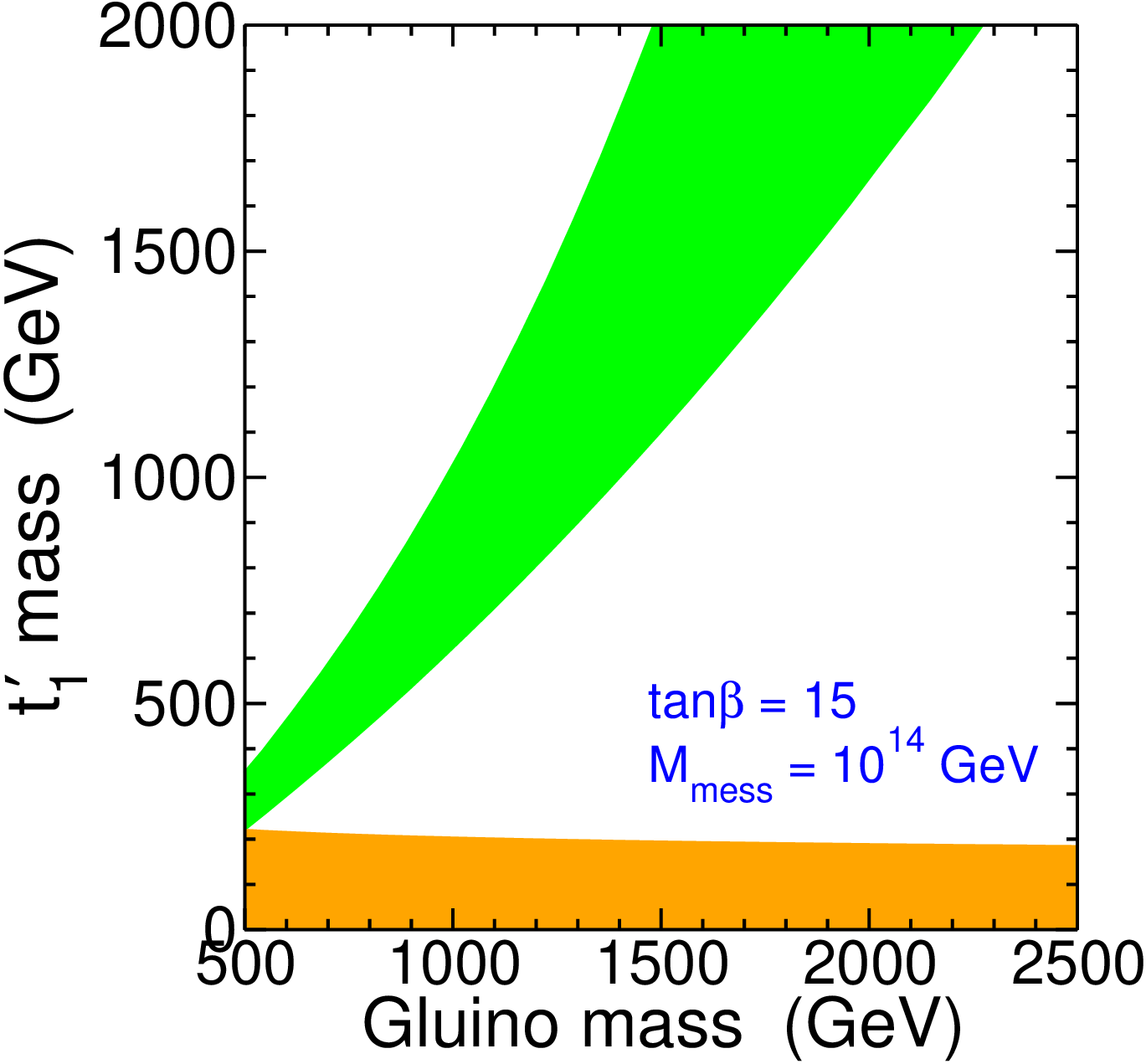}
\end{flushright}
\end{minipage}
\begin{minipage}[]{0.49\linewidth}
\begin{minipage}[]{0.04\linewidth}\end{minipage}
\begin{minipage}[]{0.90\linewidth}
\caption{\label{fig:MtpMgluino}
Regions in the $M_{t_1'}$ vs. $M_{\tilde g}$ plane 
for extended GMSB models with a minimal messenger sector
and a ${\bf 10} + {\bf \overline{10}}$ of vector-like quarks and leptons,
with $\Mmess = 10 \Lambda$ and $10^{10}$ GeV and $10^{14}$ GeV. 
The vector-like fermion mass parameters $M_Q$ and $M_U$ are taken to be unified at
the gauge coupling unification scale.
The Yukawa coupling $k$ is near its fixed point, obtained by setting $k=1$ at
the unification scale, and $\tan\beta=15$.
The green sloped funnel regions have a calculated Higgs mass satisfying 
122 GeV $<M_h<128$ GeV. In the orange shaded region,
$M_{\tau'} < 100$ GeV, assuming that $M_E = M_Q = M_U$ 
at the unification scale.
}
\end{minipage}\end{minipage}
\end{figure}

The corrections to the lightest Higgs mass are most strongly dependent on the masses of
$t_1', t_2'$ and their superpartners 
$\tilde t_{1,2,3,4}'$, with $\Delta M_h$ increasing with the hierarchy between the average
scalar and fermion masses. The masses of $\tilde t_{1,2,3,4}'$ scale with the 
supersymmetry-breaking parameter $\Lambda$, and the smaller they are, the smaller the fermion masses
$t_{1,2}'$ and $b'$ must be in order to accommodate $M_h\hmass$.
The masses of the gluino and $t_1'$ are of particular interest, since pair production of one of them
is likely to give the initial discovery signal at the LHC. 
Figure \ref{fig:MtpMgluino} shows (green sloped funnel) 
regions in the $M_{t_1'}$ vs. $M_{\tilde g}$ plane in which
122 GeV$<M_h<$128 GeV, for $\tan\beta=15$, with $k=1$ at the unification scale.
The variation in $M_{\tilde g}$ is obtained by varying $\Lambda$, and that of $M_{\tilde t_1'}$
by varying $M_Q=M_U=M_E$ at the unification scale. Three choices of the messenger scale are shown,
$M_{\rm mess} = 10\Lambda$, $10^{10}$ GeV, and $10^{14}$ GeV.
Note that, pending exclusions by direct searches for gluino and $t_1'$, 
it is easy to obtain $M_h\hmass$ in this class of models, with 
$M_{\tilde g}$ lower than 700 GeV and $M_{t_1'}$ lower than 300 GeV even 
if the messengers are light. 
Therefore, each 
new search result at LHC probes an interesting region of parameter space 
consistent with $M_h\hmass$, unlike in the usual GMSB models.

\begin{figure}[!tp]
\begin{minipage}[]{0.49\linewidth}
\includegraphics[width=\linewidth,angle=0]{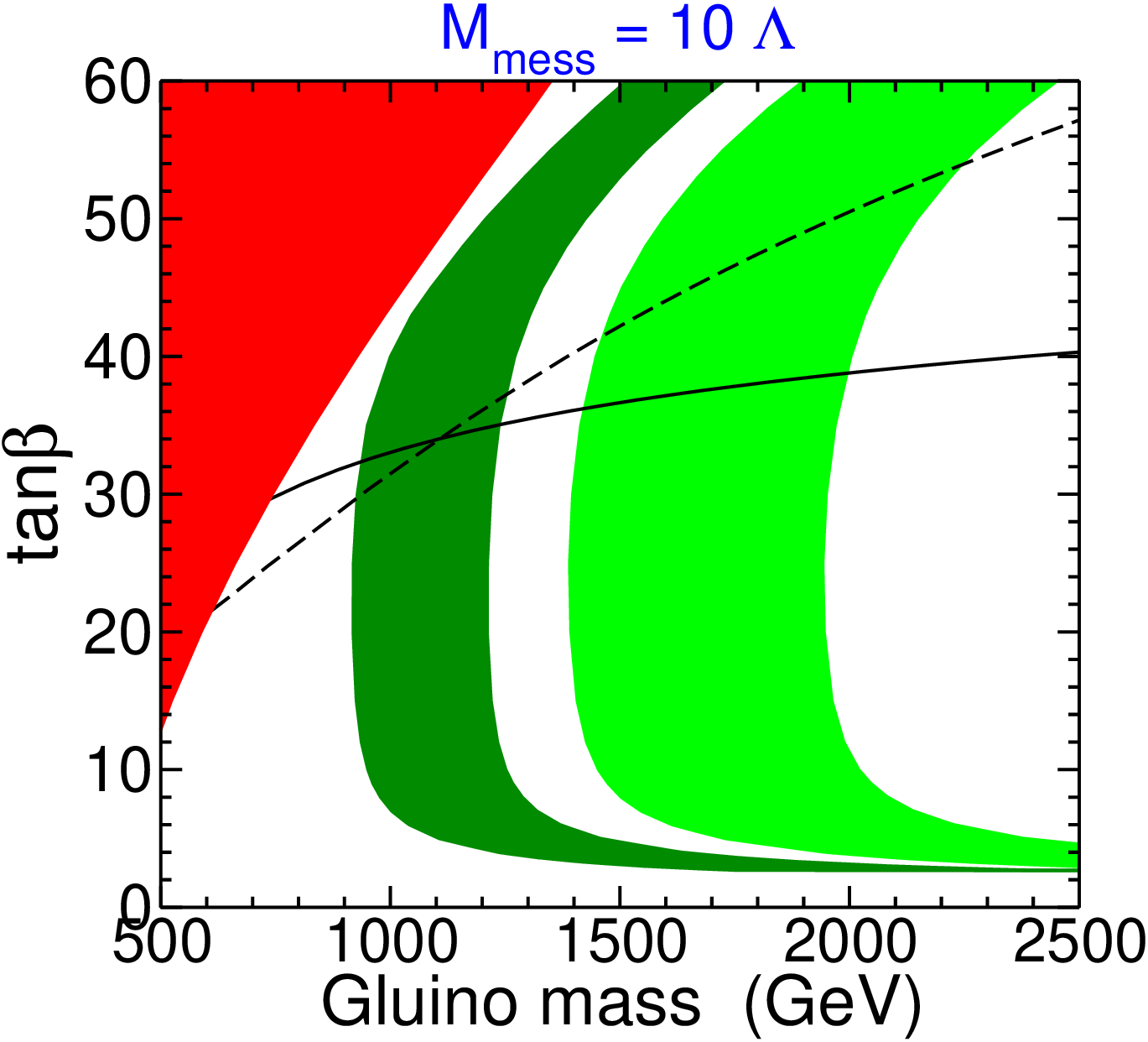}
\end{minipage}
\begin{minipage}[]{0.49\linewidth}
\begin{flushright}
\includegraphics[width=\linewidth,angle=0]{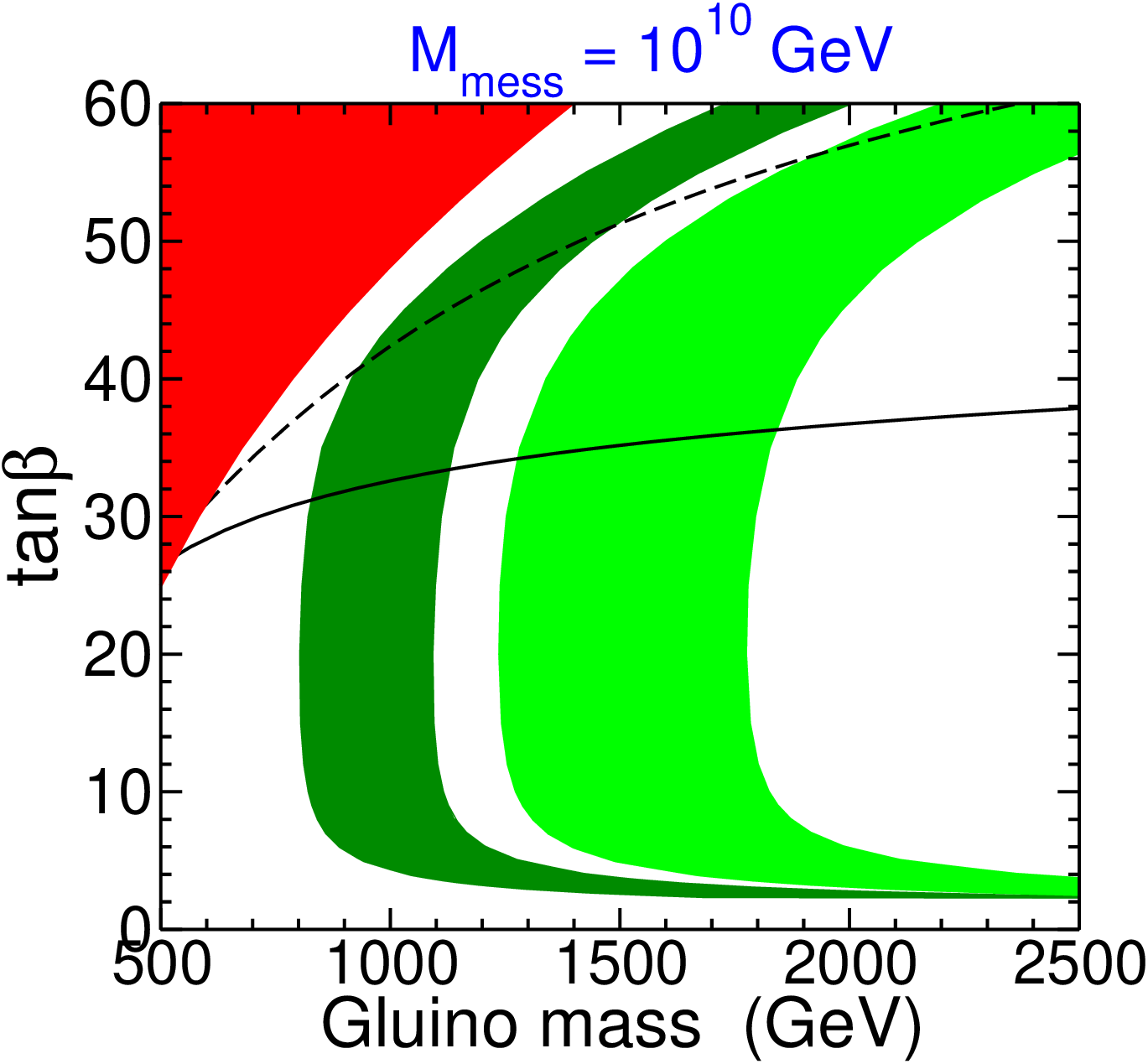}
\end{flushright}
\end{minipage}
\begin{minipage}[]{0.49\linewidth}
\begin{flushright}
\includegraphics[width=\linewidth,angle=0]{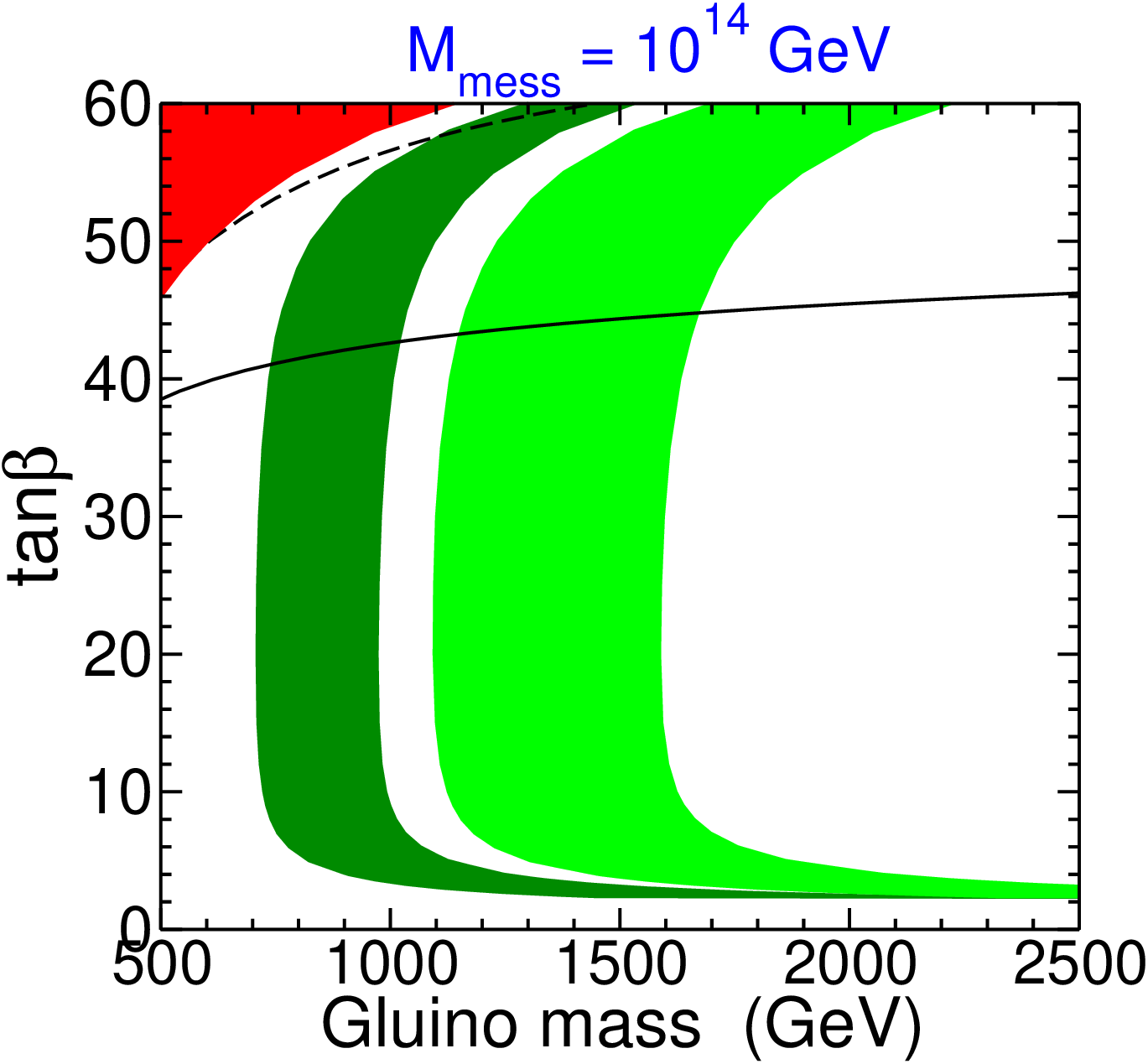}
\end{flushright}
\end{minipage}
\begin{minipage}[]{0.49\linewidth}
\begin{minipage}[]{0.04\linewidth}\end{minipage}
\begin{minipage}[]{0.90\linewidth}
\caption{\label{fig:tbMgluino}
Regions in the $\tan\beta$ vs. $M_{\tilde g}$ plane 
with $\Mmess = 10 \Lambda$ and $10^{10}$ GeV and  $10^{14}$ GeV. 
The vector-like fermion mass parameters $M_Q, M_U$ are taken to be unified at
the gauge coupling unification scale.
The Yukawa coupling $k$ is near its fixed point, obtained by setting $k=1$ at
the unification scale.
The green shaded curved regions have a calculated Higgs mass satisfying 
122 GeV $<M_h<128$ GeV, with the darker green region corresponding to 
$M_{t_1'} = 600$ GeV and the lighter green region having $M_{t_1'} = 1200$ GeV.
In the red shaded region, $M_{\tilde \tau_1} < 100$ GeV. 
Below (above) the dashed line the NLSP is a neutralino (stau).
Above the solid curves, the vacuum 
is unstable by tunneling to a state with VEVs for the staus, with a 
liftime shorter than the age of the universe.
}
\end{minipage}\end{minipage}
\end{figure}

The dependence on $\tan\beta$ is shown in Figure \ref{fig:tbMgluino}, with allowed
regions for $122 < M_h < 128$ GeV in the $\tan\beta$ and $M_{\tilde g}$ 
plane.
In each graph, the lighter green curved region furthest right  corresponds 
to the choice of $M_{t'_1}=1200\gev$ and the darker green curved region to the left of it 
corresponds to $M_{t'_1}=600\gev$.  The upper left triangular red region corresponds to $M_{\tau'}
<100\gev$.
The three graphs shown correspond to $\Mmess = 10 \Lambda$ and $10^{10}$ 
GeV and $10^{14}$ GeV, and all have $k=1$ at the 
unification scale. More details regarding underlying parameters are found 
in the figure caption.

Note that an intermediate value of $10\lsim \tan\beta\lsim 35$ enables 
a lighter gluino mass, and so lighter MSSM squark masses, than found 
for $\tan\beta$ outside of that range.  For larger $\tan\beta$, the 
corrections to $M_h$ from the tau-stau sector are negative and 
big,\footnote{The tau-stau loop contributions are larger than the 
bottom-sbottom ones, despite having a smaller Yukawa coupling and no 
color factor, because the staus are much lighter than the sbottoms.} so 
that larger supersymmetry breaking masses (indicated in the plot by 
$M_{\tilde g}$) are required. For $\tan\beta<10$, the tree-level $M_h$ is 
much smaller, requiring heavier superpartners to obtain $M_h\hmass$. 
Similar figures are found in \cite{Endo:2011xq,Endo:2012rd}, but with 
$M_Q = M_U$ at the TeV scale, rather than at the unification scale as 
chosen here. An important point \cite{Endo:2012rd} is that there is an 
upper bound on $\tan\beta$ in these models, following from the general 
bound obtained in \cite{Hisano:2010re} by requiring the standard 
electroweak-breaking vacuum to be stable (with a lifetime longer than the 
age of the universe) against tunneling to a vacuum in 
which the stau fields have VEVs. We show this bound for our models as the 
solid lines in Figure \ref{fig:tbMgluino}. We see again here in this 
figure that gluino masses easily accessible by LHC now or in the near 
future are sufficient to deliver a light Higgs boson of mass $\hmass$, 
and this can be achieved for $M_{\tilde g}\lsim 2.5\tev$ even if 
$\tan\beta$ is as low as about 3.

\begin{figure}[!tp] \begin{minipage}[]{0.49\linewidth} 
\includegraphics[width=\linewidth,angle=0]{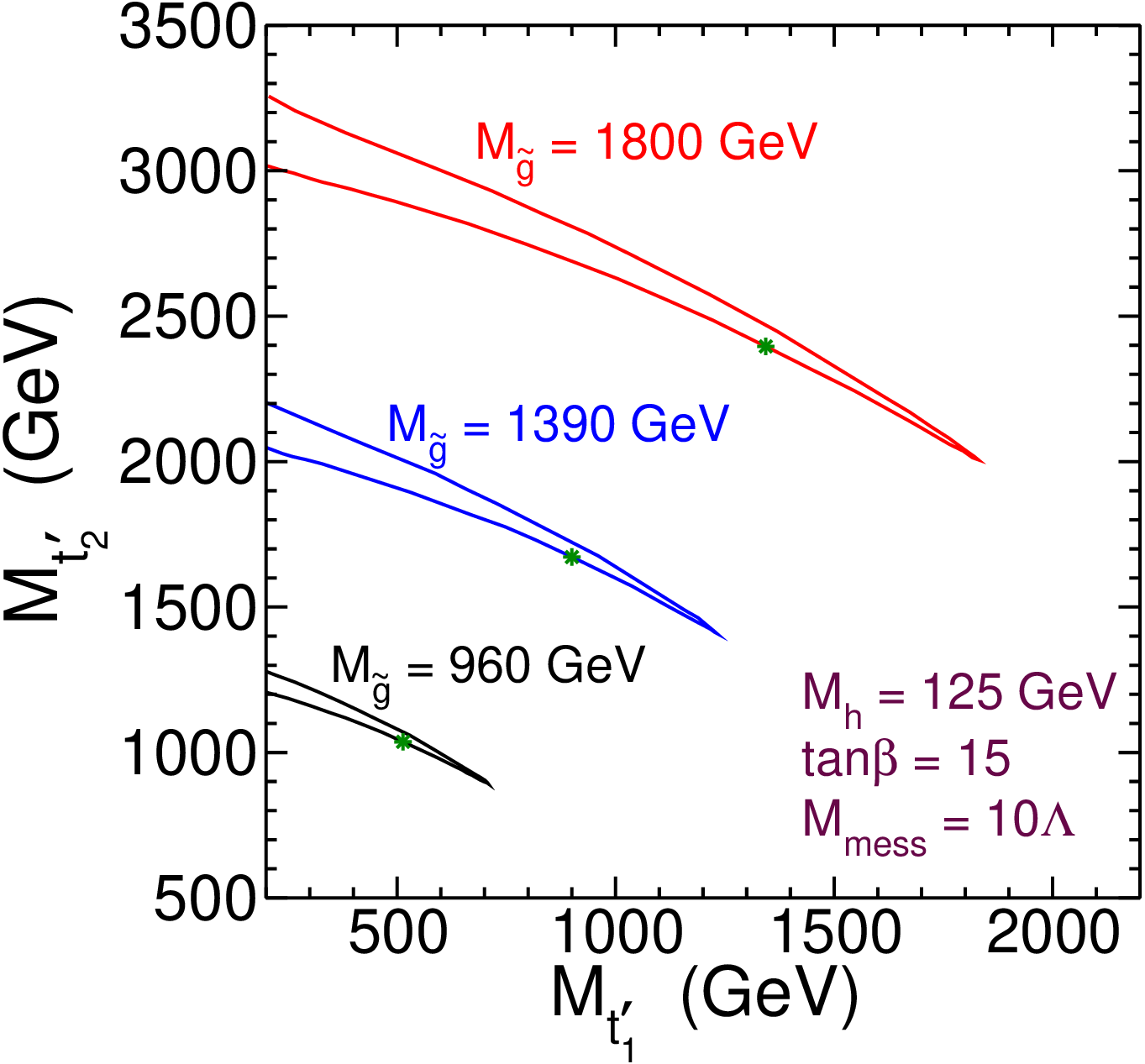} 
\end{minipage} 
\hspace{0.05\linewidth} 
\begin{minipage}[]{0.44\linewidth} 
\caption{\label{fig:t1t2} Lines in the $M_{t_1'}, 
M_{t_2'}$ plane with predicted $M_h = 125$ GeV, for models with 
$\tan\beta = 15$ and $\Mmess = 10 \Lambda$, with varying 
$M_Q/M_U$. 
The three lines correspond to $\Lambda = 120, 180$, and 240 TeV, 
corresponding to $M_{\tilde g} =$ 960, 1390, and $1800\gev$ respectively. The 
lower (upper) branch in each case corresponds to $M_Q/M_U > 1$ ($<1$) at 
the TeV scale. The green stars correspond to $M_Q = M_U$ at the gauge 
coupling unification scale.}
\end{minipage}
\end{figure}

As remarked above, $M_Q$ and $M_U$ are independent
in a general theory. Figure \ref{fig:t1t2} explores this freedom
by showing lines in the $M_{t_1'}, M_{t_2'}$ plane that predict $M_h = 125$ GeV, for models with
$\tan\beta=15$ and $M_{\rm mess} = 10\Lambda$, 
with the ratio $M_Q/M_U$ allowed to vary.
The special cases with $M_Q=M_U$ at the unification scale (as in the examples of
Figures \ref{fig:samples1}, \ref{fig:samples2}, \ref{fig:MtpMgluino}, \ref{fig:tbMgluino} above) are 
noted by green stars.
The three curves correspond to $\Lambda = 120, 180$, and 240 TeV, 
resulting in $M_{\tilde g} \approx$ 960, 1390, and 1800 respectively. (There is some small variation in
the gluino masses on each curve.) We find that for equal values of other parameters,
$M_h$ remains approximately constant for fixed values of the arithmetic mean of
$M_{t_1'}$ and $M_{t_2'}$. In particular, the geometric mean is not as good a figure of merit.
For each curve in Figure \ref{fig:t1t2}, we see that there is no minimum value of $M_{t_1'}$
from the $M_H\hmass$ constraint alone,
because one can always take a very large or small ratio of $M_Q/M_U$. However, on each curve corresponding to a fixed $M_{\tilde g}$, the requirement $M_h\hmass$ implies
a minimum value of $M_{t_2'}$, and a maximum value of $M_{t_1'}$.

\asubsection{Comment on gravitino dark matter}

\begin{figure}
\begin{minipage}[]{0.49\linewidth}
\begin{flushright}
\includegraphics[width=\linewidth,angle=0]{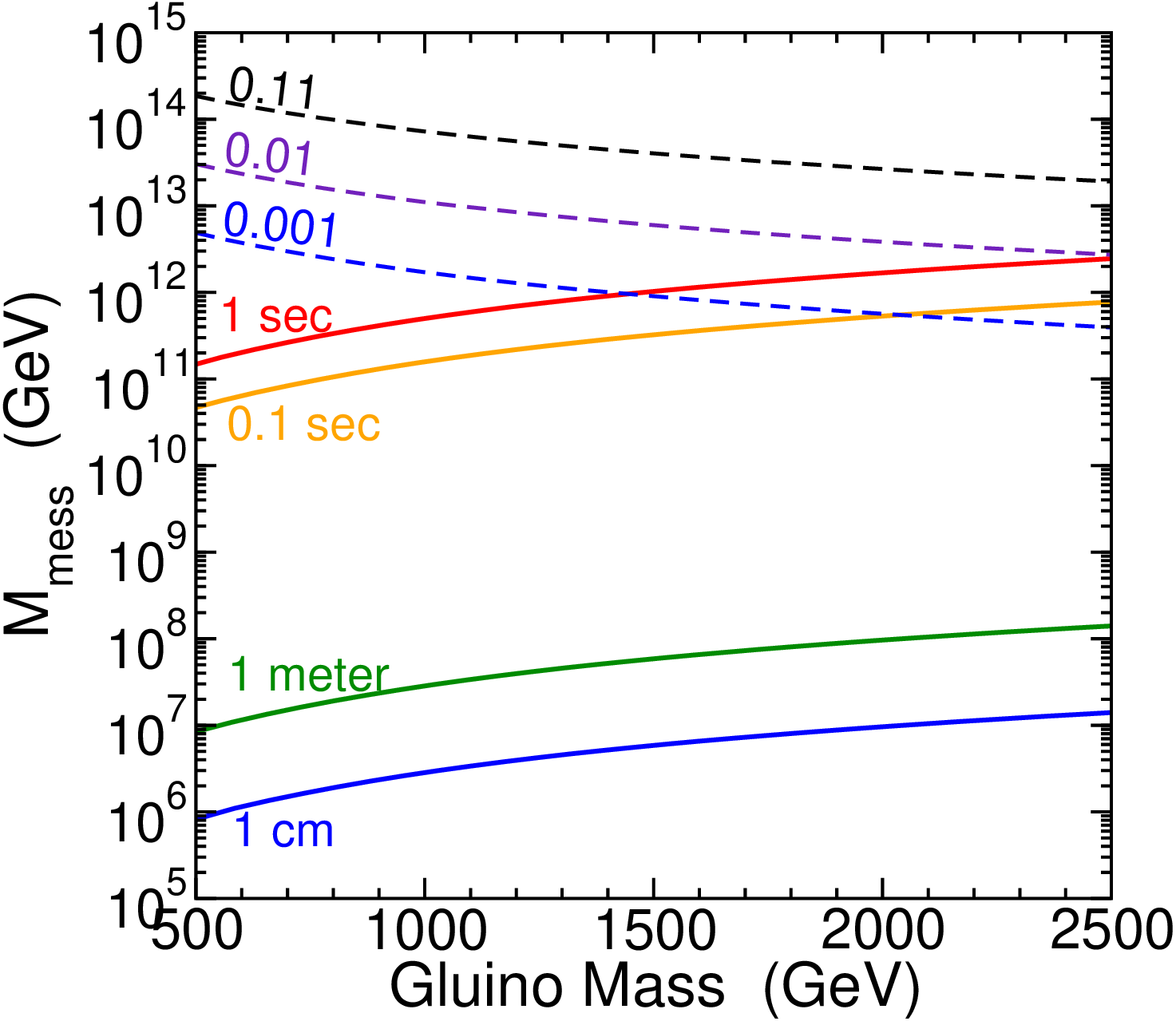}
\end{flushright}
\end{minipage}
\begin{minipage}[]{0.49\linewidth}
\begin{minipage}[]{0.04\linewidth}\end{minipage}
\begin{minipage}[]{0.90\linewidth}
\caption{\label{fig:gmsbDM}
Lines of constant NLSP lifetime $\tau_{\tilde N_1} = 1$ second, $0.1$ 
second, 1 meter, and 1 cm (solid lines), and gravitino abundance 
resulting from decays of thermal neutralino NLSPs $\Omega_{\tilde G} h^2 
= 0.11$, $0.01$, and $0.001$ (dashed lines), in the $m_{\tilde g}$ and 
$M_{\rm mess}$ 
plane. The gluino mass variation was obtained by varying $\Lambda$, with 
$\tan\beta = 15$ and $k = 1$.}
\end{minipage}\end{minipage}
\end{figure}

In GMSB models, the LSP is likely to be the gravitino $\tilde G$, with 
mass $M_{\tilde G} = \Lambda M_{\rm mess}/\sqrt{3} M_P$, where 
$M_p = 2.44 \times 10^{18}$ GeV. 
In principle, the gravitino could be a dark matter candidate. 
One possibility is the gravitino
superwimp scenario~\cite{superwimp} in which the gravitino abundance is 
assumed to be suppressed by a low reheating temperature or 
diluted by some other 
non-standard 
cosmology, followed by the bino-like neutralino LSP freezing out and then decaying 
out of equilibrium according to $\tilde N_1 \rightarrow \gamma \tilde G$, with a lifetime given approximately by
\cite{Cabibbo:1981er}
\beq
\tau_{\tilde N_1} = 7.5 \times 10^4\>\mbox{sec}\>\left (
\frac{M_{\tilde G}}{\mbox{GeV}}\right)^2 \left (\frac{\mbox{100 GeV}}{M_{\tilde N_1}} \right )^5.
\eeq
If $\tilde N_1 \rightarrow Z \tilde G$ is kinematically allowed, then this lifetime is reduced by a factor $1 + 0.3 (1 - m_Z^2/m_{\tilde 
N_1}^2)^4$ \cite{AKKMM}. 
If the gravitino is to be a significant component of the dark matter, 
this lifetime should be smaller than about 0.1 to 1 sec, in order that 
the 
successful 
predictions of primordial nucleosynthesis are not affected. 
This is in tension with a cosmologically relevant relic abundance of 
gravitinos 
from decays of thermal binos, given by
\beq
\Omega_{\tilde G} h^2 = \frac{m_{\tilde G}}{m_{\tilde N_1}} 
\Omega_{\tilde N_1}\! h^2 .
\eeq
Here 
\beq
\Omega_{\tilde N_1}\! h^2 \approx 
0.013 \frac{(1 + r)^4}{r (1 + r^2)} \left 
(\frac{m_{\tilde e_R}}{\mbox{100 GeV}} \right )^2
\left [1 + 0.07 \ln (\sqrt{r}\,\mbox{100 GeV}/m_{\tilde e_R}) \right ]
\eeq
is the relic density of binos that would be found today if they were 
stable,
given in a convenient approximation \cite{DMapprox}, 
with $r = m^2_{\tilde N_1}/m^2_{\tilde e_R}$. To 
illustrate this, we show in Figure \ref{fig:gmsbDM} 
solid lines of constant $\tau_{\tilde N_1} = 0.1$ and 1 second (relevant 
for nucleosynthesis) and 
1 cm and 1 meter (relevant for collider physics), compared to dashed lines of constant $\Omega_{\tilde G} h^2 =0.11$, $0.01$, and 
$0.001$, in the $(m_{\tilde g}, M_{\rm mess})$ plane, with the variation in gluino mass obtained by varying $\Lambda$.
It is difficult to reconcile 
gravitino dark matter with the standard picture of primordial
nucleosynthesis in this model,
without going to very large superpartner masses ($m_{\tilde g} \gg 2.5$ 
TeV), in which case the vector-like quarks 
would not be necessary and prospects for any discovery of new 
particles beyond $h^0$  
at the LHC would be exceedingly grim.\footnote{Two recent papers 
\cite{Okada:2012gf,FSY} have noted the 
complementary approach that in normal gauge mediation models, one can accommodate gravitino dark matter and $M_h\hmass$, at the cost of such 
very heavy superpartners.} Such a massive superpartner spectrum runs counter to the purpose of this paper, which aims to accommodate the $\sim 125\gev$ 
with lighter superpartners accessible to the LHC. In the scenario considered in the present paper, these considerations suggest that 
dark matter is composed mostly of 
axions or some other particles, with a negligible contribution from gravitinos, and messenger mass scales much above $10^{11}$ GeV are 
therefore apparently disfavored as indicated in Figure \ref{fig:gmsbDM}.

\section{Masses of exotic quarks and $t_1'$ decays\label{sec:massesanddecays}}
\setcounter{equation}{0}
\setcounter{figure}{0}
\setcounter{table}{0}
\setcounter{footnote}{1}

Taking into account the full superpotential of the theory 
$W_{\rm MSSM}+W_{QUE}+W_{\rm mix}$ the fermionic mass matrices for 
up-type and down-type quarks are~\cite{Martin:2009bg} 
\beq
{\cal M}_u = \begin{pmatrix} y_t v_u & \epsilon_{\scriptscriptstyle U} v_u & 0 \cr
                             0 & M_U & k' v_d \cr
                             \epsilon'_{\scriptscriptstyle U} v_u & k v_u & M_Q 
             \end{pmatrix},
\qquad\quad
{\cal M}_d = \begin{pmatrix} y_b v_d & 0 \cr
                             \epsilon_{\scriptscriptstyle D} v_d & \phantom{x}-M_Q 
             \end{pmatrix},
\eeq
with mass eigenstates $t, t_1', t_2'$ and $b, b'$ respectively. The zeros appear as a consequence of a choice of basis. As mentioned 
earlier, we assume
that $\epsu$, $\epsup$, and $\epsd$ can be treated as small perturbations
in these mass matrices. Then one always finds $M_{t_1'} < M_{b'} < M_{t_2'}$, and
the exotic quarks will decay according to 
$t_2' \rightarrow Wb',\> ht_1',\>Zt_1'$ and
$b' \rightarrow W^{(*)}t_1',\>Wt,\>Zb,\>hb$
and 
$t_1' \rightarrow Wb,\> ht,\>Zt$, when kinematically allowed. 
Formulas for these decays widths, which will be used in 
our phenomenological discussion below, can be found in 
Appendix B of~\cite{Martin:2009bg}, and in a more general framework in
the Appendix of the present paper.

In Figure~\ref{fig:MQoMU}, we plot the mass eigenvalues of the exotic quark 
states $t'_{1,2}$ and $b'$ as a function of $M_Q/M_U$ in the left panel, 
and the branching fractions of $t_1'$ vs.\ $M_Q/M_U$ in the right panel.  
\begin{figure}[!tp]
\begin{minipage}[]{0.49\linewidth}
\includegraphics[width=\linewidth,angle=0]{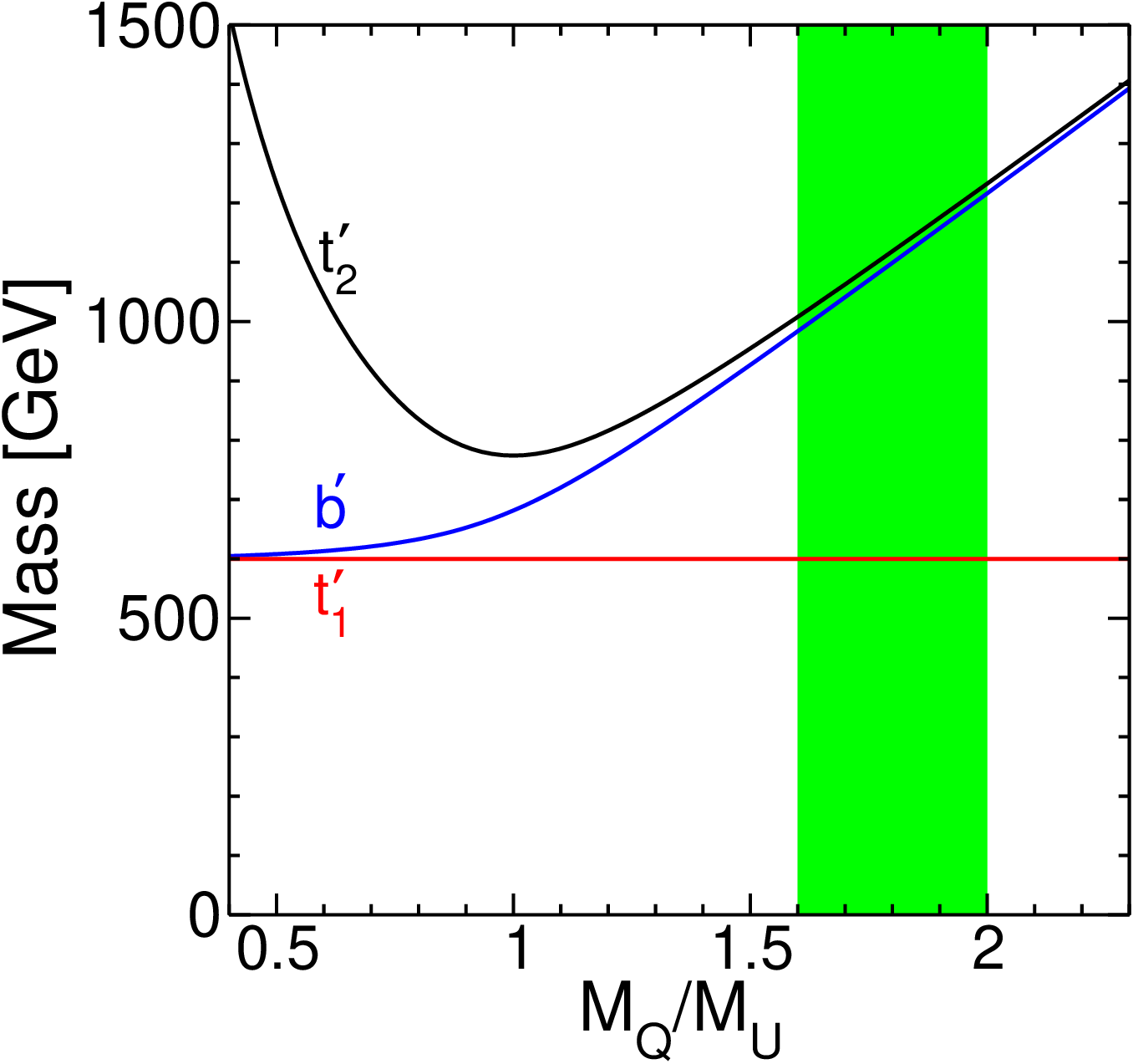}
\end{minipage}
\begin{minipage}[]{0.49\linewidth}
\includegraphics[width=\linewidth,angle=0]{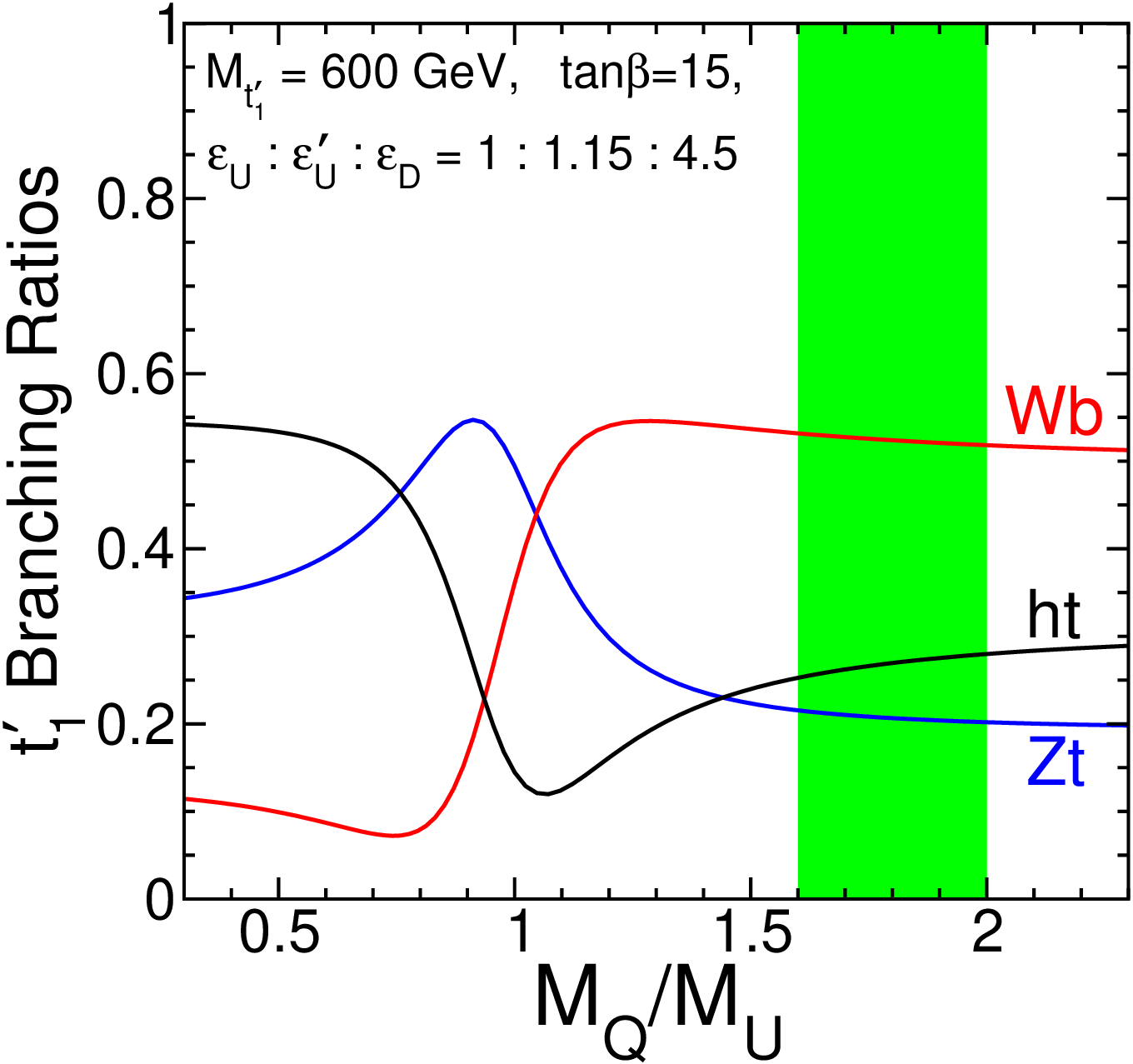}
\end{minipage}
\caption{\label{fig:MQoMU} Masses (left panel) and $t_1'$ branching 
ratios (right panel) for vector-like quarks. Here, $M_{t_1'}$ is fixed to 
600 GeV, and the ratio of mass parameters $M_Q$ and $M_U$ at the TeV 
scale is varied. For small (large) $M_Q/M_U$, the state $t_1'$ is mostly 
$SU(2)_L$ doublet (singlet). The green band shows the region obtained 
with the unification condition $M_Q=M_U$ imposed at the gauge coupling 
unification scale. The left edge of this band corresponds to $M_{\rm 
mess} = 10^{14}$ GeV, and the right edge to $\Mmess = 10 \Lambda$. 
The weak-scale parameters $\epsu$, $\epsup$, and $\epsd$ 
that describe mixing of the vector-like quarks with the top and bottom quarks 
are in the ratio $1:1.15:4.5$, which are typical approximate values predicted by requiring them to be 
unified at the gauge coupling unification scale.}
\end{figure}
Within this figure $m_{t'_1}$ is fixed to be $600\gev$. For $M_U\ll M_Q$ 
the $t'_1$ state is a nearly pure $SU(2)_L$-singlet,
and it decays into $Wb$, $ht$ and $Zt$ primarily 
through the interaction $\epsu H_u q_3 \bar U$. The dominant decay 
mode in that limit is to $Wb$ at slightly over 50\%, but $ht$ and $Zt$ final states
are non-negligible. In the opposite limit $M_Q \ll M_U$, the state
$t'_1$ is nearly pure $SU(2)_L$-doublet, and it decays mostly into
$ht$, with $Zt$ a significant subdominant mode. Note that the case
$M_Q \approx M_U$ at the TeV scale is actually in a 
transition region for the
branching ratios. These results were obtained assuming 
that $\epsu$, $\epsup$ and $\epsd$ are in the low-scale 
ratios of $1:1.15:4.5$, which are approximate results from assuming they are unified at 
the gauge coupling unification scale. The thick vertical band in 
Figure~\ref{fig:MQoMU} indicates the ratio of $M_Q/M_U$ at the TeV scale 
under the assumption that $M_Q=M_U$ at the gauge coupling unification 
scale (typically in the range $\sim 3$-$8 \times 10^{16}\gev$ for these 
models). The left edge of this band corresponds to $\Mmess=10^{14}\gev$, 
while the right edge of the band corresponds to $\Mmess=10\Lambda$.

\begin{figure}[!tp]
\begin{minipage}[]{0.495\linewidth}
\begin{flushleft}
\includegraphics[width=\linewidth,angle=0]{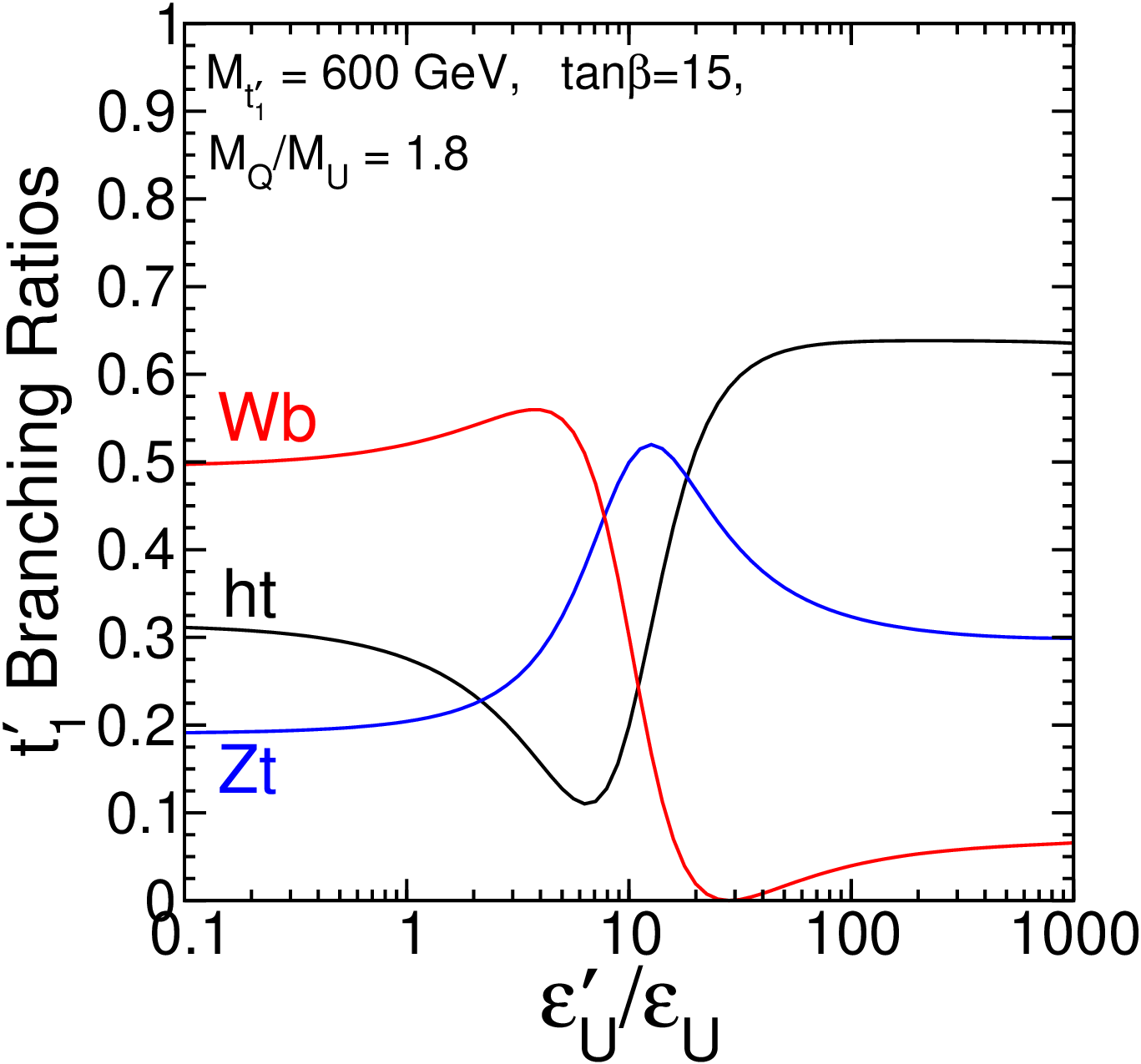}
\end{flushleft}
\end{minipage}
\begin{minipage}[]{0.495\linewidth}
\begin{flushleft}
\includegraphics[width=\linewidth,angle=0]{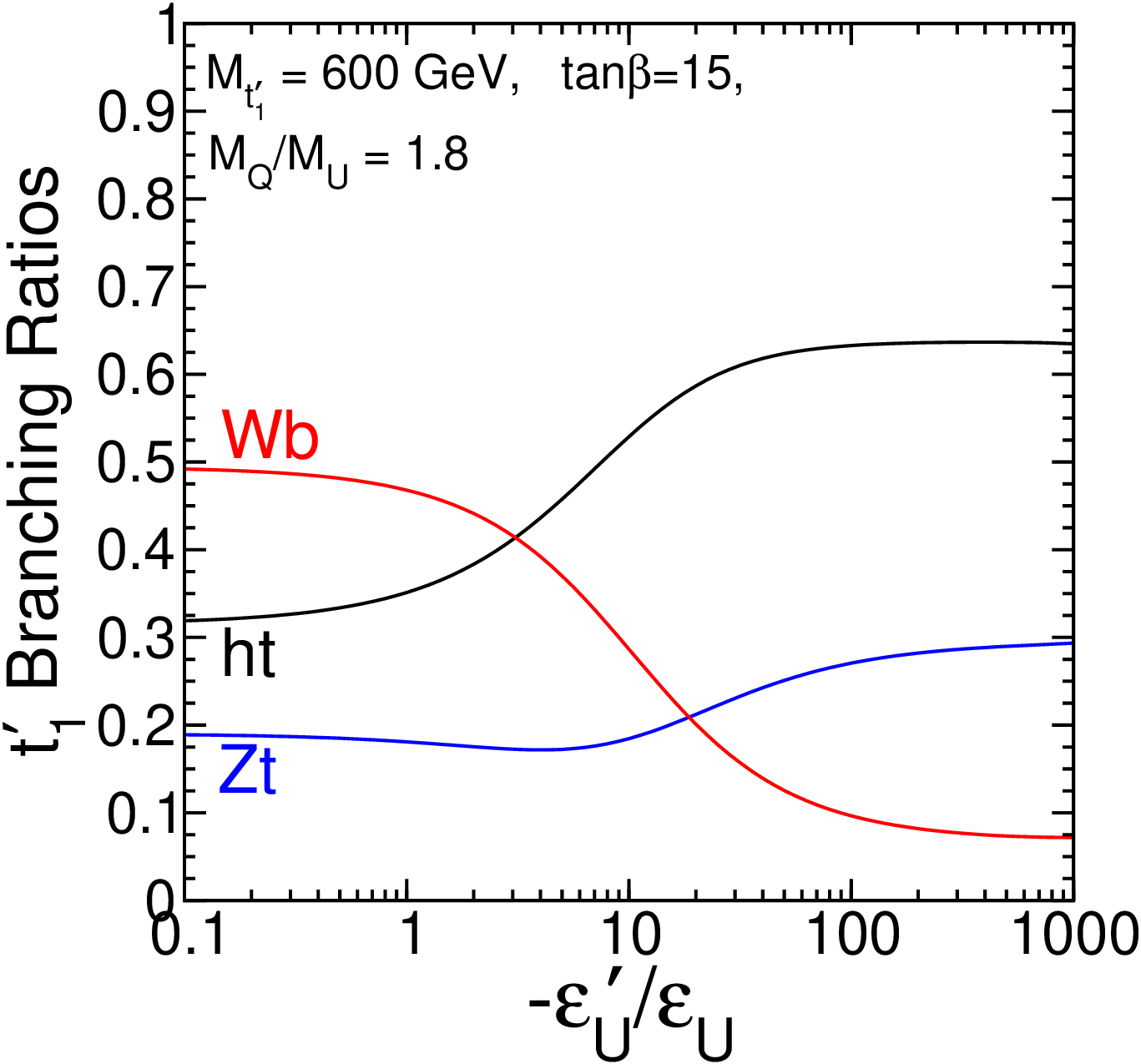}
\end{flushleft}
\end{minipage}
\begin{minipage}[]{0.495\linewidth}
\begin{flushleft}
\includegraphics[width=\linewidth,angle=0]{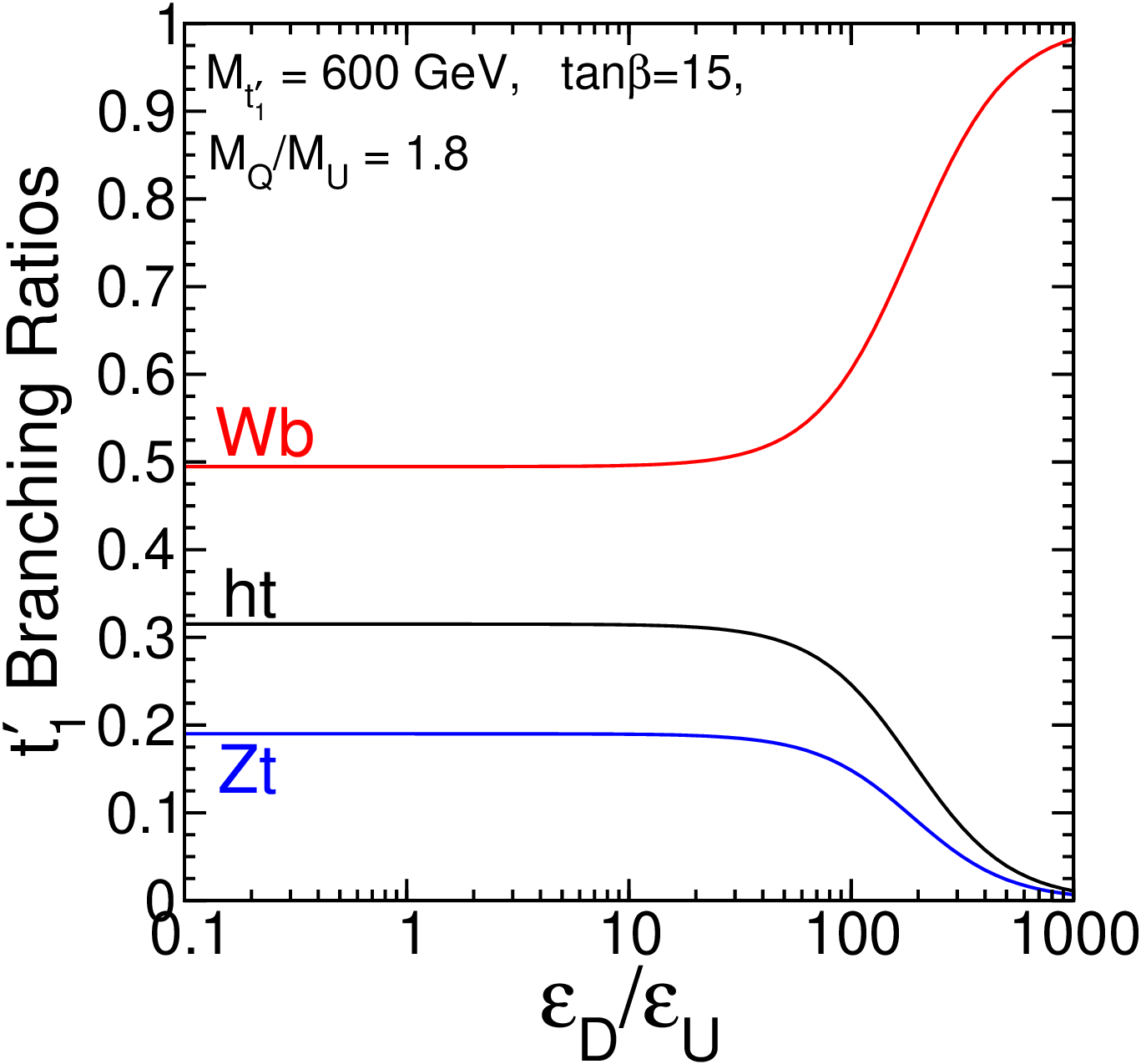}
\end{flushleft}
\end{minipage}
\begin{minipage}[]{0.495\linewidth}
\begin{flushright}
\begin{minipage}[]{0.9\linewidth}
\caption{\label{fig:BRepsvar} Branching ratios for $t_1'$ as a function of ratios
of the mixing couplings. In the top two panels, the ratio $\epsup/\epsu$ is varied
over positive (left panel) and negative (right panel) values, with $\epsd = 0$.
In the lower panel, the ratio $\epsd/\epsu$ is varied, with $\epsup=0$. (The
results do not depend on the sign of the ratio in the latter case.)}
\end{minipage}
\end{flushright}
\end{minipage}
\end{figure}

It is also interesting to consider the dependence on the mixing couplings
$\epsu$, $\epsup$, and $\epsd$, because the relation eq.~(\ref{eq:unifiedeps})
may not hold. This is illustrated in Figure \ref{fig:BRepsvar}, in which
we hold fixed $M_Q/M_U = 1.8$, and vary $\epsup/\epsu$ with $\epsd=0$,
and $\epsd/\epsu$ with $\epsup=0$. When the ratio $|\epsup/\epsu|$ is less than a few,
and when
$\epsd/\epsu \lsim 50$, one recovers
results similar to the unification-motivated results given in Figure \ref{fig:MQoMU}.
This is because in that case the effects of $\epsu$ are dominant because 
of the $SU(2)_L$-singlet nature of $t_1'$. However, for larger values of 
$\epsup$, one enters a ``$W$-phobic" regime for $t_1'$ in which the $ht$ 
final state can dominate with B$(Wb)$ very small. Conversely, for 
$\epsd$ very large, one goes over into the charged-current dominated case 
that B$(Wb) = 1$, which coincides with the prediction for a sequential 
$t'$, the subject of most experimental searches. Clearly, it is crucial 
that experimental searches go beyond this case, to take into account and 
hopefully exploit the $ht$ and $Zt$ final states.

\begin{figure}[!tp]
\begin{minipage}[]{0.49\linewidth}
\includegraphics[width=\linewidth,angle=0]{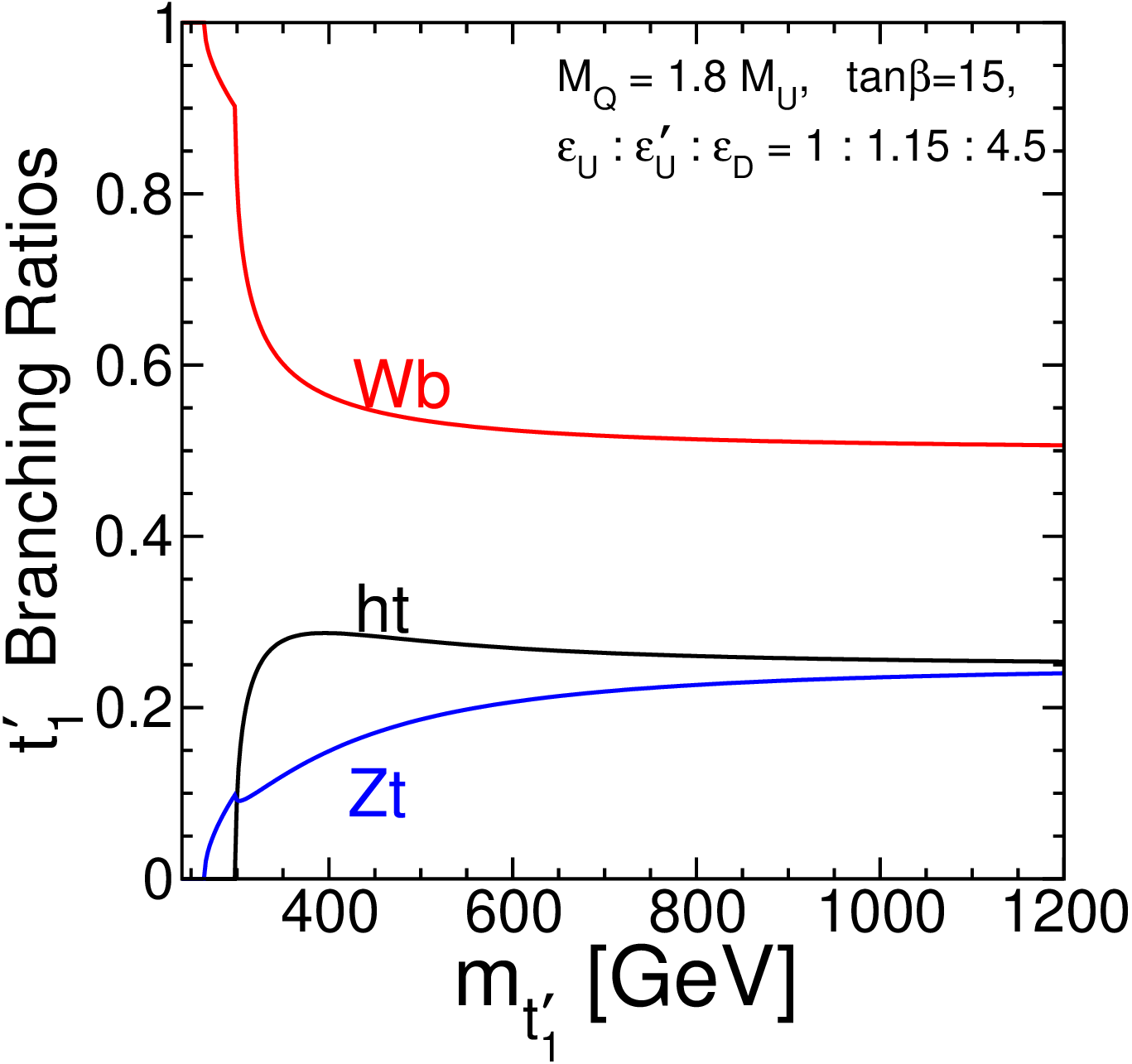}
\end{minipage}
~~~~
\begin{minipage}[]{0.45\linewidth}
\caption{\label{fig:BRtprime} Branching ratios for $t_1'$ as a function of its 
mass, obtained with 
$M_Q = 1.8 M_U$ and $\epsu : \epsup : \epsd = 1 : 1.15: 4.5$.}
\end{minipage}
\end{figure}

The dependence of these branching ratios on the magnitude of the $t_1'$ mass is mild 
provided that it is well above the $W,Z,h$ masses. This is illustrated in
Figure~\ref{fig:BRtprime}, which shows the branching fractions of $t'_1$ as 
a function of its mass, keeping fixed $k=1$ and using the unified
boundary conditions $M_Q = 1.8 M_U$ and 
$\epsu : \epsup : \epsd = 1: 1.15: 4.5$. 
For low 
$t_1'$
masses $\lsim 400$ GeV, the branching fractions show some
variation, but with higher $t_1'$ mass they
asymptote to B$(Wb)=50\%$, B$(ht)=25\%$, and B$(Zt)=25\%$,
but with $\mbox{B}(ht)>\mbox{B}(Zt)$ for finite masses relevant to the LHC.

\section{Precision tests from mixing with third-family fermions\label{sec:precisiontests}}

The introduction of an additional $b'$ quark that mixes with the third generation $b$ quark can 
induce a tree-level shift in the $Z$ boson coupling to the right-handed $b$ quark mass eigenstate compared to the SM. 
Such a shift is very severely constrained by the measurement of $R_b$ at LEP~\cite{Bamert:1996px},
with
\beq
R_b^{expt}=0.21629\pm 0.00066
\eeq
from \cite{LEPEWWG}.
The SM computed best fit value is \cite{LEPEWWGupdate,LEPEWWG}
\beq
R_b^{SM}=0.21579 \pm 0.00013.
\eeq
Thus, the $3\sigma$ range of allowed shifts in $R_b$ compared to the SM value is
\beq
-0.0015<\delta R_b < 0.0025,
\eeq
where $\delta R_b\equiv R_b-R_b^{SM}$.
From eqs.~(\ref{eq:ZLHdown}) and (\ref{eq:ZRHdown}) in the Appendix, and relating the
coupling conventions in \cite{LEPEWWGupdate,LEPEWWG} to ours by $g^b_L \equiv (c_W/g) g^Z_{d^\dagger_3 d_3}$
and $g^b_R \equiv - (c_W/g) g^Z_{\bar d^\dagger_3 \bar d_3}$,
we see that the tree-level
shifts in the couplings are
\beq
\delta g^b_L = 0,
\qquad\qquad
\delta g^b_R =-\frac{1}{2}|R'_{43}|^2 \approx -\frac{\epsd^2v^2\cos^2\beta}{2 m^2_{b'}},
\label{eq:gRtheory}
\eeq
which shows that the mixing always reduces the magnitude of the right-handed
$b$ quark couplings to the $Z$ boson. With this definition the resulting shift in 
$R_b$ is~\cite{Bamert:1996px}
\beq
R_b=R_b^{SM}(1+0.645\, \delta g_R^b)
\eeq
which implies that the $3\sigma$ range of allowed $\delta g_R^b$ is
\beq
\label{eq:gRshift}
-0.011 < \delta g_R^b < 0.018.
\eeq
Thus the requirement that $R_b$ is in 3-$\sigma$ agreement with experiment 
gives a constraint on $\epsd$, $\tan\beta$ and $m_{b'}$.  From eqs.~(\ref{eq:gRtheory}) 
and (\ref{eq:gRshift})  we find the requirement that
\beq
\label{eq:epsDlimit}
|\epsd| < 0.42\,  \tan\beta\, \left( \frac{m_{b'}}{500\gev}\right)\sqrt{1+\frac{1}{\tan^2\beta}}.
\eeq

A similar analysis follows from considering shifts in ${\cal A}_b$ and $A_{FB}^b$. The 
current~\cite{LEPEWWG,LEPEWWGupdate} experimental situation is that
\beq
{\cal A}_b^{expt}=0.923\pm 0.020~~~{\rm and}~~~A_{FB}^{b,expt}=0.0992\pm 0.0016,
\eeq
whereas the SM computed best fit values are
\beq
{\cal A}_b^{SM}=0.9346\pm 0.0001~~~{\rm and}~~~A_{FB}^{b,SM}=0.1033\pm 0.0008.
\eeq
Let us focus on $A^b_{FB}$, as the SM prediction is $2.8\sigma$ too high compared to the measurement (see table 8.4 of~\cite{LEPEWWG}).

From the definition $A_{FB}\equiv (g^2_L-g_R^2)/(g^2_L+g^2_R)$ one can compute the shift in $A^b_{FB}$ from a shift in $g_R^b$ to be
\beq
A_{FB}^b=A_{FB}^{b,SM}(1-1.7\, \delta g_R^b)~~~\Longrightarrow~~~
\delta A_{FB}^{b}=-0.18 \, \delta g_R^b.
\eeq
Since $\delta g_R^b<0$, this implies that the shift in the prediction of $A_{FB}^b$ is always positive, 
increasing the tension between theory and experiment. If we therefore 
assume that the $b-b'$ mixing is no 
more than a $1\sigma$ effect in the ``wrong" direction (i.e., $\delta A_{FB}^b < 0.0016$ 
from $b-b'$ mixing), this puts 
a limit on $\delta g_R^b$ that translates to exactly the same formula as eq.~(\ref{eq:epsDlimit}) 
except that $0.42$ is replaced by $0.38$. 
Thus, the constraints on $b'$ mixing are not very severe as long as 
$m_{b'}$ is greater than a few hundred GeV or $\tan\beta$ is not small.

Another way to constrain the mixing of SM third-family quarks with the exotic quarks
is through the CKM matrix element $V_{tb}$. Here, we cannot assume unitarity of the CKM matrix,
since it will not be in general [see eq.~(\ref{eq:ourCKMmatrix}) in the Appendix]. If the $\epsd$
coupling is present simultaneously with the $\epsu$ or $\epsup$ couplings, then the situation is
complicated by the fact that the $W$ boson will have small couplings to right-handed
SM quarks as well as left-handed quarks. 
For the sake of illustration, consider the case that only $\epsu$ is important,
and suppose that the SM Yukawa coupling matrices are such that if $\epsu$ were exactly
0, then 
$V_{tb}$ would be very close to 1 (as one finds in the SM with CKM unitarity assumed),
so that all mixing of the first two families 
with the third family and the vector-like quarks can be neglected. With those assumptions, from 
eq.~(\ref{eq:ourCKMmatrix}) 
we obtain 
\beq
1 - V_{tb} \approx 0.06\, \epsu^2 \sin^2\beta \left (\frac{\mbox{500 GeV}}{M_U}\right )^2,
\eeq
This can be compared to the values obtained from single top production 
without assuming CKM unitarity, $V_{tb} = 0.88 \pm 0.07$ (from Tevatron \cite{singletop})
and $V_{tb} = 1.04 \pm 0.09$ (from CMS \cite{CMSsingletop}). 
Thus, even if $\epsu$ is near unity and $t_1'$ is not much heavier than its experimental bound,
the CKM constraint does not impact the model.

Next, we consider the implications of $\tau-\tau'$ mixing. This mixing will induce a positive 
shift in the $g_L^\tau \equiv (g/c_W) g^Z_{e^\dagger_3e_3}$ coupling to the $Z$ boson,
while $g_R^\tau \equiv -(g/c_W) g^Z_{\bar e^\dagger_3 \bar e_3}$ is unaffected.
From eqs.~(\ref{eq:ZLHleptons}) and (\ref{eq:ZRHleptons}),
\beq
\delta g_L^\tau = \frac{1}{2} |U_{43}|^2 \approx \frac{\epse^2 v^2 \cos^2\beta}{2 m^2_{\tau'}},
\qquad\qquad
\delta g_R^\tau = 0.
\eeq
An important effect that results from this shift is an alteration in the $A_\tau$ observable. From the definition $A_\ell=(g_L^2-g_R^2)/(g_L^2+g_R^2)$, the shift in $\delta A_\tau$ from a shift in $\delta g_L^\tau$ is
\beq
\label{eq:Atau}
A_\tau=A_\tau^{SM}(1-23\, \delta g_L^\tau)
\eeq
which demonstrates the high sensitivity to changes in the $\tau$ lepton couplings to the $Z$.

The experimental and theoretical values~\cite{LEPEWWG} of $A_\tau$ are 
\beq
A^{expt}_\tau(P_\tau)=0.1465\pm 0.0033~~~{\rm and}~~~
A^{SM}_\tau(P_\tau)=0.1480\pm 0.0011.
\eeq
Keeping the prediction to within $3\sigma$ of the experimental measurement requires that
$-0.0120<\delta A_\tau < 0.0090$. Since $\delta A_\tau$ is always negative from the $\tau'$ mixing, the lower limit is the applicable constraint. From eq.~(\ref{eq:Atau}) we see that $\delta g^\tau_L<0.0033$, or
\beq
|\epse| < 0.23\, \tan\beta\, \left( \frac{m_{\tau'}}{500\gev}\right)\, \sqrt{1+\frac{1}{\tan^2\beta}}.
\label{eq:Ztauconstraint}
\eeq
This requirement is not terribly constraining, especially considering that the SM $\tau$ Yukawa coupling 
$y_\tau=0.01$ is much smaller than the general constraints on $\epsilon_E$ when $m_{\tau'}>100\gev$. 

Finally, one can attempt to constrain the $\tau$-$\tau'$ mixing through the $\tau$ decay measurement.
The analysis of \cite{SwainTaylor} corresponds to $|U_{43}|^2 < 0.0053$ in the notation of the Appendix
of the present paper, which therefore implies the same constraint as
eq.~(\ref{eq:Ztauconstraint}) but with 
$0.21$ replacing $0.23$. However, this is a 1-$\sigma$ constraint.
Also, this assumes that the PMNS matrix is unitary, and that mixing in the electron and muon 
sectors is absent, which need not hold \cite{Lacker:2010zz}. In any case, there is no impact
on the coupling $\epse$ in this model unless $\tan\beta$ is small, and the $\tau'$ is light.

\section{LHC phenomenology\label{sec:colliderpheno}}
\setcounter{equation}{0}
\setcounter{figure}{0}
\setcounter{table}{0}
\setcounter{footnote}{1}

The exotic quarks could in principle have a significant effect on the production and decay
of the lightest Higgs boson. For an additional chiral fourth family, which relies 
entirely on Yukawa couplings for its large masses, 
there is a very large positive effect on the production cross-section
\cite{sigmahfourthfamily,KPST}, 
in strong conflict with the current limits \cite{ATLAScombined,CMScombined}. 
However, in the vector-like 
model under present consideration, the situation
is very different. The corrections to the $hgg$ and $h\gamma\gamma$ 
effective interactions can be found from the 
general formulas in \cite{Ishiwata:2011hr}. Applying these, we find that
for the case $k' \ll k \approx 1$, these corrections are totally negligible. 
Even if $k'$ is sizable, the
corrections to $gg\rightarrow h \rightarrow \gamma\gamma$ are quite modest, at most at the 5\% level for 
$M_{t_1'} = 500$ GeV and $\tan\beta = 5$ and $k=1$, $k'=0.7$, and can have either sign depending on
the relative phases in the $t_1', t_2', b'$ sector mass matrices. For larger $\tan\beta$, the size of 
the effect decreases. We conclude that at least for LHC physics in the short term, the loop effects of 
the exotic quarks on Higgs production and decay are probably too small to hope to observe.
 
The model under consideration differs from
other variants of the MSSM in that there are two distinct paths to a new physics discovery.
First, we may discover the odd $R$-parity superpartners of the SM
states. Second, we have the exotic quark and lepton states. These two possibilities are essentially 
decoupled, and it is unclear which of them 
should provide the initial discovery of physics beyond the SM, since the masses
and decays are negotiable within the general model framework. We will begin by commenting
on features of the superpartner phenomenology at LHC, making the comparison to other standard searches.
 
\asubsection{Superpartner signals\label{subsec:superpartnersignals}}

\begin{figure}[t]
\begin{minipage}[]{0.49\linewidth}
\includegraphics[width=\linewidth,angle=0]{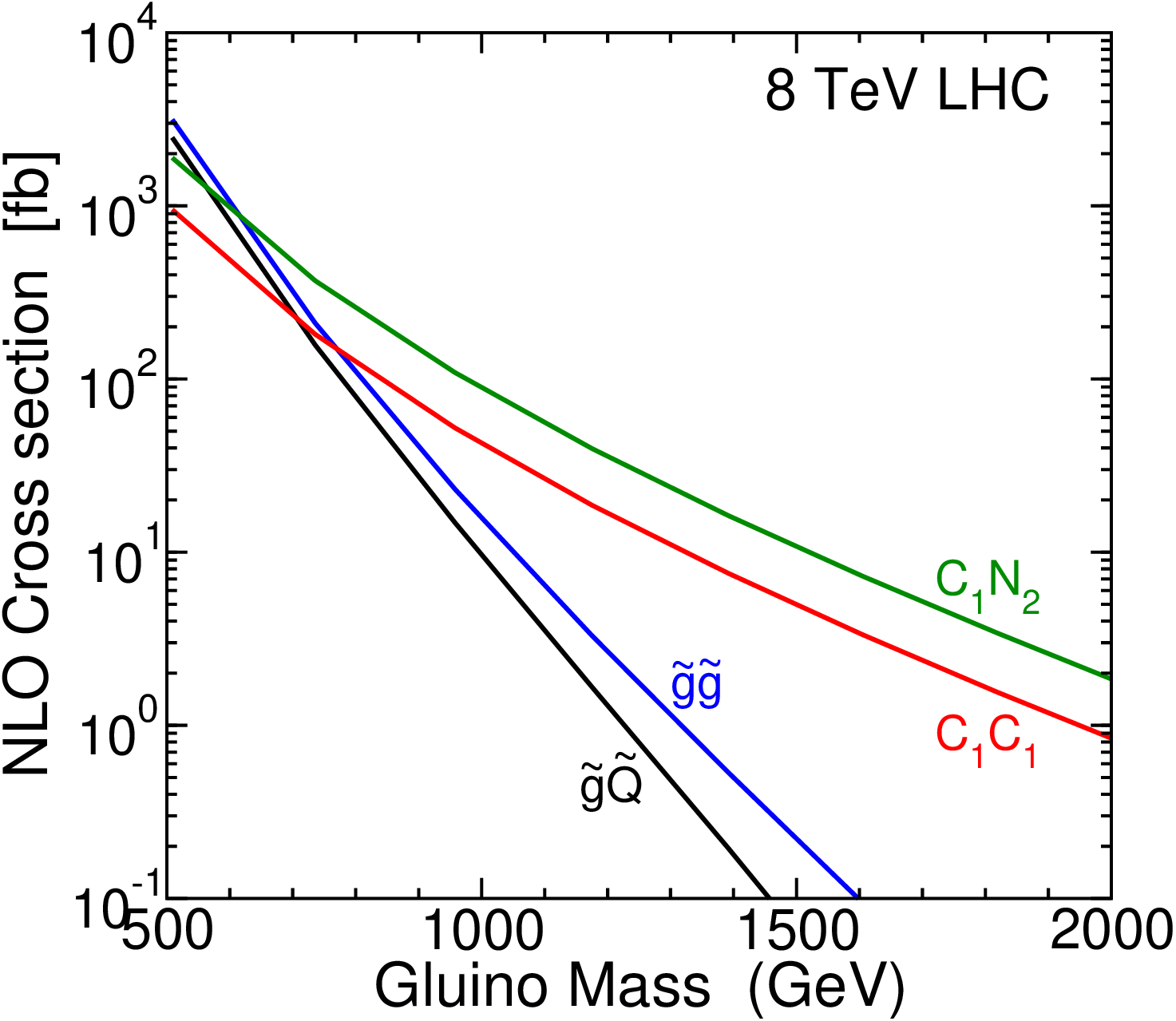}
\end{minipage}
\begin{minipage}[]{0.49\linewidth}
\begin{flushright}
\includegraphics[width=\linewidth,angle=0]{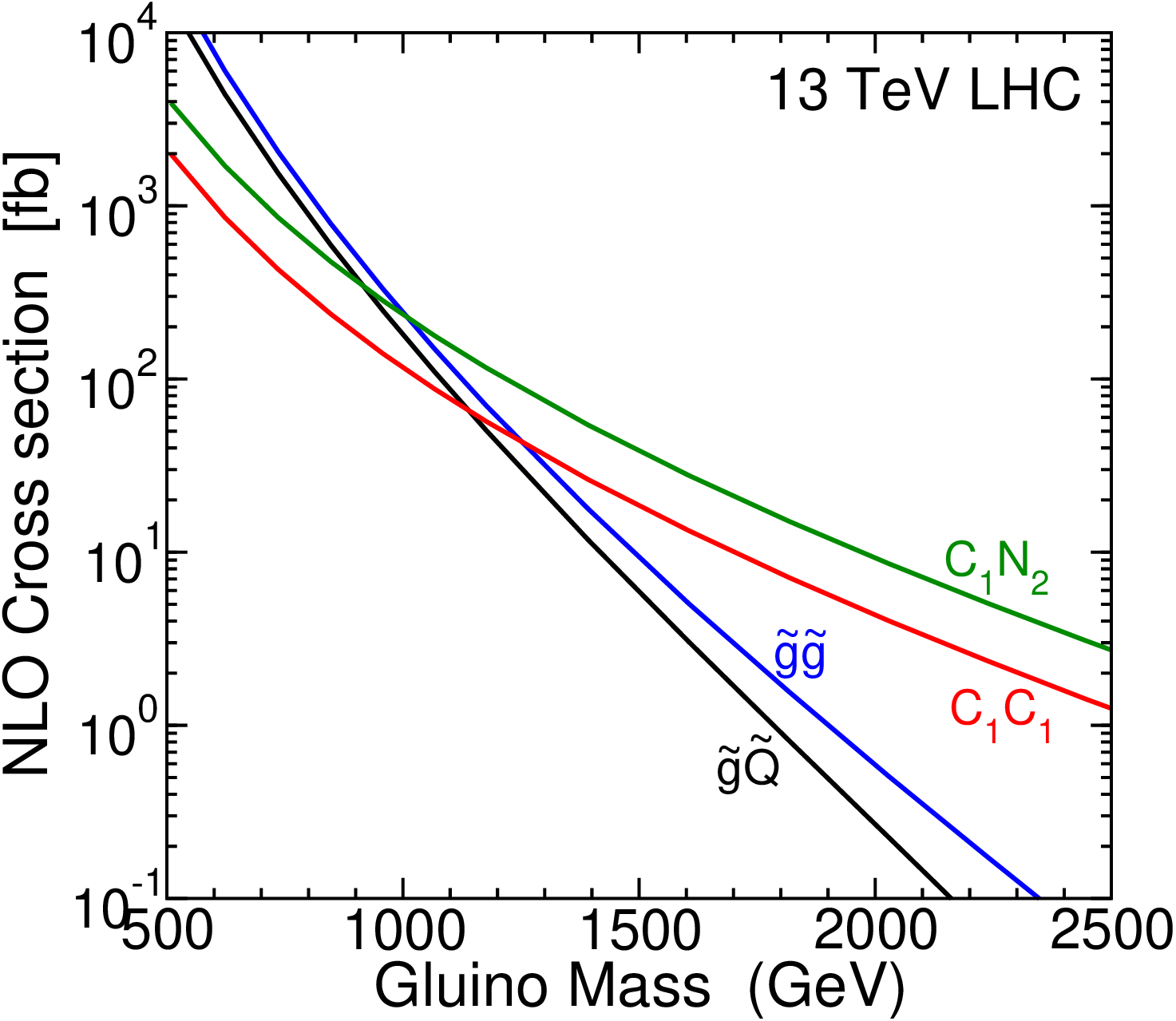}
\end{flushright}
\end{minipage}
\caption{\label{fig:gmsbNLO}NLO production cross-sections for $\tilde g\tilde g$
and $\tilde g \tilde Q + \tilde g \tilde{\overline{Q}}$ and $\tilde C_1^\pm \tilde N_1$ and $\tilde C_1 \tilde C_1$,
as a function of the gluino mass, for GMSB models with extra vector-like quarks
in this paper with $\Mmess = 10 \Lambda$ and $\tan\beta=15$,
in $pp$ collisions with $\sqrt{s} = 8$ and 13 TeV.}
\end{figure}

If the NLSP is a neutralino $(\tilde N_1)$ that is stable on detector-crossing time scales, 
the resulting phenomenology is very similar to 
``standard supersymmetry" signatures (e.g., mSUGRA). The squarks are comparatively heavy, with up and down squarks, which play the most important role in LHC production, between about 1.6 and 2.3 times
heavier than the gluino
(see for example Figures \ref{fig:samples1}, \ref{fig:samples2}). Therefore, 
the discovery potential comes mostly from gluino pair 
production, gluino+squark production, or the production of wino-like charginos and neutralinos, 
followed in each case by decays to jets, leptons and large missing energy. 
The production cross-sections
computed to next-to-leading order by {\tt Prospino} \cite{prospino}
are shown in Figure \ref{fig:gmsbNLO}
for the most important processes $pp \rightarrow \tilde g\tilde g$ and
$\tilde g \tilde Q + \tilde g \tilde{\overline{Q}}$ and
$\tilde C_1^\pm \tilde N_2$ and $\tilde C_1^+ \tilde C_1^-$.
Here we used a model line
with $\Mmess = 10 \Lambda$, $\tan\beta=15$, $k=1$, and $\mu>0$, but the dependence on these 
particular assumptions is mild, with the exception of 
$\tilde g \tilde Q + \tilde g \tilde{\overline{Q}}$, which becomes smaller
for a given $M_{\tilde g}$ if $\Mmess$ is larger. Although the gluino+gluino  
and gluino+squark pair production
cross-section are smaller than the chargino-neutralino rates
for $M_{\tilde g} \gsim 650$ GeV at $\sqrt{s} = 8$ TeV, and for
$M_{\tilde g} \gsim 1050$ GeV at $\sqrt{s} = 13$ TeV, the gluino and squark signals 
should have higher acceptances due to more visible energy. However, any attempts to probe much beyond
$M_{\tilde g}= 1$ TeV at $\sqrt{s}=8$ TeV may have to rely on chargino/neutralino production
rather than gluino/squark production.

The branching ratios of the gluino are, for the typical low-$\Mmess$
model in Figure \ref{fig:samples1}:
\beq
\tilde g &\rightarrow& 
jj \tilde C_1\>\>(\mbox{38\%}),
\qquad t b \,\tilde C_1\>\>(\mbox{17\%}),
\qquad jj \tilde N_2\>\>(\mbox{19\%}),
\qquad b b \tilde N_2\>\>(\mbox{6\%}),\nonumber\\
&& t t \tilde N_2\>\>(\mbox{3\%}),
\qquad jj \tilde N_1\>\> (\mbox{12\%}),
\qquad b b \tilde N_1\>\> (\mbox{1\%}),
\qquad t t \tilde N_1\>\> (\mbox{4\%}),
\label{eq:gluinoBRlo}
\eeq
from SDECAY \cite{SDECAY}, where $j$ denotes a jet from a $u,d,s,c$ quark, and 
the notation omits the distinction between quarks and antiquarks. 
Up and down squarks essentially always decay to a gluino and a very energetic jet. 
The wino-like
charginos and neutralinos decay almost entirely through the lighter stau,
which then decays as $\tilde \tau_1 \rightarrow \tau \tilde N_1$ with a branching ratio of 100\%:
\beq
\tilde C_1 &\rightarrow& \tilde \tau_1 \nu \rightarrow \tau \nu \tilde N_1 \>\>\>(\mbox{96\%}),
\qquad W \tilde N_1\>\>(\mbox{4\%}),
\\
\tilde N_2 &\rightarrow& \tilde \tau_1 \tau \rightarrow \tau^+\tau^- \tilde N_1\>\,\>(\mbox{96\%}),
\qquad h \tilde N_1\>\>(\mbox{4\%}),
\eeq
Thus a high proportion of events will have 2, 3, or 4 taus in the final state, manifested 
either as hadronic tau jets or softer $e,\mu$.
This is an important difference
compared to mSUGRA, where comparable models with such heavy squarks have large $m_0$ and therefore
also have heavy staus, and so cannot 
produce such a predominance of taus in the final state. 

\begin{figure}[t]
\begin{minipage}[]{0.55\linewidth}
\includegraphics[width=3.1in,angle=0]{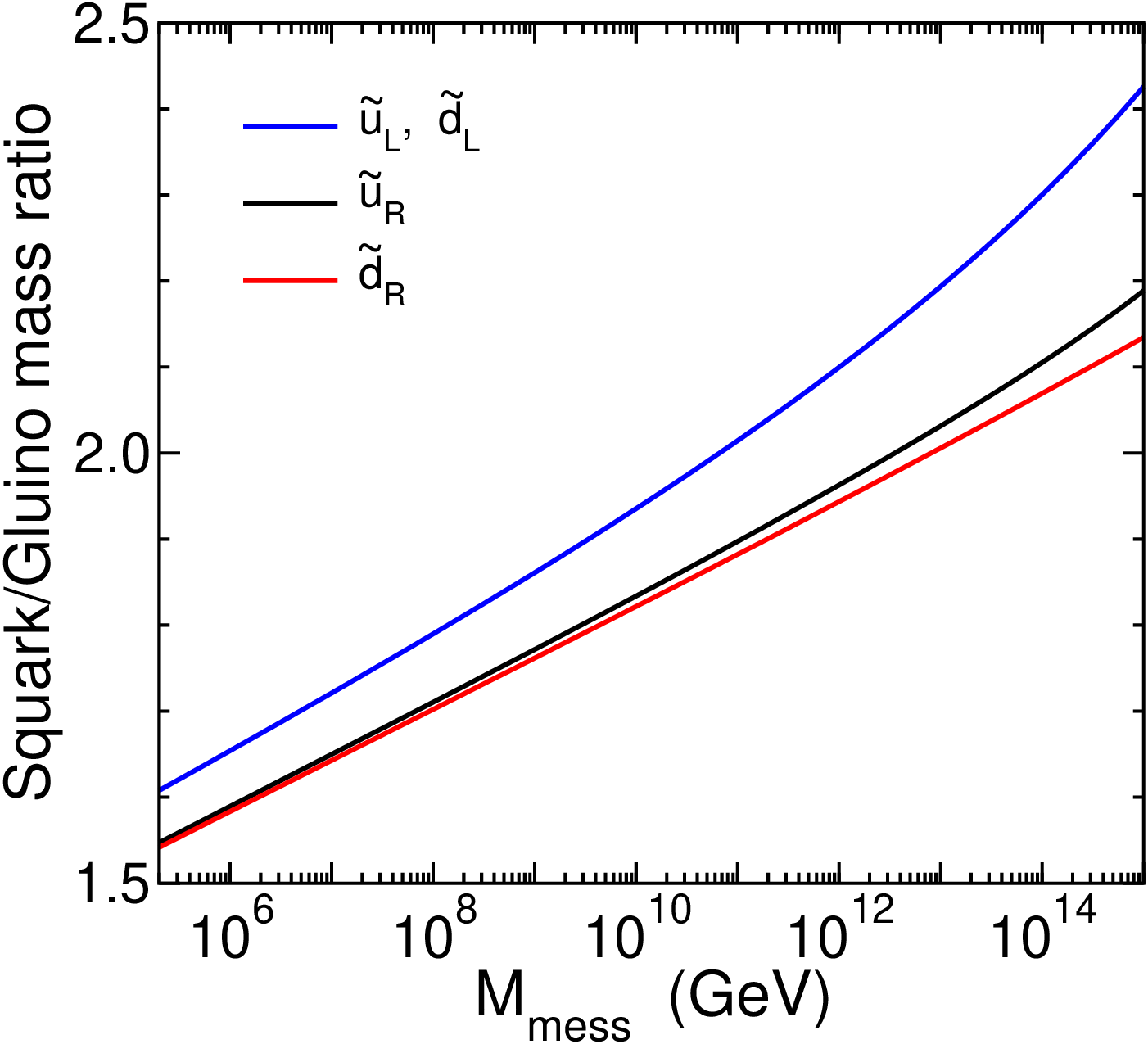}
\end{minipage}
~~~
\begin{minipage}[]{0.4\linewidth}
\caption{\label{fig:squarkgluinomassratio}The ratio of the first-family squark masses to the
mass of the gluino, as a function of $M_{\rm mess}$. Here, $\tan\beta = 15$, $\mu>0$, and
$\Lambda = 160$ TeV.}
\end{minipage}
\end{figure}

In contrast, models with higher $\Mmess$ will have 
$M_{\tilde \tau_1} > M_{\tilde C_1} \approx M_{\tilde N_2}$, 
as illustrated by the example in Figure \ref{fig:samples2}, implying a much lower tau multiplicity.
In that example model, we have for the gluino decays
\beq
\tilde g &\rightarrow& 
jj \tilde C_1\>\>(\mbox{33\%}),
\qquad t b \,\tilde C_1\>\>(\mbox{18\%}),
\qquad jj \tilde N_2\>\>(\mbox{16\%}),
\qquad b b \tilde N_2\>\>(\mbox{6\%}),\nonumber\\
&& t t \tilde N_2\>\>(\mbox{3\%}),
\qquad jj \tilde N_1\>\> (\mbox{13\%}),
\qquad b b \tilde N_1\>\> (\mbox{2\%}),
\qquad t t \tilde N_1\>\> (\mbox{9\%}),
\eeq
similar to eq.~(\ref{eq:gluinoBRlo}), with a slightly higher average number of $b$ jets.
However, the wino-like charginos and neutralinos decay very differently than in the low-$M_{\rm mess}$ case:
\beq
\tilde C_1 &\rightarrow& W \tilde N_1\>\>(\mbox{100\%}),
\\
\tilde N_2 &\rightarrow& h \tilde N_1\>\>(\mbox{97\%}),
\qquad Z \tilde N_1\>\>(\mbox{3\%}).
\eeq
This means that over 40\% of gluino pair production events, and almost all $\tilde C_1 \tilde N_2$
events, will have a Higgs boson in them. For such models with $M_{\tilde \tau_1} > M_{\tilde C_1} 
\approx M_{\tilde N_2}$, the signals are sufficiently similar to mSUGRA ones with large $m_0$
that one can safely approximate the limits by those obtained by ATLAS and CMS for the same gluino mass
and heavier squarks.
The ratios of squark masses to the gluino mass are shown for our model in Figure \ref{fig:squarkgluinomassratio}.
These squark/gluino mass ratios correspond approximately to CMSSM models with
$m_0/M_{1/2}$ ranging from about 3.4 (for low $\Mmess$) to 5.2 (for high $\Mmess$).
At present, the LHC limits for these large $m_0$ cases 
imply only $M_{\tilde g} \gsim 850$ GeV from  
\cite{ATLASSUSYlargem0,CMSSUSYlargem0}. A direct comparison is hindered somewhat
by the fact that the LHC collaborations unfortunately choose to present results for the CMSSM
in terms of the unphysical input variables $(m_0, M_{1/2})$ 
rather than physical gluino and squark masses.

\begin{figure}[t]
\begin{minipage}[]{0.55\linewidth}
\includegraphics[width=3.1in,angle=0]{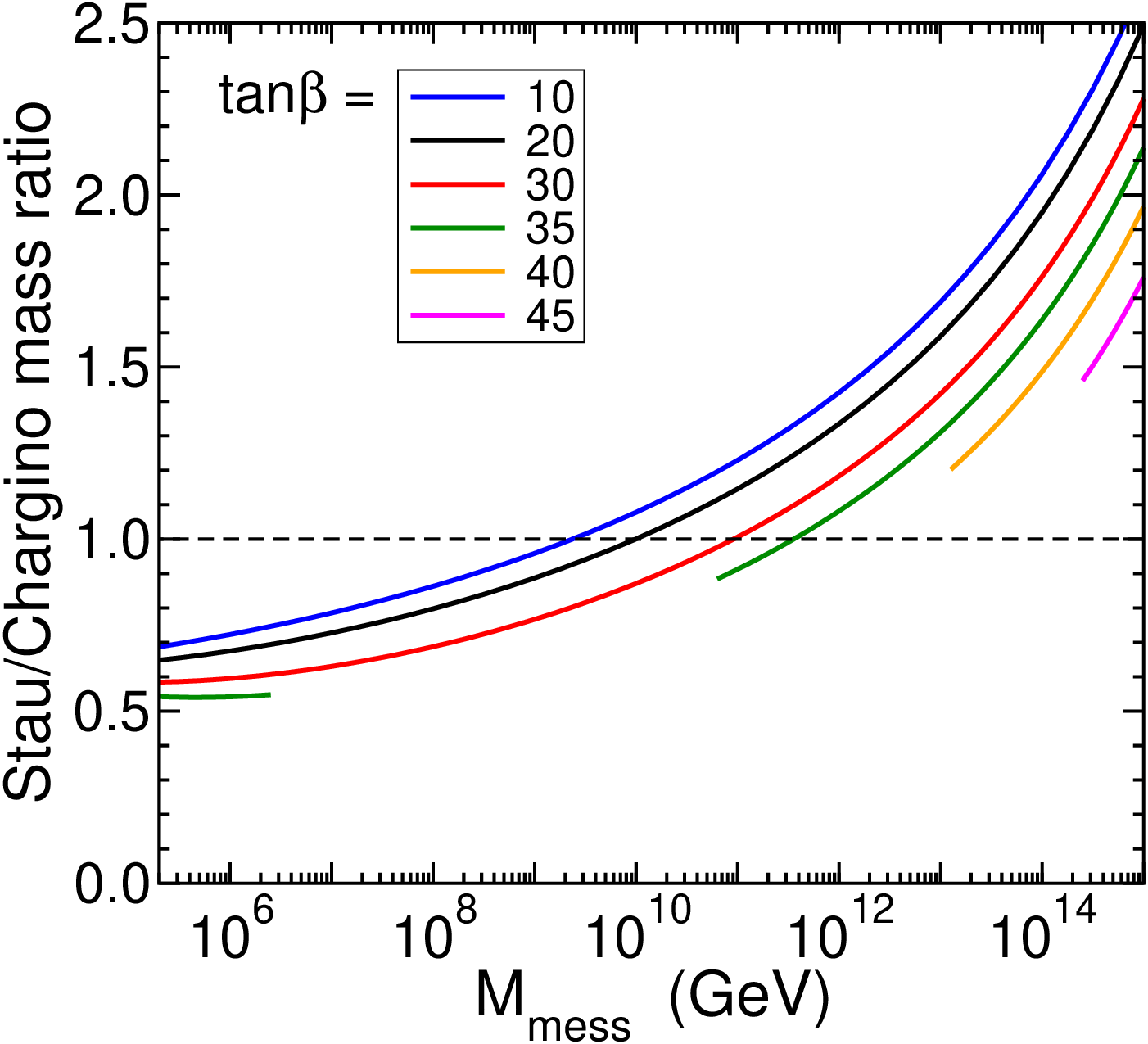}
\end{minipage}
~~~
\begin{minipage}[]{0.4\linewidth}
\caption{\label{fig:staucharginomassratio}The ratio $M_{\tilde \tau_1}/M_{\tilde C_1}$ of the lighter stau and chargino
masses, as a function of $M_{\rm mess}$, for $\tan\beta = 
10,20,30,35,40,45$. Here,
$\Lambda = 160$ TeV and $\mu>0$. The missing parts of the lines for $\tan\beta=35,40,45$
are due to the vacuum stability requirement.}
\end{minipage}
\end{figure}

Because of the importance of the 
transition in parameter space between the cases that $\tilde \tau_1$ is lighter or heavier 
than the wino-like neutralinos and charginos, we show in 
Figure \ref{fig:staucharginomassratio} how the ratio 
$M_{\tilde \tau_1}/M_{\tilde C_1}$ behaves as a function of $M_{\rm mess}$ for various values of
$\tan\beta$. 
For $M_{\tilde \tau_1}/M_{\tilde C_1} \lsim 1$, the decays $\tilde C_1 \rightarrow \tilde \tau_1 \nu$
and $\tilde N_2 \rightarrow \tilde \tau_1 \tau$ dominate; otherwise, decays to $W$ and $h$ dominate.
Depending on $\tan\beta$, we see from  Figure \ref{fig:staucharginomassratio} that the 
transition between these two regimes occurs at an intermediate scale of a few times $10^9$ 
to a few times $10^{11}$ GeV.

If the decay $\tilde N_1 \rightarrow \gamma \tilde G$ is prompt, then the above event topologies will
be supplemented by two energetic isolated photons, for which SM backgrounds are quite low.
This would increase the discovery potential dramatically, and would probably guarantee that the discovery would happen in the $\tilde C_1 \tilde N_2$ production channel, due to its larger cross-section. Because the NLSP decay width is proportional to $1/F^2$, where
$F  \gsim \Lambda \Mmess$, we see from Figure \ref{fig:staucharginomassratio} 
that the prompt neutralino NLSP decay 
signal should be $\tau\tau\tau\gamma\gamma + \ETmiss$, where $\tau$ can be either 
a softer lepton or a hadronic tau jet. 

Another possibility is that the NLSP is the lighter stau, which can only 
occur in our model framework if $\tan\beta$ is large. (However, 
$\tan\beta$ cannot be too large, and $M_{\rm mess}$ must be low, given 
the constraints on vacuum stability evident in Figure 
\ref{fig:tbMgluino}.) In that case, all superpartner decays chains will 
terminate in $\tilde \tau_1 \rightarrow \tau \tilde G$, where $\tilde G$ 
is the goldstino (gravitino). In each decay chain from a gluino, 
chargino, or neutralino parent, lepton flavor conservation dictates that 
there is another $\tau$ produced. This means that if the NLSP stau decay 
is prompt, essentially all supersymmetric events will have at least 4 
taus, while if it is not prompt, one has at least 2 taus and 2 
quasi-stable staus which can be detected as slow-moving heavy charged 
particles. However, the parameter space in which there is a stau NLSP is 
limited, as one runs into the constraint from stability of the vacuum 
noted in \cite{Hisano:2010re,Endo:2012rd}. For example, in the models of 
Figure \ref{fig:tbMgluino}, one sees that this requires a low $\Mmess$, 
and $M_{\tilde g} < 1100$ GeV, and $21 < \tan\beta < 34$.

\asubsection{Search for $t_1'$ at the LHC\label{subsec:tprimesearches}}

\begin{figure}[t]
\begin{minipage}[]{0.55\linewidth}
\includegraphics[width=3.4in,angle=0]{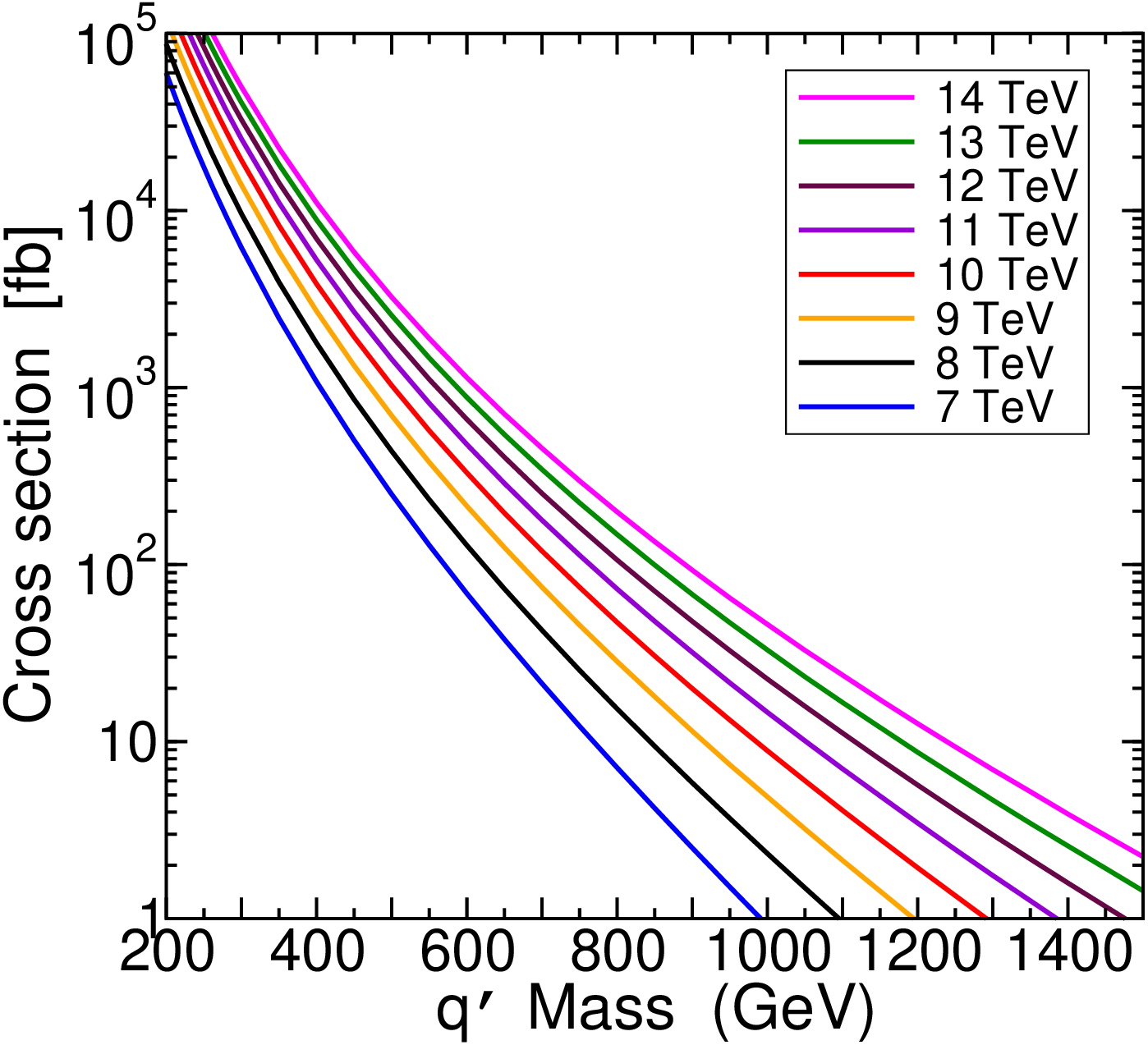}
\end{minipage}
~~~
\begin{minipage}[]{0.4\linewidth}
\caption{\label{fig:topprimeCS}The production cross-section
in $pp$ collisions for an exotic quark-antiquark pair
$\sigma(q'\bar q')$ as a function of its mass $M_{q'}$,
for $\sqrt{s} = 7,8,9,10,11,12,13,14$ TeV, obtained using {\tt HATHOR} \cite{HATHOR}.
}
\end{minipage}
\end{figure}

The production cross-sections for generic exotic heavy quarks at the LHC are shown in
Figure \ref{fig:topprimeCS}, for various $\sqrt{s}$ values.
The collider phenomenology of the $t'_1$ depends crucially on whether it decays promptly or not.
If the mixing between the exotic quarks and the SM quarks is very small, then there is
a chance that $t_1'$ could be stable on time scales relevant for collider detectors. Assuming first the
unification ratio $M_Q/M_U=1.8$, so that $t_1'$ is mostly $SU(2)_L$ singlet, we find a lifetime 
for $t'_1$ of  
\beq
c\tau = \left (\frac{\mbox{1000 GeV}}{M_{t_1'}}\right )
        \left (\frac{10^{-7}}{\epsilon} \right )^2
\times
\left \{ \begin{array}{ll}
\mbox{0.94 mm}& {\rm for}\>\, \epsilon = \epsu \sin\beta,\>\>\>\epsup = \epsd = 0, 
\\[-5pt]
\mbox{290 mm}& {\rm for}\>\, \epsilon = \epsup \sin\beta,\>\>\>\epsu = \epsd = 0,
\\[-5pt]
\mbox{340 mm} & {\rm for}\>\, \epsilon = \epsd \cos\beta,\>\>\>\epsu = \epsup = 0.
\end{array}
\right.
\label{eq:ctausinglettopp}
\eeq
For simplicity, 
we have taken the limit $M_{t_1'}^2 \gg M_h^2$ in eq.~(\ref{eq:ctausinglettopp}). 
For masses closer to the weak boson masses, kinematic factors
increase the lifetime somewhat. To illustrate the opposite limit of the $t_1'$ being mostly an
$SU(2)_L$ doublet, consider $M_Q = 0.5 M_U$, which results in
\beq
c\tau = \left (\frac{\mbox{1000 GeV}}{M_{t_1'}}\right )
        \left (\frac{10^{-7}}{\epsilon} \right )^2
\times
\left \{ \begin{array}{ll}
\mbox{270 mm}& {\rm for}\>\, \epsilon = \epsu \sin\beta,\>\>\>\epsup = \epsd = 0, 
\\[-5pt]
\mbox{1.8 mm}& {\rm for}\>\, \epsilon = \epsup \sin\beta,\>\>\>\epsu = \epsd = 0,
\\[-5pt]
\mbox{2.0 mm} & {\rm for}\>\, \epsilon = \epsd \cos\beta,\>\>\>\epsu = \epsup = 0.
\end{array}
\right.
\label{eq:ctaudoublettopp}
\eeq
If we require for the definition of prompt decays that $c\tau< 1$ mm,
then we need only either $\epsu$ or $\epsup$ to be greater than a few times $10^{-7}$ to 
ensure prompt decays.
Note that the $\epsd$ contribution to the inverse lifetime is suppressed by $\cos^2\beta$. 

\begin{figure}[t]
\begin{minipage}[]{0.49\linewidth}
\begin{center}
\includegraphics[width=0.9\linewidth,angle=0]{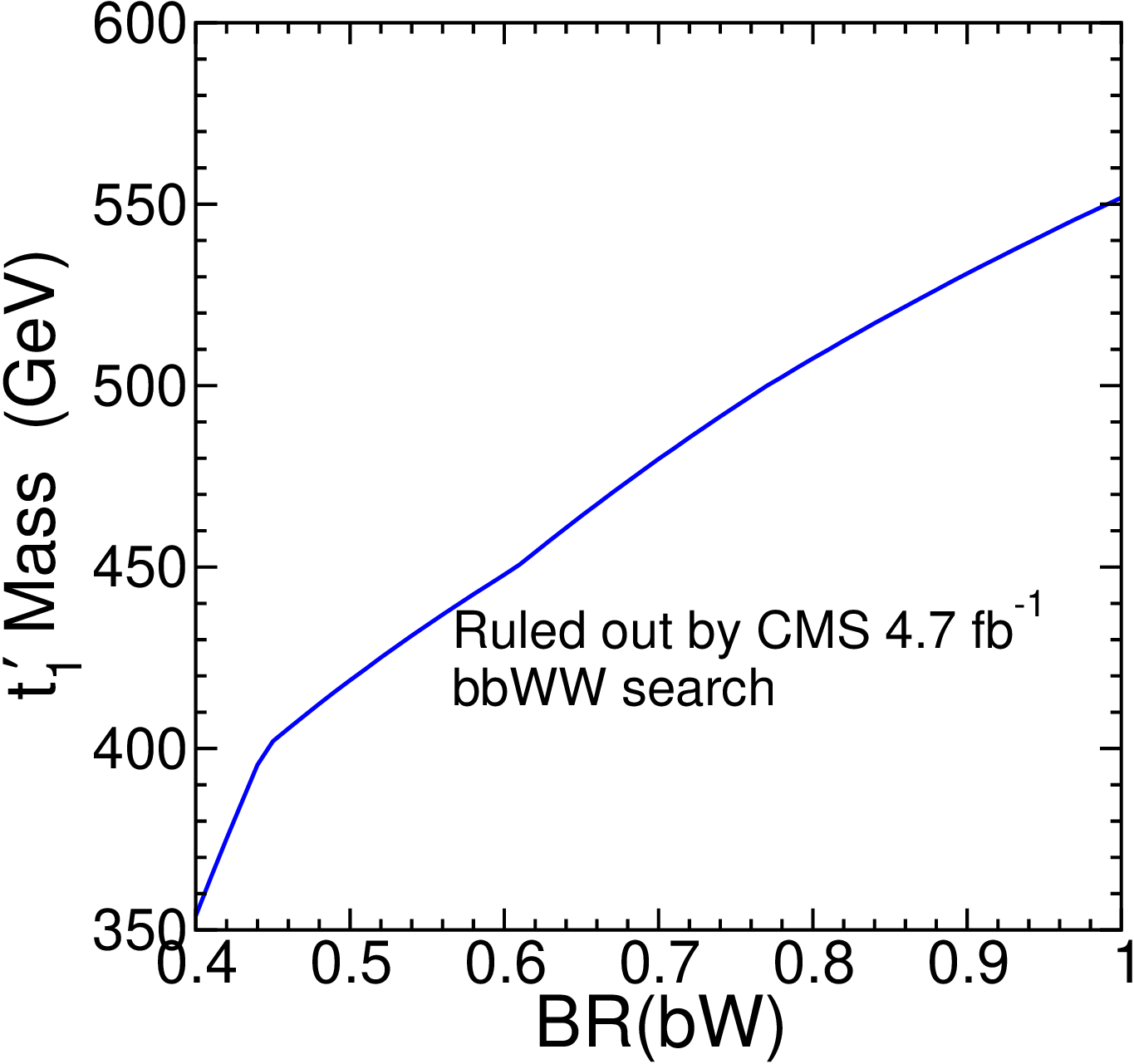}
\end{center}
\end{minipage}
\begin{minipage}[]{0.49\linewidth}
\begin{minipage}[]{0.04\linewidth}\end{minipage}
\begin{minipage}[]{0.90\linewidth}
\caption{\label{fig:mtpBlimits}Limits on $m_{t'_1}$ vs.\ B$(t'_1\to bW)$, 
from
the CMS search for the final state $bbWW$ based
on 4.7$^{-1}$ fb reported in~\cite{Halk:2012}.}
\end{minipage}\end{minipage}
\end{figure}

Let us first assume the case of $t'_1$ decaying promptly. The LHC 
experiments have several analyses based on the production of heavy top-like quarks.
The most stringent direct search bounds are from CMS, but are limited to the 
extreme cases that either the $Wb$ or the $Zt$ final state dominates. For
$B(t'\rightarrow Wb) = 1$, CMS obtains $M_{t'} > 557$ GeV using 4.7 fb$^{-1}$ \cite{CMStprimeWb},
and for $B(t' \rightarrow Zt) = 1$, they obtain $M_{t'} > 475$ GeV using 1.14 fb$^{-1}$ \cite{CMStprimeZt}.
However, in our case the branching ratios are split among the final states depending on
the mixing couplings, as seen in Figures \ref{fig:MQoMU}-\ref{fig:BRtprime}.
In much of parameter space, where
$M_{t'_1}>$ few hundred GeV and $M_Q>M_U$, we find $B(t'_1\to bW)\approx 50\%$.
Therefore, requiring $t'_1\bar t'_1\to bWbW$ means a reduction by a factor of 4 in
total cross-section applicable for the analysis. Taking this factor into
account, and comparing the cross-section limits at LHC as derived
in~\cite{Halk:2012} with the total direct rate in our theory assuming
$B(t'_1\to bW)=50\%$, we extrapolate to find the current limit to be
$M_{t'_1}\gsim 420$, even without using the other final states. This is consistent
with another recent analysis~\cite{RaoWhiteson}.
In Figure~\ref{fig:mtpBlimits} we show the
limits as a function of $B(t'_1\to bW)$ based on the $t'_1\bar t'_1\to
bWbW$ limits only.
Of course,
a more general search using all three final states $Wb$, $Zt$ and $ht$ will find a 
stronger limit. 
In ref.~\cite{RaoWhiteson} a reanalysis of these direct search 
limits together with a reinterpretation of 
an ATLAS search \cite{ATLASbprimeWt} for 
$b'\rightarrow Wt$ in terms of $t'$ pair production is argued to give a bound $M_{t_1'} > 415$ GeV,
for any combination of the three branching ratios for $t' \rightarrow Wb, Zt, Zh$. 
Going forward, the detector collaborations should strive to incorporate all
three final states in their search strategies as much as possible, in order to maximize the
model-independent 
reach in the $t_1'$ mass. 
For any value of $M_{t_1'}$, the mixing couplings can be chosen in such a 
way that any of the $Wb$, $Zt$, or $ht$ is the dominant decay mode, and 
they may all be comparable to each other. This should be kept in mind in 
the planning and interpretation of hadron collider searches. Even if 
$t'\rightarrow Wb$ has the largest branching ratio, searches with mixed 
final states $(t' \rightarrow Wb)(t'\rightarrow Zt)$ or $(t' \rightarrow 
Wb)(t'\rightarrow ht)$ may give the strongest signal, exploiting the 
presence of $Z \rightarrow \ell^+\ell^-$ and 2, 3, or 4 $b$-tagged jets, 
or even $h \rightarrow b \overline b$ or $h \rightarrow \tau^+\tau^-$ 
with a ``known" invariant mass of $\hmass$. This is especially important 
given that there are other, completely different, new physics models that 
predict exotic quarks within the reach of the LHC 
\cite{LittleHiggs}-\cite{Geller:2012wx}, which can span the possible 
branching ratios into these three final states. It would be especially 
interesting to observe and study events with $h\rightarrow b\bar b$ or $h 
\rightarrow \tau^+\tau^-$ in $t_1'$ production, since these decay modes 
are quite difficult to observe at the LHC from direct Higgs production.

If the $t'_1$ is stable, it can be searched for as a strongly interacting 
heavy stable charged particle. The implications for the search are 
expected to be similar to that of a quasi-stable top squark when, for 
example, it is the NLSP and the decay to gravitino is very suppressed and 
the lifetime is greater than the size of the detector, $c\tau>\ell_{\rm 
detector}$. The search strategy~\cite{CMS-11-022} relies on first 
identifying large $dE/dx$ energy depositions in the inner tracker due to 
the massive stable charged particle traversing it. This combined with the 
requirement of high $p_T$ is the so-called tracker method of discovery. 
In addition the excellent timing of the muon system enables a 
time-of-flight cut, since a massive particle will have smaller velocity 
usually than a muon and thus takes more time to reach the outer muon 
chambers. The combination of these two methods, tracker and 
time-of-flight, yields powerful constraints from the $\sqrt{s}=7\tev$ 
data. With $4.7\xfb$ of integrated luminosity, we can compare the 
cross-section vs.\ mass limits of~\cite{Halk:2012} to the cross-section 
computation in Figure~\ref{fig:topprimeCS}, and from extrapolation of these 
results conclude that there is a limit of quasi-stable $t'_1$ mass of 
$m_{t'_1}\gsim 950\gev$. We estimate that more than twice this sensitivity 
could be achieved at $14\tev$ LHC with more than $10\xfb$ of integrated luminosity.

\asubsection{Search for $b'$ at the LHC\label{subsec:bprimesearches}}

In addition to the $t_1'$, the ${\bf 10}+\overline{\bf 10}$ model has
exotic quarks $b'$ and $t_2'$. It is of particular interest to ask what are the sensitivities to 
$b'$ production at the LHC~\cite{Gopalakrishna:2011ef}, 
since its mass may be nearly that of the $t'_1$ fermion when $M_Q < M_U$, 
as seen in Figure \ref{fig:MQoMU}. 
Given that its production rate is nearly the same as that of a similar mass $t'_1$, 
due to QCD contributions dominating, we must ask how the LHC would find this state, which is almost
pure $SU(2)_L$-doublet in both its right- and left-handed components.

\begin{figure}[t]
\begin{minipage}[]{0.49\linewidth}
\begin{center}
\includegraphics[width=\linewidth,angle=0]{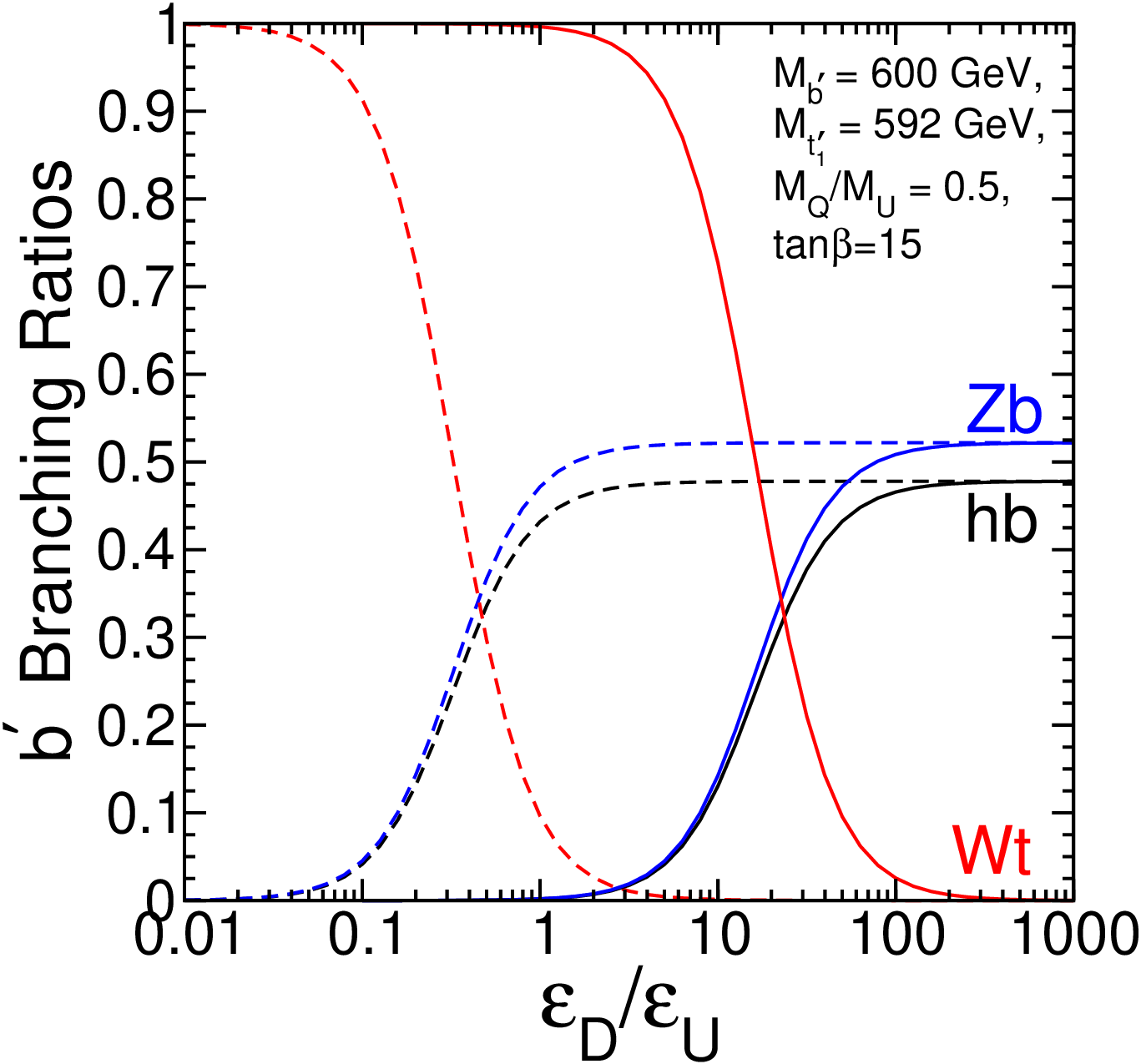}
\end{center}
\end{minipage}
\begin{minipage}[]{0.49\linewidth}
\begin{minipage}[]{0.04\linewidth}\end{minipage}
\begin{minipage}[]{0.90\linewidth}
\caption{\label{fig:BRbprime} Branching ratios for a nearly pure $SU(2)_L$ doublet $b'$
into $Wt$, $Zb$, and $hb$, 
as a function of $\epsd/\epsu$. Here $M_Q = 600$ GeV and $M_U = 1200$ GeV with $k=1$, so that
$m_{b'} = 600$ GeV and $m_{t_1'} = 592$ GeV. The solid lines have $\epsup = 1.1 \epsu$, and the
dashed lines have $\epsup = 0$.
The three-body decay $b' \rightarrow t_1' f \overline f$
through an off-shell $W$ boson is highly suppressed by kinematics, and is assumed to have a small
branching ratio.}
\end{minipage}\end{minipage}
\end{figure}

The $b'$ can have two-body decays through the mixing parameters $\epsu$, $\epsup$, or $\epsd$ to 
possible final states $Wt$, $Zb$ and $hb$.
Again, we are assuming that the exotic fermions couple only to the third generation weak eigenstates in 
order to tame potential flavor problems in the theory. The decay widths are calculated (in the more
general case of arbitrary mixing with the SM quarks) in the 
Appendix, eqs.~(\ref{eq:dgendecaytoW})-(\ref{eq:gendecaytoh}).
The relative fraction of the decays into $Wt$ 
versus $Zb$ and $ht$ depends to a large extent on the ratio
$\epsd/(\epsup\tan\beta)$. 
If this ratio is smaller than 1, or if $\epsd/\epsu$ is small,  
then $b'$ yields mostly $Wt$, if kinematically accessible. If the ratio is larger than 1, then the 
$b'$ yields mostly $Zb$ and $hb$.  
The branching ratios are shown in Figure \ref{fig:BRbprime} for a nearly pure doublet
$b'$, as a function of $\epsd/\epsu$ with $\epsup = 1.1 \epsu$ and $\epsu$ = 0.

\begin{figure}[t]
\begin{minipage}[]{0.55\linewidth}
\includegraphics[width=2.9in,angle=0]{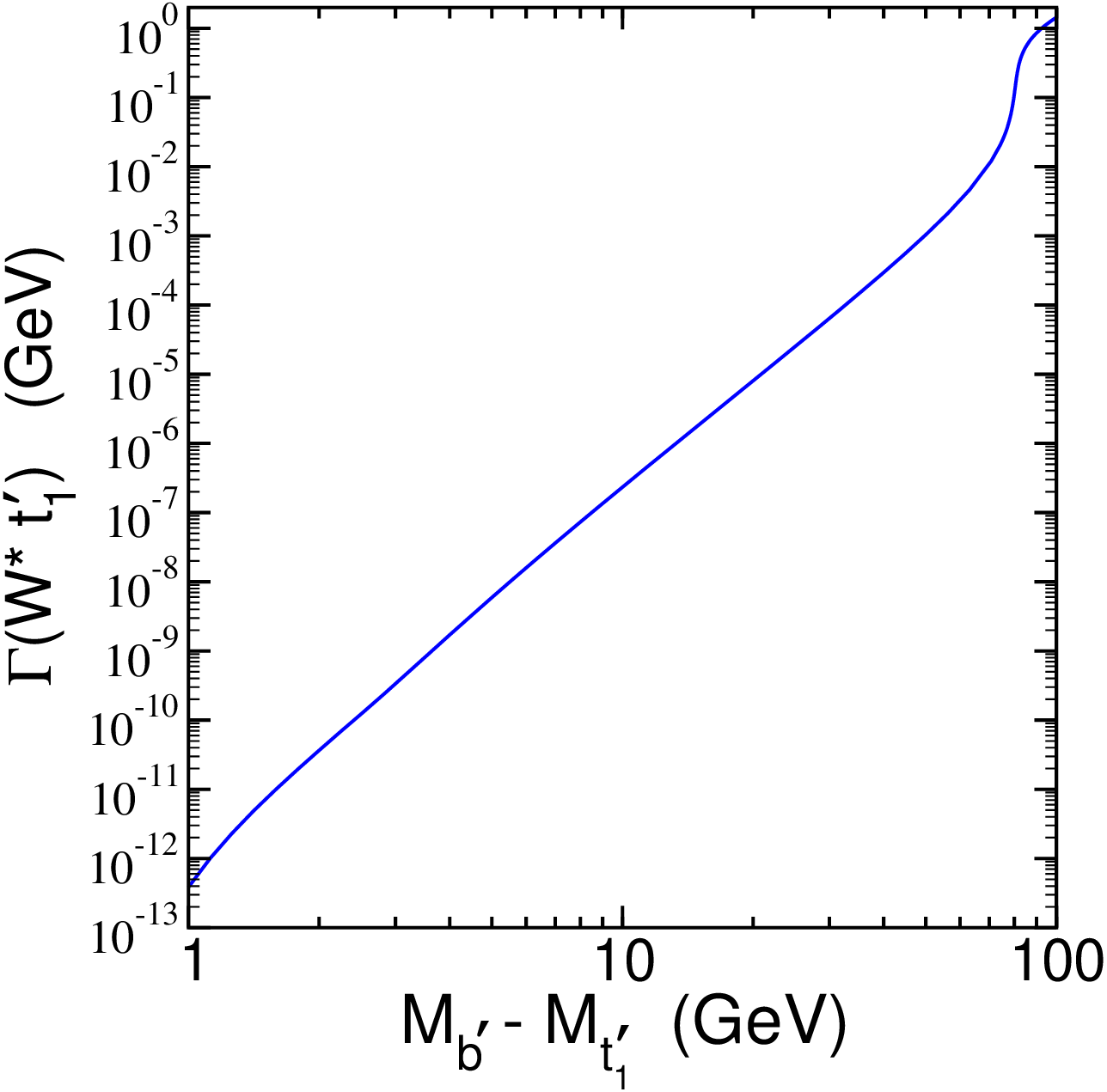}
\end{minipage}
~~~
\begin{minipage}[]{0.4\linewidth}
\caption{\label{fig:bprime3bodywidth}The decay width $\Gamma(b'\rightarrow W^{(*)}t_1')$
as a function of the mass difference $M_{b'} - M_{t_1'}$, assuming that $b'$ and
$t_1'$ form a nearly unmixed $SU(2)_L$ doublet, for $M_{b'} = 500$ GeV. 
(The results for the range 400 GeV $<M_{b'}<$ 1000 GeV are visually nearly indistinguishable from the line shown on this graph.)}
\end{minipage}
\end{figure}

It is also necessary to consider the flavor-preserving decay $b'\rightarrow W^{(*)}t_1'$.
If $M_{b'} > M_{t_1'} + M_W$, then this is an on-shell two-body decay and it will dominate. However,
for the case that $b'$ is mostly doublet, the decay will be three-body with the $W$ boson off shell.
The formula for this decay width is found in the 
Appendix, eq.~(\ref{eq:bprime3bodywidth}). In Figure \ref{fig:bprime3bodywidth}, we show this
width for the idealized case that $b'$ has pure doublet couplings to $W$ and $t_1'$,
as a function of the mass difference $M_{b'} - M_{t_1'}$, which is the most crucial parameter.
For comparison, the two-body flavor-violating decay widths are approximately:
\beq
\Gamma_{b'} &=& \mbox{0.1 GeV} \left (\frac{M_{b'}}{\mbox{1000 GeV}}\right ) 
\left (\frac{\epsilon}{0.1}\right)^2
\eeq
for the cases $\epsilon =\epsup \sin\beta$, $\epsd = \epsu = 0$
and $\epsilon = \epsd \cos\beta$, $\epsu = \epsup = 0$,
and
\beq
\Gamma_{b'} = 9 \times 10^{-5}\>\mbox{GeV} 
\left (\frac{M_{b'}}{\mbox{1000 GeV}}\right ) 
\left (\frac{\mbox{1000 GeV}}{M_U}\right )^4
\left (\frac{\epsu \sin\beta}{0.1}\right)^2
\eeq
for $\epsup = \epsd = 0$. Thus, the decay $b' \rightarrow W^{(*)} t_1'$ may or may not dominate in this
case, with a strong dependence on both the mass difference and the mixing couplings. 

If $b'$ mostly decays into $Wt$ the current limits arise from a search by 
CMS~\cite{CMSbprimeWt} based on 4.9 fb$^{-1}$ of 
integrated luminosity, resulting in a limit $M_{b'} > 611$ GeV if $B(b'\rightarrow Wt) = 1$.
For a $b'$ quark decaying only into $Zb$, there is an ATLAS search \cite{ATLASbprimeZb}
based on 2.0 fb$^{-1}$ which results in $M_{b'} > 400$ GeV. In our case, we see from Figure \ref{fig:BRbprime}
that $B(b'\rightarrow Zb) = 0.5$ is a more likely scenario, in which case
the limit from \cite{ATLASbprimeZb} is about 360 GeV. However, the ATLAS analysis only
uses $Z \rightarrow e^+e^-$, so improvements can be expected both from using $\mu^+\mu^-$ and
more integrated luminosity. As in the case of $t_1'$, it would be useful to exploit the 
other decay modes in a comprehensive search strategy 
that allows the branching ratios to vary. In particular, the decay $b' \rightarrow hb$
will lead to a nice signal in which there are at least 4 potentially taggable $b$-jets. For example
$pp\rightarrow b'\bar b' \rightarrow (Zb)(hb) \rightarrow \ell^+\ell^-bbbb$ should make for a
background-free signal.

\asubsection{Search for $\tau'$ at the LHC\label{subsec:tauprimesearches}}

\begin{figure}[t]
\begin{minipage}[]{0.55\linewidth}
\includegraphics[width=3.2in,angle=0]{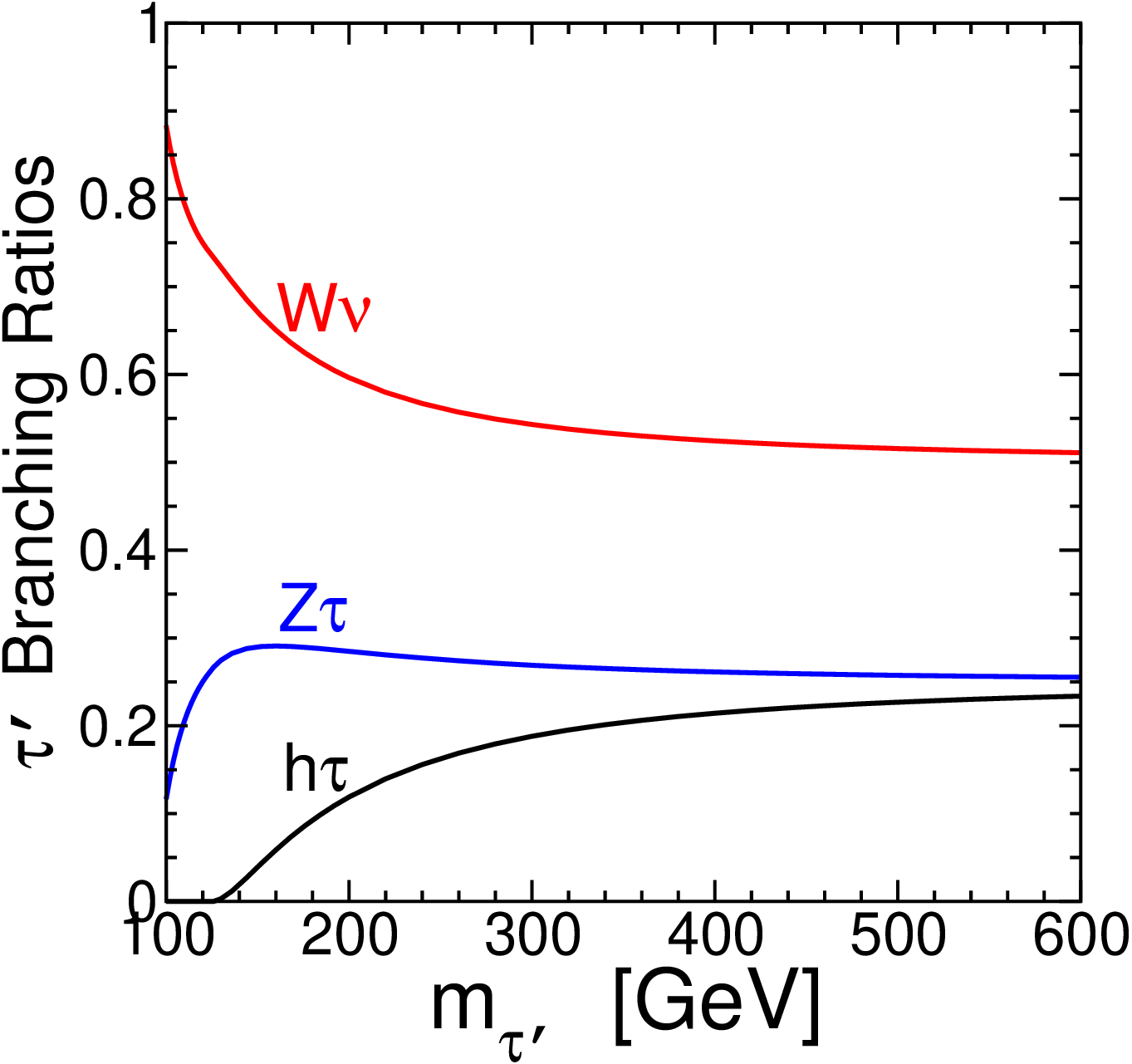}
\end{minipage}
~~~
\begin{minipage}[]{0.4\linewidth}
\caption{\label{fig:BRtaup}The branching ratio of
$\tau'$ into the final states $W\nu$, $Z\tau$, and $h\tau$, as a function of its mass,
assuming $M_h = 125$ GeV.}
\end{minipage}
\end{figure}

\begin{figure}[t]
\begin{minipage}[]{0.49\linewidth}
\begin{center}
\includegraphics[width=\linewidth,angle=0]{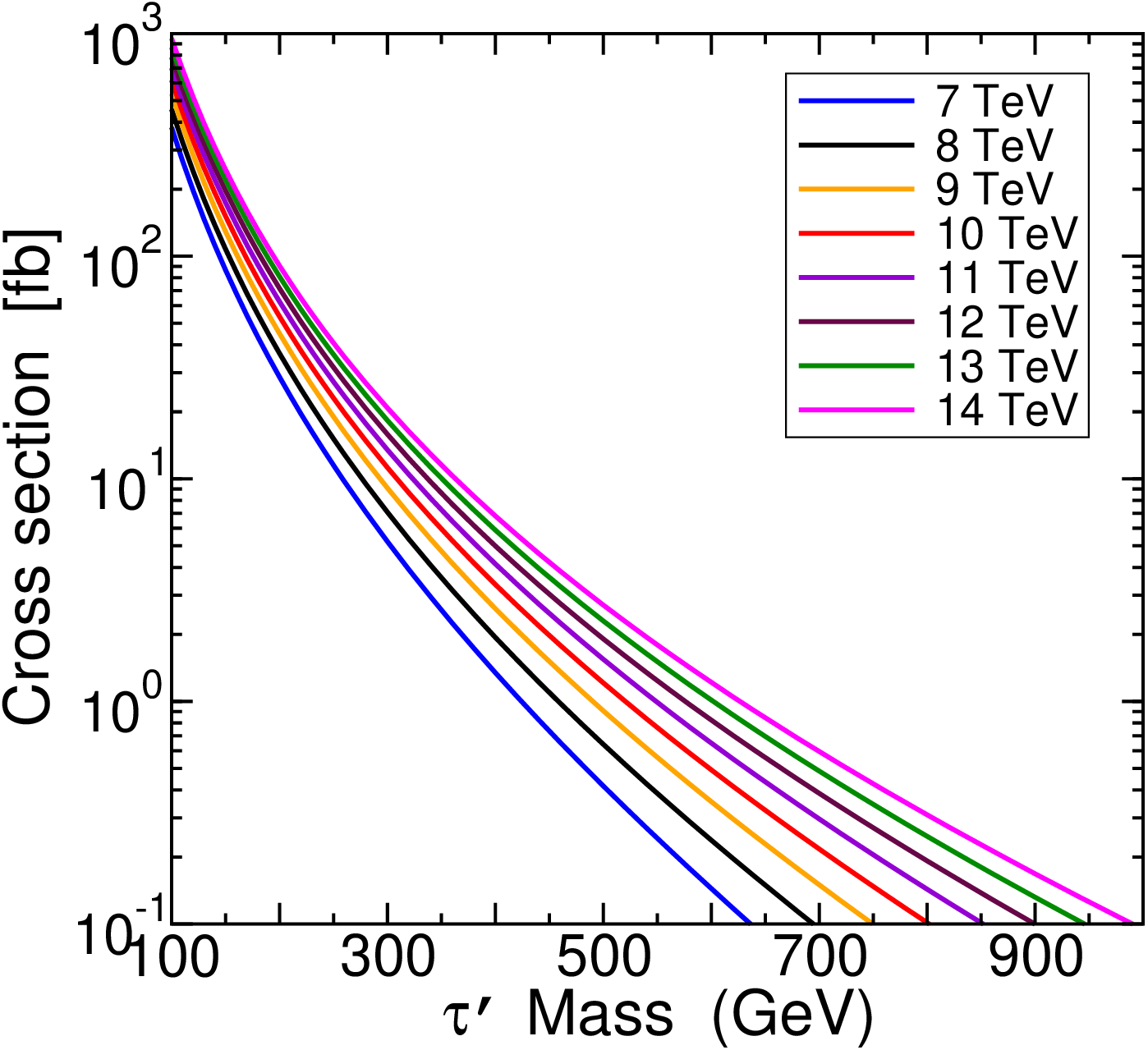}
\end{center}
\end{minipage}
\begin{minipage}[]{0.49\linewidth}
\begin{minipage}[]{0.04\linewidth}\end{minipage}
\begin{minipage}[]{0.90\linewidth}
\caption{\label{fig:tauprimecs} The total production cross-section
$\sigma(\tau^{\prime +}\tau^{\prime -})$ as a function of $m_{\tau'}$,
for $\sqrt{s} = $7, 8, 9, 10, 11, 12, 13, 14 TeV.}
\end{minipage}\end{minipage}
\end{figure}

The spectrum of the model we are considering also has an exotic lepton, the $\tau'$, whose 
quantum numbers are those of a right-handed electron with its vector 
complement. If the $\tau'$ decays promptly, it will be 
difficult to find.  Assuming that mixing is only with the $\tau$, the branching ratios to
final states $W\nu$, $Z\tau$ 
and $h\tau$ are shown in Figure \ref{fig:BRtaup}. 
The total width is determined by the $\epse$ coupling in 
eq.~(\ref{eq:WQUEmix}), but the branching ratios depend only on the $\tau'$ mass.  
The production cross-section is rather low for this state, 
being electroweak strength, as is shown in Figure~\ref{fig:tauprimecs}.
However, one can produce unique signatures such as $\ell^+\tau^-h+\ETmiss$ 
that could be exploited at the LHC to simultaneously find the Higgs boson 
and the $\tau'$. A full exploration of these 
prospects will be pursued in another publication.

If the $\tau'$ is stable, it can be searched for as a weakly 
interacting heavy stable charged particle. The lifetime depends only on
the mass and on the mixing coupling $\epse$, with
\beq
c\tau = \left (\frac{\mbox{1000 GeV}}{M_{\tau'}}\right )
        \left (\frac{10^{-7}}{\epse \cos\beta} \right )^2 \mbox{1.0 mm}.
\label{eq:ctautaup}
\eeq
We have taken the formal limit of $M_{\tau'} \gg M_h, M_Z, M_W$ here for simplicity,
and kinematic effects will lengthen $c\tau$ by a factor of a few if
the $\tau'$ mass is not far above 100 GeV. Note that there is also an enhancement
of the lifetime proportional to $1/\cos^2\!\beta$, so that the $\tau'$
could have a measurable decay length with $\epse$ as large as a few
times $10^{-5}$ if $\tan\beta$ is large. While this may seem quite small, it is larger than the
electron Yukawa coupling in the SM.

The implications for the 
search are like that of a quasi-stable stau boson NLSP. The search 
strategies are very similar to that described in the $t'$ section above, 
and so we shall not repeat it here. The result is that with $4.7\xfb$ of 
integrated luminosity, we can compare the cross-section vs.\ mass limits 
of~\cite{Halk:2012} to the cross-section computation in 
Figure~\ref{fig:tauprimecs}, and conclude that there is a limit of 
quasi-stable $\tau'$ mass of $m_{\tau'}\gsim 450\gev$. We estimate 
that with $100\xfb$ of integrated luminosity at a $14\tev$ LHC phase, the 
reach for quasi-stable $\tau'$ can extend up to nearly $1\tev$, which is well 
within the range of $\tau'$ masses expected for $M_h\hmass$ assuming that
$M_E = M_Q = M_U$ at the unification scale, as illustrated by the examples of
Figures \ref{fig:samples1} and \ref{fig:samples2}.

\section{Conclusion\label{sec:conclusion}}
\setcounter{equation}{0}
\setcounter{figure}{0}
\setcounter{table}{0}
\setcounter{footnote}{1}

A minimal GMSB model, with one $SU(5)$ ${\bf 5}+\overline{\bf 5}$ 
messenger pair, can explain a Higgs mass of $\hmass$ with even a sub-TeV 
gluino. This is accomplished by adding to the spectrum ${\bf 
10}+\overline{\bf 10}$ vector-like states, which then couple to the Higgs 
boson via the superpotential of eq.~(\ref{eq:WQUE}). The resulting 
radiative corrections can easily add 10 GeV or more to the light Higgs 
boson mass, which is crucial to achieve the $\hmass$ naturally, without 
requiring superpartners to be well above $1\tev$ or invoking ad hoc 
non-GMSB stop mixing. We have paid special attention to cases inspired by 
unification of masses, $M_Q=M_U$, and mixing couplings, 
$\epsu=\epsup=\epsd$, and we have characterized its parameter space. In 
this case, it is generic that the lightest exotic quark of the spectrum, 
$t'_1$, is mostly that with quantum numbers similar to right-handed top 
quark, with particular decay branching fractions.

The most obvious implication for this scenario is the existence of 
low-scale supersymmetry that should reveal itself at the LHC in the 
coming years. The searches for superpartners should follow the usual 
searches for GMSB models, which implies the existence of standard 
supersymmetry missing energy signatures with the addition of extra 
photons (or taus) if the NLSP is a neutralino (or stau) and decays promptly.
The signals may feature also either 
the presence of the lightest Higgs boson $h$ 
or a high 
multiplicity of taus due to wino decays 
in 
many events, depending on the messenger scale.
If the decays 
of the NLSP are not prompt, the collider phenomenology will be similar to 
that of standard scenarios with the neutralino being the 
LSP (i.e., stable NLSP on detector time scales), or there will be stable 
charged particle tracks from a quasi-stable charged NLSP stau.

The scenario under consideration in this paper yields additional 
phenomenological implications due to the existence of the $t'_{1,2}$ and 
$b'$ and $\tau'$ exotic fermion states. In previous sections we have 
explained that these states can also yield quasi-stable charged particle 
tracks, with sensitivity being nearly $1\tev$ already for $t'_1$ and 
nearly $0.5\tev$ for $\tau'$. If the decays are prompt, the limits are 
reduced. In that case the $t'_1$ pair-production signal is probably the 
most telling one for our scenario. We estimate sensitivity to the $t'_1$ 
mass to be higher than $1800\gev$ at 14 TeV LHC with more than $10\, {\rm 
fb}^{-1}$ of integrated luminosity. If seen with the properties described 
in the previous sections, the signal would point to the existence of 
extra vector-like quarks that lift the Higgs boson mass to $\hmass$.

\section*{Appendix: Exotic quark and lepton couplings to $W,Z,h$
and decay widths}
\label{appendixBR} \renewcommand{\theequation}{A.\arabic{equation}}
\setcounter{equation}{0}
\setcounter{footnote}{1}
\setcounter{subsubsection}{0}

This Appendix is devoted to a systematic description of the interactions 
of quarks and leptons to the massive weak bosons $W,Z,h$, allowing for arbitrary flavor violation, and to formulas
for the corresponding flavor-violating fermion decays. 

In the quark sector, we promote the third-family mixing parameters
$\epsu$, $\epsup$, and $\epsd$ to couplings 
$\epsilon^{\scriptscriptstyle U}_i$, 
$\epsilon^{\scriptscriptstyle U\prime}_i$, and 
$\epsilon^{\scriptscriptstyle D}_i$
respectively, where the index $i=1,2,3$ indicates the three SM generations. The masses 
for up-type and down-type quarks in the gauge-eigenstate basis are then 
respectively $5\times 5$ and $4\times 4$ matrices:
\beq
{\cal M}_u = \begin{pmatrix} y^u_{ij} v_u & \epsilon^{\scriptscriptstyle U}_i v_u & 0 \cr
                             0 & M_U & k' v_d \cr
                             \epsilon^{\scriptscriptstyle U\prime}_j v_u & k v_u & M_Q 
             \end{pmatrix},
\qquad\quad
{\cal M}_d = \begin{pmatrix} y^d_{ij} v_d & 0 \cr
                             \epsilon^{\scriptscriptstyle D}_j v_d & \phantom{x}-M_Q 
             \end{pmatrix},
\eeq
where $y^u_{ij}$ and $y^d_{ij}$ are the $3\times 3$ MSSM Yukawa couplings 
for the ordinary quarks, and the 0 entries appear by a choice of basis. 
One can now obtain 
the gauge-eigenstate two-component left-handed fermions\footnote{We use 
the two-component fermion notations of \cite{DHM}. The four-component 
Dirac fields are $\begin{pmatrix} u_i \cr \overline u_i^\dagger 
\end{pmatrix}$ and $\begin{pmatrix} d_i \cr \overline d_i^\dagger 
\end{pmatrix}$.} 
by applying unitary rotation matrices $L$, $R$, $L'$ and $R'$ on 
the mass eigenstates $u_i = (u,c,t,t_1',t_2')$ and 
$\overline u_i = (\overline u,\overline c, \overline t, \overline 
t_1',\overline t_2')$, and $d_i = (d,s,b,b')$ and $\overline u_i = 
(\overline d,\overline s, \overline b, \overline b')$, so that
\beq
L^T {\cal M}_u R &=& \mbox{diag}(m_u, m_c, m_t, m_{t_1'}, m_{t_2'}),
\\
L^{\prime T} {\cal M}_d R' &=& \mbox{diag}(m_d, m_s, m_b, m_{b'}).
\eeq
The first index of each of $L,R,L',R'$ is a gauge eigenstate index, and the second is a mass eigenstate index.\footnote{The notation used in \cite{Martin:2009bg} had a similar appearance but
different index orderings.}
Then the interaction Lagrangian for couplings of $W,Z,h$ to the quarks can be written as
\beq
-{\cal L}_{\rm int} &=& 
W_\mu^+ \Bigl (
g^W_{u_i^\dagger d_j}\, u^{\dagger}_i \sigmabar^\mu d_j \,+\, 
g^W_{\overline d^\dagger_i \overline u_j}\, \overline d^\dagger_i \sigmabar^\mu \overline u_j
\Bigr ) 
+ W_\mu^- \Bigl (
g^W_{d_j^\dagger u_i}\, d^{\dagger}_j \sigmabar^\mu u_i \,+\, 
g^W_{\overline u^\dagger_j \overline d_i}\, \overline u^\dagger_j \sigmabar^\mu \overline d_i
\Bigr )
\nonumber \\ &&
+ Z_\mu \Bigl (
g^Z_{u^\dagger_i u_j} \, u^\dagger_i \sigmabar^\mu u_j
\, + \,
g^Z_{\overline u^\dagger_i \overline u_j} \, \overline u^\dagger_i \sigmabar^\mu \overline u_j
\, + \,
g^Z_{d^\dagger_i d_j} \, d^\dagger_i \sigmabar^\mu d_j
\, + \,
g^Z_{\overline d^\dagger_i \overline d_j} \, \overline d^\dagger_i \sigmabar^\mu \overline d_j
\Bigr )
\nonumber \\ &&
+ \Bigl ( y^{h}_{u_i \overline u_j} h^0  u_i \overline u_j 
+ y^{h}_{d_i \overline d_j} h^0  d_i \overline d_j + {\rm c.c.} \Bigr ) ,
\eeq
where the couplings for the $W$ boson are
\beq
g^W_{u_i^\dagger d_j} = \bigl (g^W_{d_j^\dagger u_i} \bigr)^* 
&=& \frac{g}{\sqrt{2}} \biggl (
\sum_{k=1}^3 L_{ki}^* L'_{kj} + L_{5i}^* L'_{4j} \biggr ),
\label{eq:WLHquarks}
\\
g^W_{\overline d^\dagger_i \overline u_j} = \bigl ( 
g^W_{\overline u^\dagger_j \overline d_i} \bigr )^*
&=& \frac{g}{\sqrt{2}} R_{4i}^{\prime *} R_{5j},
\label{eq:WRHquarks}
\eeq
and the couplings for the $Z$ boson are
\beq
g^Z_{u^\dagger_i u_j} &=& \frac{g}{c_W} \biggl [
\Bigl (\frac{1}{2} - \frac{2}{3} s_W^2 \Bigr ) \delta_{ij} - \frac{1}{2} L^*_{4i} L_{4j} \biggr ],
\label{eq:ZLHup}
\\
g^Z_{\overline u^\dagger_i \overline u_j} &=&
\frac{g}{c_W} \biggl (
\frac{2}{3} s_W^2 \delta_{ij} - \frac{1}{2} R^*_{5i} R_{5j} \biggr ),
\label{eq:ZRHup}
\\
g^Z_{d^\dagger_i d_j} &=& \frac{g}{c_W}\Bigl ( -\frac{1}{2} + \frac{1}{3} s_W^2 \Bigr )  \delta_{ij} ,
\label{eq:ZLHdown}
\\
g^Z_{\overline d^\dagger_i \overline d_j} &=&
\frac{g}{c_W} \biggl (
-\frac{1}{3} s_W^2 \delta_{ij} + \frac{1}{2} R^{\prime *}_{4i} R'_{4j} \biggr ),
\label{eq:ZRHdown}
\eeq
and the couplings for the lightest Higgs scalar boson are
\beq
y^{h}_{u_i \overline u_j} &=& \frac{1}{\sqrt{2}} \cos\alpha \Bigl (
L_{ki} R_{nj} y^u_{kn} + L_{ki} R_{4j} \epsilon^{\scriptscriptstyle U}_k 
+ L_{5i} R_{nj} \epsilon^{\scriptscriptstyle U\prime}_n 
+ L_{5i} R_{4j} k \Bigr ) - \frac{1}{\sqrt{2}}\sin\alpha\, L_{4i} R_{5j} k',\phantom{xxx}
\\
y^{h}_{d_i \overline d_j} &=& -\frac{1}{\sqrt{2}} \sin\alpha \,\Bigl (
L'_{ki} R'_{nj} y^d_{kn} + L'_{4i} R'_{nj} \epsilon^{\scriptscriptstyle D}_n
\Bigr ).
\eeq
The couplings to the heavier neutral Higgs bosons $H^0$ and $A^0$ are obtained by the replacements
$(\cos\alpha, \sin\alpha) \rightarrow (\sin\alpha, -\cos\alpha)$ and $(i \cos\beta, -i\sin\beta)$
respectively.

Note that in the couplings of the $W$ boson in eq.~(\ref{eq:WLHquarks}), 
the role of the SM CKM matrix is played by the restriction to the $i,j=1,2,3$
subspace of the $5\times 4$ matrix
\beq
K_{ij} =  \sum_{k=1}^3 L_{ki}^* L'_{kj} + L_{5i}^* L'_{4j}.
\label{eq:ourCKMmatrix}
\eeq
Clearly, neither the full matrix $K_{ij}$ nor its restriction is unitary. 
(In the standard notation of \cite{RPP}, our $K_{11}$ is $V_{ud}$,
our $K_{23}$ is $V_{cb}$, etc.) Also, there is a nonzero coupling of the $W$ boson to right-handed
quarks in eq.~(\ref{eq:WRHquarks}), unlike in the SM. However, 
these flavor-violating effects do decouple as 
$\epsilon^{\scriptscriptstyle U}_i$, $\epsilon^{\scriptscriptstyle U\prime}_i$ and $\epsilon^{\scriptscriptstyle D}_i$ are taken to zero 
or as $M_Q$ and $M_U$ are taken very large.
Similarly, tree-level flavor-changing neutral currents of the $Z$ boson couplings appear as 
the three terms with explicit reference to the
exotic quarks' gauge-eigenstate indices $4,5$ in eqs.~(\ref{eq:ZLHup}), 
(\ref{eq:ZRHup}), and (\ref{eq:ZRHdown}).

The widths of kinematically allowed flavor-changing 
two-body decays of quarks involving weak bosons are given by 
\beq
\Gamma (u_i \rightarrow W d_j) &=& \frac{M_{u_i}}{32\pi}
   \lambda^{1/2}(1, r_W, r_j) \Bigl \{ [1 + r_j - 2 r_W + (1-r_j)^2/r_W] (
   |g^W_{u^\dagger_i d_j}|^2 + |g^W_{\overline u_i^\dagger \overline d_j}|^2)
\nonumber \\ &&
   + 12 \sqrt{r_j}\, {\rm Re} [g^W_{u^\dagger_i d_j} g^W_{\overline u_i^\dagger \overline d_j}]
   \Bigr \} ,
\\
\Gamma (d_i \rightarrow W u_j) &=& \frac{M_{d_i}}{32\pi}
   \lambda^{1/2}(1, r_W, r_j) \Bigl \{ [1 + r_j - 2 r_W + (1-r_j)^2/r_W] (
   |g^W_{u^\dagger_j d_i}|^2 + |g^W_{\overline u_j^\dagger \overline d_i}|^2)
\nonumber \\ &&
   + 12 \sqrt{r_j}\, {\rm Re} [g^W_{u^\dagger_j d_i} g^W_{\overline u_j^\dagger \overline d_i}]
   \Bigr \} , 
\label{eq:dgendecaytoW}   
\\
\Gamma (q_i \rightarrow Z q_j) &=& \frac{M_{q_i}}{32\pi}
   \lambda^{1/2}(1, r_Z, r_j) \Bigl \{ [1 + r_j - 2 r_Z + (1-r_j)^2/r_Z] (
   |g^Z_{q^\dagger_i q_j}|^2 + |g^Z_{\overline q_i^\dagger \overline q_j}|^2)
\nonumber \\ &&
   + 12 \sqrt{r_j}\, {\rm Re} [g^Z_{q^\dagger_i q_j} g^Z_{\overline q_i^\dagger \overline q_j}]
   \Bigr \} ,
\label{eq:gendecaytoZ}
   \\
\Gamma (q_i \rightarrow h q_j) &=& \frac{M_{q_i}}{32\pi}
   \lambda^{1/2}(1, r_h, r_j) \Bigl \{ [1 + r_j - r_h] (
   |y^h_{q_i \overline q_j}|^2 + |y^h_{q_j \overline q_i}|^2)
   + 4 \sqrt{r_j}\, {\rm Re} [y^h_{q_i \overline q_j} y^h_{q_j \overline q_i}]
   \Bigr \} ,\phantom{xxxx}
\label{eq:gendecaytoh}
\eeq
with $q=u$ or $d$, and
$\lambda(x,y,z) = x^2 + y^2 + z^2 - 2 x y - 2 x z - 2 y z$, and 
$r_X = M_X^2/M_{q_i}^2$.
The special cases considered in the text above are $t_1' \rightarrow Wb, Zt, ht$,
and $b'\rightarrow Wt, Zb, hb$, 
both obtained by taking $i=4$ and $j=3$, with the mixing of exotic 
quarks to SM quarks restricted to the third family.
The $t_1'$ decays were also discussed in \cite{Martin:2009bg} (using a different notation).

In the case of a $b'$ with $M_{b'} < M_{t_1'} + M_W$, 
there may be a competition between the two-body decays
above and the flavor-preserving three-body decay through an off-shell $W$ boson
to SM fermions.
In the approximation that flavor mixing between the exotic fermions and
the SM leptons and first and second-family quarks is neglected, we obtain
\beq
\Gamma(b' \rightarrow t_1'\bar f f') &=&
\frac{M_Q g^2 |V_{ff'}|^2}{1536\pi^3} 
\left [
(|g^W_{t_1^{\prime\dagger} b'}|^2 + |g^W_{\overline t_1^{\prime \dagger} \overline b'}|^2) F_1 +
12 \sqrt{r_{t_1'}}\, {\rm Re} [g^W_{t_1^{\prime\dagger} b'} g^W_{\overline t_1^{\prime \dagger} \overline b'}] F_2 \right ],
\label{eq:bprime3bodywidth}
\eeq
where $V_{ff'}$ is the standard CKM matrix for quarks ($f=u,c$ and $f'=d,s$) 
and 
is the Pontecorvo--Maki--Nakagawa--Sakata (PMNS) 
matrix for leptons ($f=$ neutrinos and $f'=e,\mu,\tau$), and
\beq
F_i &=& \int_{(\sqrt{r_f} + \sqrt{r_{f'}})^2}^{(1 - \sqrt{r_{t_1'}})^2} dx
\frac{\lambda^{1/2}(1,x,r_{t_1'})\lambda^{1/2}(1,r_f/x,r_{f'}/x)}{(x - r_W)^2 + \gamma r_W} f_i
\eeq
with $\gamma = \Gamma_W^2/M^2_{b'}$ and $r_X = M^2_X/M^2_{b'}$, and
\beq
f_1 &=& \left \{ x (1 + r_{t_1'} - x) \lambda(x,r_f,r_{f'}) + 
[(1- r_{t_1'})^2 - x^2]
[x^2 + x (r_f + r_{f'}) - 2 (r_f - r_{f'})^2]\right \}/x^2
\phantom{xxx}
\nonumber
\\
&&
+3 \left (\frac{1}{2 r_W^2} - \frac{1}{x r_W} \right )
[(1 - r_{t_1'})^2 - x (1 + r_{t_1'})][x (r_f + r_{f'}) - (r_f - r_{f'})^2],
\\
f_2 &=& x - r_f - r_{f'} + \left (\frac{1}{r_W} - \frac{x}{2 r_W^2} \right )
[x (r_f + r_{f'}) - (r_f - r_{f'})^2].
\eeq
This formula is also valid (and smoothly approaches) the two-body decay width  when the $W$ boson is on-shell,
in the narrow-width approximation $\gamma \ll r_W$,
\beq
\frac{1}{(x - r_W)^2 + \gamma r_W} &\rightarrow & \frac{\pi}{\sqrt{\gamma r_W}} \delta(x-r_W).
\eeq
The $F_1$ kinematic part of this result was obtained in \cite{Barger:1984jc}.

In the charged lepton sector, the $4\times 4$ mass matrix is
\beq
{\cal M}_e = \begin{pmatrix}
y^e_{ij} v_d & \epsilon^{\scriptscriptstyle E}_i v_d \cr
0 & M_E
\end{pmatrix}
\eeq
where $\epsilon^{\scriptscriptstyle E}_i$ is a mixing coupling, with 
$i,j=1,2,3$. The gauge eigenstate two-component fields are
related by unitary rotations $U,V$ acting on the mass eigenstate basis 
$(e,\mu,\tau,\tau')$ and $(\overline e,\overline \mu,\overline 
\tau,\overline \tau')$ in such a way that 
\beq U^T {\cal M}_e V = 
\mbox{diag}(m_e, m_\mu, m_\tau, m_{\tau'}). \eeq 
We assume that there are 
3 light Majorana mass eigenstate neutrinos $\nu_{1,2,3}$, related to the 
gauge eigenstates $\nu_{e,\mu,\tau}$ by a unitary PMNS matrix $N$ 
according to
\beq
\begin{pmatrix}
\nu_e \\[-6pt] \nu_\mu \\[-6pt] \nu_\tau \end{pmatrix}
 = N 
\begin{pmatrix}
\nu_1 \\[-6pt] \nu_2 \\[-6pt] \nu_3 
\end{pmatrix}.
\eeq
The weak boson interactions with mass-eigenstate leptons are
\beq
-{\cal L}_{\rm int} &=& 
W_\mu^+ \Bigl (
g^W_{\nu_i^\dagger e_j}\, \nu^{\dagger}_i \sigmabar^\mu e_j 
\Bigr ) 
+ W_\mu^- \Bigl (
g^W_{e_j^\dagger \nu_i}\, e^{\dagger}_j \sigmabar^\mu \nu_i 
\Bigr )
\nonumber \\ &&
+ Z_\mu \Bigl (
g^Z_{\nu^\dagger_i \nu_j} \, \nu^\dagger_i \sigmabar^\mu \nu_j
\, + \,
g^Z_{e^\dagger_i e_j} \, e^\dagger_i \sigmabar^\mu e_j
\, + \,
g^Z_{\overline e^\dagger_i \overline e_j} \, \overline e^\dagger_i \sigmabar^\mu \overline e_j
\Bigr )
+ \Bigl ( y^{h}_{e_i \overline e_j} h^0  e_i \overline e_j + {\rm c.c.} \Bigr ),\phantom{xx}
\eeq
where the couplings are
\beq
g^W_{\nu_i^\dagger e_j} = \bigl (g^W_{e_j^\dagger \nu_i} \bigr)^* 
&=& \frac{g}{\sqrt{2}} \sum_{k=1}^3 N_{ki}^* U_{kj},
\label{eq:Wleptons}
\\
g^Z_{\nu^\dagger_i \nu_j} &=& \frac{g}{2c_W} \delta_{ij},
\label{eq:Zneutrinos}
\\
g^Z_{e^\dagger_i e_j} &=& \frac{g}{c_W}
\Bigl [ \Bigl ( -\frac{1}{2} + s_W^2 \Bigr )  \delta_{ij} + \frac{1}{2} U^*_{4i} U_{4j} \Bigr ],
\label{eq:ZLHleptons}
\\
g^Z_{\overline e^\dagger_i \overline e_j} &=&
-\frac{g}{c_W} s_W^2 \delta_{ij},
\label{eq:ZRHleptons}
\\
y^{h}_{e_i \overline e_j} &=& -\frac{1}{\sqrt{2}} \sin\alpha \,\Bigl (
U_{ki} V_{nj} y^e_{kn} + U_{ki} V_{4j} \epsilon^{\scriptscriptstyle 
E}_k
\Bigr ).
\eeq
Note that unlike in the SM with 3 massive Majorana neutrinos, the effective PMNS
matrix ${\cal N}_{ij} = \sum_{k=1}^3 N_{ki}^* U_{kj}$ 
is not unitary in general. The other change from the
SM prediction comes from the 
left-handed coupling to the $Z$ boson in eq.~(\ref{eq:ZLHleptons}).
This deviation from lepton universality is small in the limits that $\epsilon^{\scriptscriptstyle E}_i$ is small
or $M_E$ is large.

The resulting two-body decay widths for $\tau'$ are
\beq
\Gamma (\tau' \rightarrow W \nu_j) &=& \frac{M_{\tau'}}{32\pi}
    (1 - r_W)^2 (2 + 1/r_W) |g^W_{\nu^\dagger_j e_4}|^2, 
\\
\Gamma (\tau' \rightarrow Z e_j) &=& \frac{M_{\tau'}}{32\pi}
   (1 - r_Z)^2 (2 + 1/r_Z) |g^Z_{e^\dagger_4 e_j}|^2 ,
\\
\Gamma (\tau' \rightarrow h e_j) &=& \frac{M_{\tau'}}{32\pi}
   (1 - r_h)^2 (
   |y^h_{e_4 \overline e_j}|^2 + |y^h_{e_j \overline e_4}|^2),
\eeq
where the $e_j = e,\mu,\tau$ lepton mass is neglected for kinematic purposes, and the
first decay should be summed over $j=1,2,3$ when the neutrinos are not observed. In the numerical
example in this paper and in \cite{Martin:2009bg}, the special case is taken in which 
$\epsilon^{\scriptscriptstyle E}_j$ coupling is only 
non-zero for $j=3$, so that electrons and muons do not mix with the $\tau'$, and only the decays
$\tau' \rightarrow W\nu, Z\tau, h \tau$ occur.
   
\bigskip \noindent 
{\it Acknowledgments:} 
The work of SPM was supported in part by the National Science Foundation grant 
number PHY-1068369. 
%


\end{document}